\documentclass{elsart}
\usepackage[square,comma]{natbib}
\usepackage{graphicx}
\usepackage{pxfonts}
\usepackage{lineno}
\usepackage{amssymb}

\journal{}

\begin{document}

\thispagestyle{empty}
\begin{Large}
\textbf{DEUTSCHES ELEKTRONEN-SYNCHROTRON}

\textbf{\large{Ein Forschungszentrum der Helmholtz-Gemeinschaft}\\}
\end{Large}

DESY 12-082

May 2012

\begin{eqnarray}
\nonumber &&\cr \nonumber && \cr \nonumber &&\cr
\end{eqnarray}
\begin{eqnarray}
\nonumber
\end{eqnarray}
\begin{center}
\begin{Large}
\textbf{Conceptual design of an undulator system for a dedicated
bio-imaging beamline at the European X-ray FEL}
\end{Large}
\begin{eqnarray}
\nonumber &&\cr \nonumber && \cr
\end{eqnarray}

\begin{large}
Gianluca Geloni,
\end{large}
\textsl{\\European XFEL GmbH, Hamburg}
\begin{large}

Vitali Kocharyan and Evgeni Saldin
\end{large}
\textsl{\\Deutsches Elektronen-Synchrotron DESY, Hamburg}
\begin{eqnarray}
\nonumber
\end{eqnarray}
\begin{eqnarray}
\nonumber
\end{eqnarray}
ISSN 0418-9833
\begin{eqnarray}
\nonumber
\end{eqnarray}
\begin{large}
\textbf{NOTKESTRASSE 85 - 22607 HAMBURG}
\end{large}
\end{center}
%\end{widetext}
\clearpage
\newpage

\begin{frontmatter}



\title{Conceptual design of an undulator system for a dedicated bio-imaging beamline at the European X-ray FEL}


\author[XFEL]{Gianluca Geloni\thanksref{corr},}
\thanks[corr]{Corresponding Author. E-mail address: gianluca.geloni@xfel.eu}
\author[DESY]{Vitali Kocharyan}
\author[DESY]{and Evgeni Saldin}

\address[XFEL]{European XFEL GmbH, Hamburg, Germany}
\address[DESY]{Deutsches Elektronen-Synchrotron (DESY), Hamburg,
Germany}

\begin{abstract}
We describe a future possible upgrade of the European XFEL
consisting in the construction of an undulator beamline dedicated to
life science experiments. The availability of free undulator tunnels
at the European XFEL facility offers a unique opportunity to build a
beamline optimized for coherent diffraction imaging of complex
molecules, like proteins and other biologically interesting
structures. Crucial parameters for such bio-imaging beamline are
photon energy range, peak power, and pulse duration. Key component
of the setup is the undulator source. The peak power is maximized in
the photon energy range between $3$ keV and $13$ keV by the use of a
very efficient combination of self-seeding, fresh bunch and tapered
undulator techniques. The unique combination of ultra-high peak
power of $1$ TW in the entire energy range, and ultrashort pulse
duration tunable from $2$ fs to $10$ fs, would allow for single shot
coherent imaging of protein molecules with size larger than $10$ nm.
Also, the new beamline would enable imaging of large biological
structures in the water window, between $0.3$ keV and $0.4$ keV. In
order to make use of standardized components, at present we favor
the use of SASE3-type undulator segments. The number segments, $40$,
is determined by the tapered length for the design output power of
$1$ TW. The present plan assumes the use of a nominal electron bunch
with charge of $0.1$ nC. Experiments will be performed without
interference with the other three undulator beamlines. Therefore,
the total amount of scheduled beam time per year is expected to be
up to $4000$ hours.
\end{abstract}

%\begin{keyword}
%%edge radiation \sep near-field \sep electron-bunch diagnostics
%%\sep x-ray free-electron laser (XFEL)
%%\PACS 41.60.Cr \sep 42.25.-p \sep 41.75.-Ht
%\end{keyword}
%
\end{frontmatter}

% main text

%\linenumbers

%1.

\section{\label{sec:intro} Introduction}

Structural biology aims at the understanding of the biological
function of proteins by studying their three-dimensional structure.
The major method for determining such macromolecular
three-dimensional structure is X-ray crystallography
\cite{DREN,RUPP}. Requirements on the crystal samples set limits to
structural studies of biological systems with atomic resolution. In
fact, many molecules fail to form crystals. The development of XFELs
promises to open up new areas in life science by allowing structure
determination without the need for crystallization. In fact, as
suggested in \cite{HAJD}, sufficiently short and intense pulses from
X-ray lasers may allow for the imaging of single protein molecules.

This article describes a possible future upgrade of the European
XFEL. We present a study for a dedicated beamline for
single-biomolecular imaging. The main idea is to use one of the free
undulator tunnels of the European XFEL for providing the user
community with the three SASE1, SASE2 and SASE3 lines, with the
addition of a fourth beamline where the combination of $1$ TW peak
power in the energy range between $3$ keV and $13$ keV, and
tunability of the pulse duration from $2$ fs to $10$ fs will provide
significantly better conditions for single shot coherent imaging of
protein molecules than at other XFEL facilities.

The European XFEL equipped with a dedicated bio-imaging beamline
represents a development half way in between a proof-of-principle
and a fully dedicated bio-imaging XFEL facility with high-rate of
protein structure determination. The main advantage of the new
beamline is the operation point at $1$ TW with $2$ fs-$5$ fs long
pulses in the particular energy range between $3$ keV and $5$ keV,
where the diffraction signal is strong. Operation at $1$ TW in this
peculiar energy range will not be accessible to other XFEL
facilities at least until the end of the next decade. These
characteristics would enable new exciting possibilities for coherent
imaging of protein molecules with size larger than $10$ nm.

%2.

\section{\label{sec:aims} Scientific aims}

Understanding the biological function of molecules such as proteins,
requires the knowledge of their atomic structure. Today, X-ray
crystallography is a major technique to determine the atomic
structure of proteins. However, there is still a serious limiting
factor: in fact, many molecules fail to form crystals, which
prevents X-ray protein crystallography from fully realizing its
potential \cite{DREN,RUPP}. In particular, it is known that about
$40 \%$ of biomolecules, in particular membrane proteins, do not
crystallize. Hence, the scattering signal is too small to be useful.
Increasing the number of photons in the pulse without decreasing its
duration does not help, since radiation damage severely limits
structural studies of single biomolecules.

The development of XFELs promises to open up new area in life
science by allowing structure determination of
difficult-to-crystallize proteins. Single biomolecules are injected
into a pulsed X-ray beam. Imaging of a single biomolecule can then
be performed through coherent diffraction experiments \cite{HAJD}.
The elastically scattered photons are recorded by a detector array
placed downstream. The scattered pattern contains information in the
reciprocal space, from which the object may be reconstructed
\cite{FINE}. The advantage of using an XFEL source is in the
ultrashort time scale of the XFEL pulse. The scattering pattern can
be recorded before the molecule is destroyed. This effect is known
as diffraction before destruction \cite{NEUT}. Simulations and
experiments suggest that radiation damage will not significantly
disrupts the initial positions of atoms in the first few
femtoseconds of exposure and that the object may be reconstructed
\cite{SEIB}.

Focusing optics is used to closely match the size of the X-ray beam
to the size of the sample, and to maximize the signal level. The
FWHM focal spot size should be about $5-10$ times larger than the
sample size. For example, a spot size of $100$ nm is good for sample
sizes smaller than $10$ nm. The total number of different proteins
smaller than $10$ nm is about $10^{5}$. Structure determination of
many of these (about $80000$) has been successfully performed
\cite{BANK}. The largest number of biologically interesting
structures are in the size range between $10$ nm and $600$ nm. The
total number of different structures of this kind is over $10^{9}$.
Only very few of these structures have been determined \cite{HAJ2}.

Let us estimate the minimum number of required photons per XFEL
pulse for successful imaging of single biomolecules. An incident
photon flux of $10^{21}$ photons$/$mm$^2$ is sufficient for
obtaining a reasonably good image of a protein molecule like a
lysozyme (14 kD) with a size of about $6$ nm at photon energy of
$12$ keV \cite{IKED}. The  "diffraction-before-destruction" method
requires a pulse duration shorter than $10$ fs. The number of
photons per pulse required for a typical experiments is about $2
\cdot 10^{12}$. An important problem parameter is the number of
scattered photons per effective pixel, which is proportional to
$\lambda^2$, where $\lambda$ is the radiation wavelength. As a
result, lower photon energies result in a stronger diffraction
signal. However,  a decrease in the photon energy limits the
resolution. A balance may be reached in the photon energy range
between $3$ keV and $5$ keV.

The main application of the beamline proposed in this work is for
single shot imaging of individual protein molecules. It will provide
coherent X-ray radiation with properties superior to the existing
and planned XFEL sources. The X-ray beam will be delivered in
ultrashort pulses with a duration between $2$ fs and $10$ fs, with
$1$ TW peak power, and within a very wide photon energy range
between $3$ keV and $13$ keV, which covers the K-edge energies from
sulfur to selenium. The beamline is also designed to provide photon
energies down to water window at $0.3$ keV.  We believe that our
developments will open exciting possibilities for structure
determination of protein molecules with size larger than $10$ nm and
in the $3$ keV - $5$ keV photon energy range of operation.

%3.

\section{\label{sec:concept} Basic concept of the bio-imaging beamline}

For the realization of the bio-imaging beamline we propose to use
the same undulator technology optimized for the generation of soft
X-rays at the European XFEL. The installation and commissioning of
the new beamline can take place gradually. In the beginning, the new
beamline would just extend the soft X-ray (SASE3) beamline and take
advantage of the long XTD4 tunnel. An additional undulator composed
by $19$ cells would be added into the $300$ m-long XTD4 tunnel. This
undulator would extend the existing SASE3 line, composed by $21$
cells, and have the same period of SASE3, in order to obtain a total
cell number of $40$. With this, SASE3 would be de facto converted
into a bio-imaging beamline. Then, a new soft X-ray beamline,
identical to the SASE3 baseline, could be installed in the free
$150$ m-long XTD3 tunnel.

The combination of self-seeding and tapered undulator techniques
would allow to meet the design output peak power of $1$ TW. The
bio-imaging beamline would be equipped with two different
self-seeding setup, one to provide monochromatization in the soft
X-ray range, between $0.3$ keV and $1.7$ keV, and one to provide
monochromatization in the hard X-ray range between $8$ keV and $13$
keV. A border region exists between the soft and the hard X-ray
range. In fact, the X-ray optics for the self-seeding scheme adopted
in one range is not suitable for the other, and vice versa.
Self-seeding schemes with crystal monochromators work above $7$ keV,
where diamond crystals can be used. Crystals with the right lattice
parameters and with sufficiently narrow linewidth are difficult to
be obtained. Diffraction gratings with single metal layer coating
(Au, Pt) have good first order efficiency at grazing angles up to
$2$ keV, but at higher energies their throughput is usually too low.
Such gratings are conceived for the soft- X-ray self-seeding setup
at the LCLS. A combination of these two self-seeding setups with a
tunable gap, $40$ cells-long undulator would allow to meet the
design goal of $1$ TW for photon energies ranging between $0.3$ keV
and $1.7$ keV, and between $8$ keV and $13$ keV.

However, the users of the bio-imaging beamline mainly wish to
investigate their samples in the energy range between $3$ keV and
$5$ keV, where the diffraction signal is strong. Finding a solution
suitable for this spectral range is major challenge for self-seeding
designers. A promising approach in this direction consists in using
a fresh bunch technique in combination with self-seeding and
tapering techniques. We will show how installation of an additional
chicane behind the soft X-ray self-seeding setup enables an output
peak power around $1$ TW for photon energy ranges between $3$ keV
and $5$ keV. The combination self-seeding, fresh bunch and tapered
techniques has the advantage that it does not need any significant
change in  the grating monochromator design proposed by the LCLS
crew, which is still under active investigation
\cite{FENG,FENG2,COCC0}. It should be noted that the duration of the
X-ray pulse is important for bio-imaging experiments. In particular,
for energies smaller than $5$ keV, the pulse duration should be
shorter than $5$ fs. We will show how the pulse duration can be
tuned within $2$ fs and $10$ fs, still operating with nominal
electron bunch distributions.

The two design electron energies for the new beamline will be $10$
GeV and $17$ GeV, at a design charge of 0.1 nC. When the bio-imaging
beamline will run at $10$ GeV electron energy, it will be necessary
to run all the FEL lines at the same energy, although this may not
be the preferred mode of operation for SASE1 and SASE2 users.
However, this electron energy is still one of the nominal electron
energies for the European XFEL, and is compatible with the operation
of SASE1 and SASE2 in the X-ray energy range up to $12$ keV.

After conversion of the SASE3 beamline as described in this report
there will be several future upgrade options for a dedicated
bio-imaging beamline. To avoid interference with other beamlines,
the photon energy range of the self-seeding setup with grating
monochromator can be extended up to $5$ keV, and the minimal design
electron energy can be increased up to $14$ GeV. A promising
approach in this direction consists in using diffraction gratings
with very shallow groove depths, capable of working at very small
angles of incidence. Recently, a holographic laminar grating has
been realized and characterized in the energy range between $3$ keV
and $5$ keV at the incidence angle of $0.5$ degree, showing up to
$10 \%$ efficiency \cite{COCC}.

\subsection{Setup description}

\begin{figure}[tb]
\includegraphics[width=1.0\textwidth]{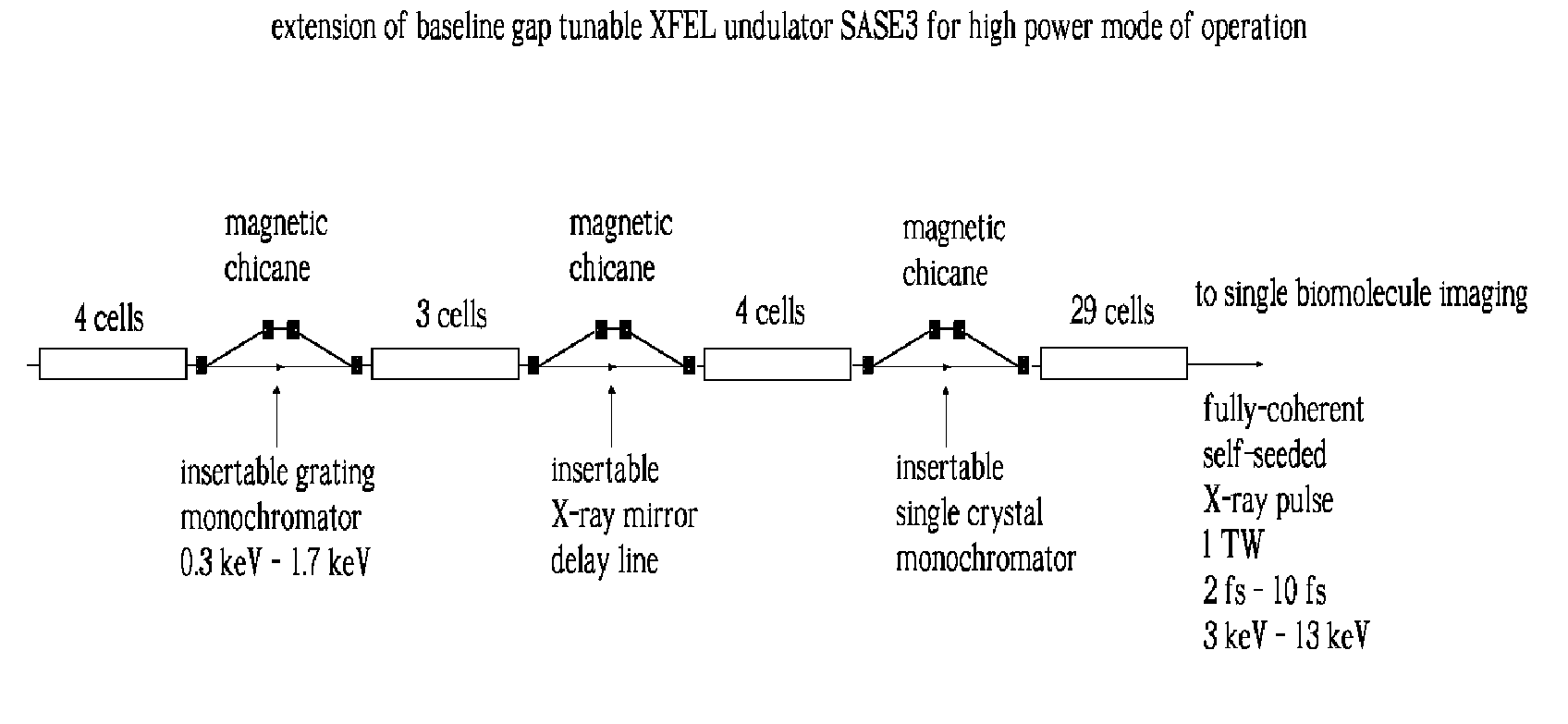}
\caption{Design of the undulator system for the bio-imaging
beamline. The method exploits a combination of self-seeding, fresh
bunch, and undulator tapering technique. Each magnetic chicane
accomplishes three tasks by itself. It creates an offset for
monochromator or X-ray mirror delay line installation, it removes
the electron microbunching produced in the upstream undulator, and
it acts as a magnetic delay line.} \label{biof3}
\end{figure}
%

%4
\begin{figure}[tb]
\includegraphics[width=1.0\textwidth]{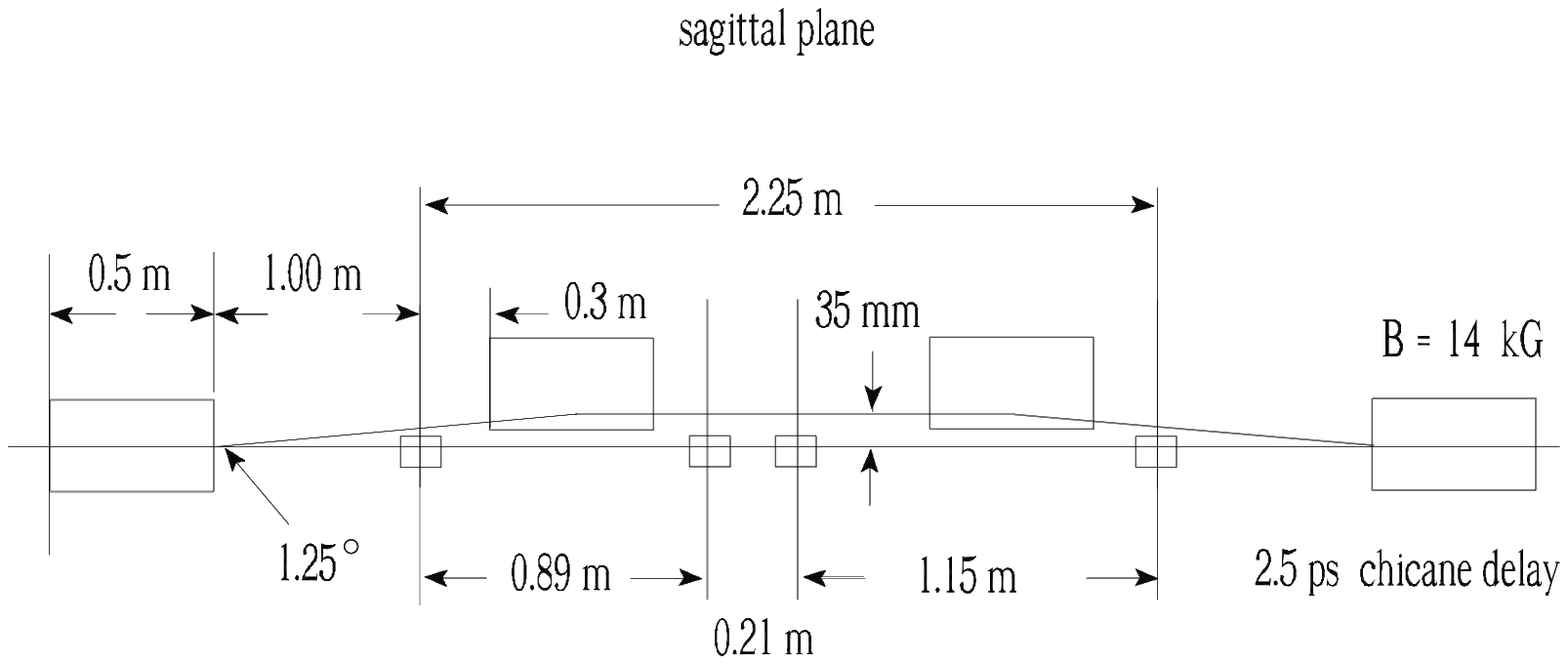}
\caption{Plan view of the self-seeding setup with compact grating
monochromator originally proposed at SLAC \cite{FENG,FENG2}.}
\label{biof4}
\end{figure}
%

%5
\begin{figure}[tb]
\includegraphics[width=1.0\textwidth]{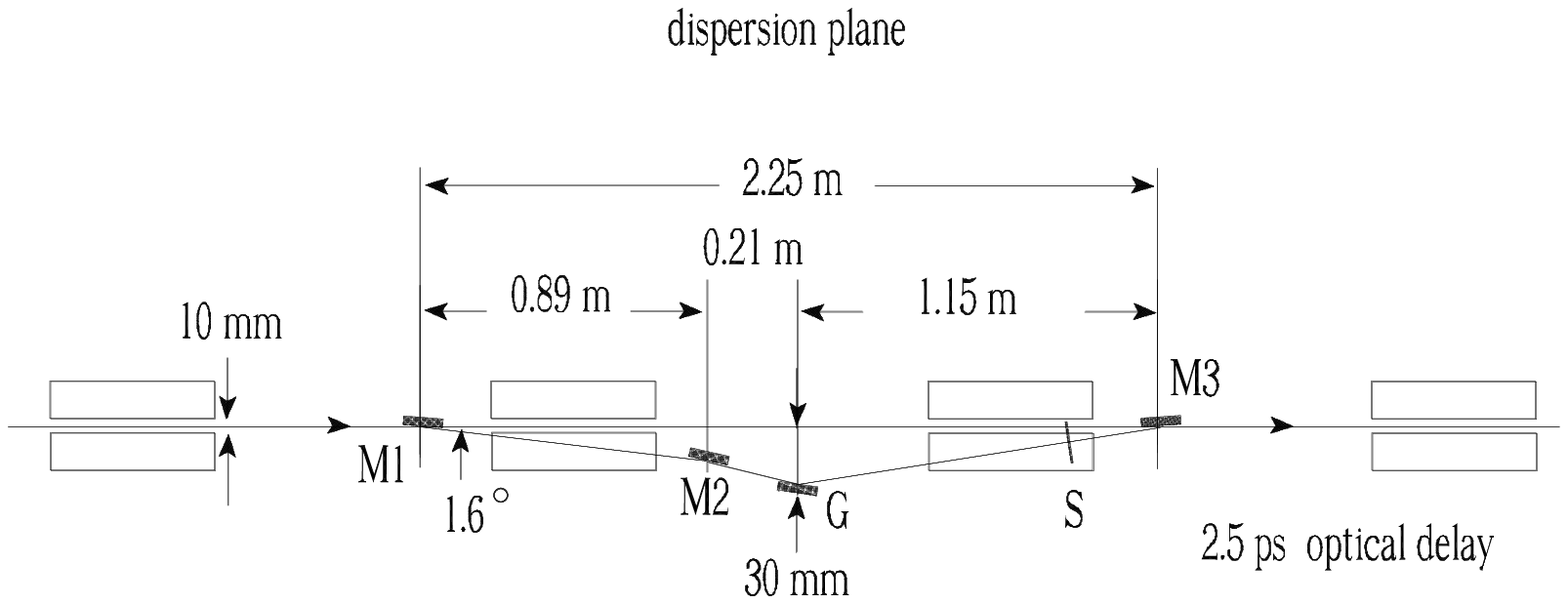}
\caption{Elevation view of the self-seeding setup with compact
grating monochromator originally proposed at SLAC
\cite{FENG,FENG2}.} \label{biof5}
\end{figure}
%

%6
\begin{figure}[tb]
\includegraphics[width=1.0\textwidth]{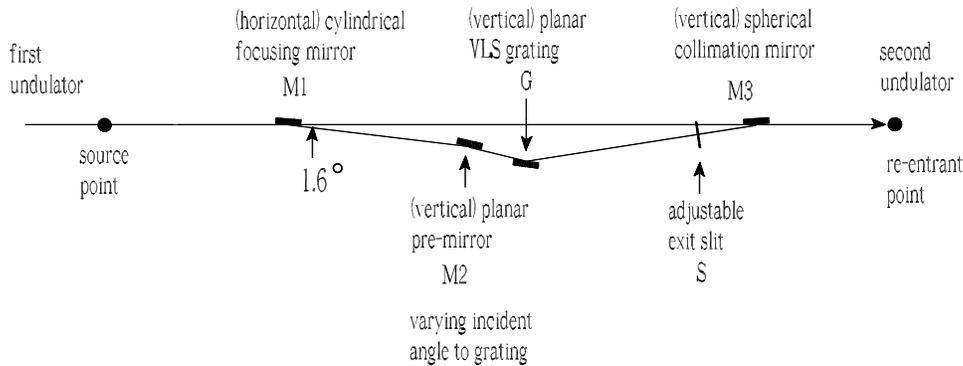}
\caption{Optics for the compact grating monochromator originally
proposed at SLAC \cite{FENG,FENG2} for the soft X-ray self-seeding
setup.} \label{biof6}
\end{figure}
%

%1
\begin{figure}[tb]
\includegraphics[width=1.0\textwidth]{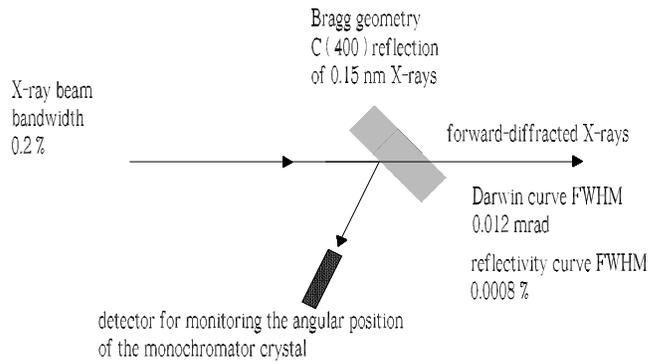}
\caption{X-ray optics for compact crystal monochromator originally
proposed in \cite{OURY5b} for a hard X-ray self-seeding setup.}
\label{biof1}
\end{figure}
%
%2
\begin{figure}[tb]
\includegraphics[width=1.0\textwidth]{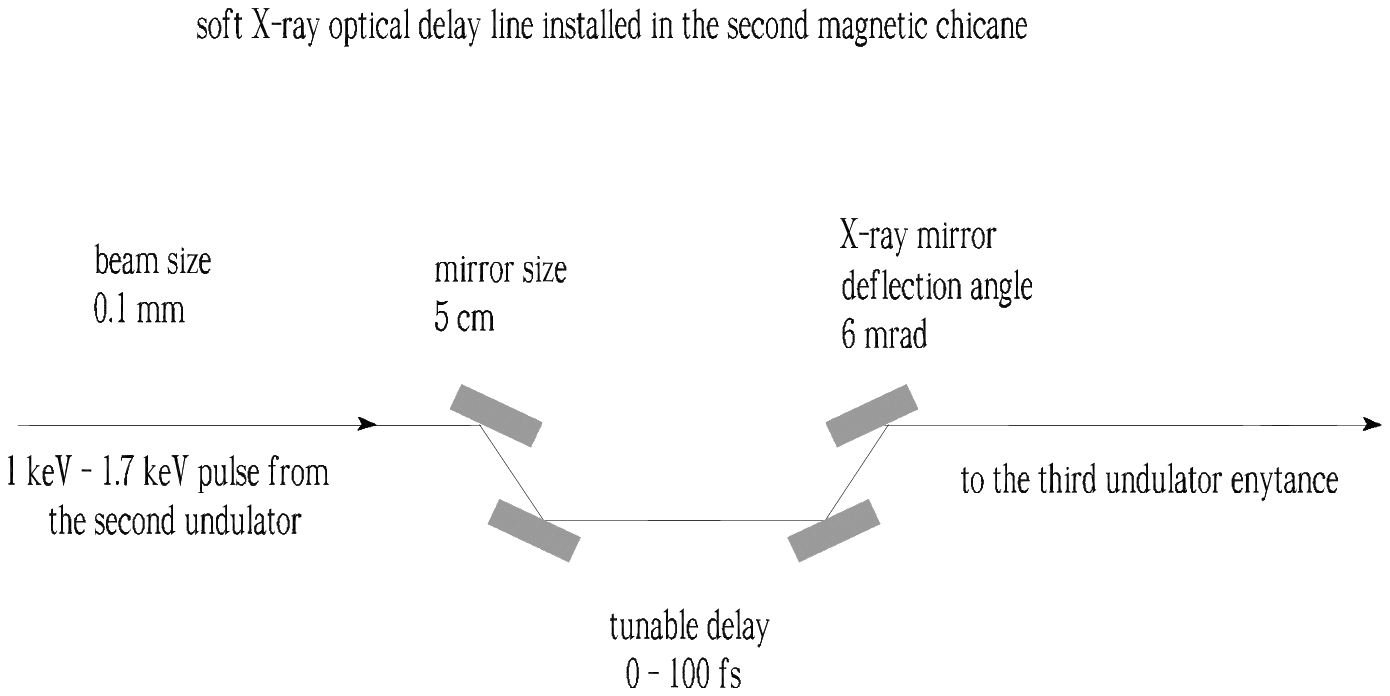}
\caption{X-ray optical system for delaying the soft X-ray pulse with
respect to the electron bunch. The X-ray optical system can be
installed within the second magnetic chicane of the fresh bunch
setup. } \label{biof2}
\end{figure}
Self-seeding is a promising approach to significantly narrow the
SASE bandwidth and to produce nearly transform-limited X-ray pulses
\cite{SELF}-\cite{WUFEL2}. In its simplest configuration the
self-seeding setup in the hard X-ray regime consists of two
undulators separated by photon monochromator and an electron bypass
beamline, typically a 4-dipole chicane. The two undulators are
resonant at the same radiation wavelength. The SASE radiation
generated by the first undulator passes through the narrow-band
monochromator, thus generating a transform-limited pulse, which is
then used as a coherent seed in the second undulator. Chromatic
dispersion effects in the bypass chicane smear out the microbunching
in the electron bunch produced by the SASE lasing in the first
undulator. Electrons and monochromatized photon beam are recombined
at the entrance of the second undulator, and the radiation is
amplified by the electron bunch in the second undulator, until
saturation is reached. The required seed power at the beginning of
the second undulator must dominate over the shot noise power within
the gain bandpass, which is order of a few kW.

Despite the unprecedented increase in peak power of the X-ray pulses
for SASE X-ray FELs (see e.g. \cite{LCLS2}), some applications,
including single biomolecule imaging, require still higher photon
flux. The most promising way to extract more FEL power than that at
saturation is by tapering the magnetic field of the undulator
\cite{TAP1}-\cite{LAST}. Also, a significant increase in power is
achievable by starting the FEL process from a monochromatic seed
rather than from noise \cite{OURY3}-\cite{WUFEL2}. Tapering consists
in a slow reduction of the field strength of the undulator in order
to preserve the resonance wavelength, while the kinetic energy of
the electrons decreases due to the FEL process. The undulator taper
could be simply implemented as a step from one undulator segment to
the next. The magnetic field tapering is provided by changing the
undulator gap.

The proposed setup is not simple as discussed above and is composed
of four undulators separated by three magnetic chicanes as shown in
Fig. \ref{biof3}. These undulators consist of $4$, $3$, $4$ and $29$
undulator cells, respectively. Each magnetic chicane is compact
enough to fit one undulator segment. The installation of chicanes
does not perturb the undulator focusing system.  The implementation
of the self-seeding scheme for soft X-ray and hard X-ray would
exploit the first and the third magnetic chicane, respectively.
Both self-seeding setups should be compact enough to fit  one
undulator module.

%3

For soft X-ray self-seeding, the monochromator usually consists of a
grating \cite{SELF}. Recently, a very compact soft X-ray
self-seeding scheme has appeared, based on a grating monochromator
\cite{FENG,FENG2}. The proposed monochromator is composed of only
three mirrors and a rotational VLS grating. It is equipped with an
exit slit only. A preliminary design of the grating monochromator
adopts a constant focal-point mode in order to fix the slit
location. The delay of the photons is about $2.5$ ps. The
monochromator is continuously tunable in the photon energy range
between $0.3$ keV and $1.7$ keV.  The resolution is about $5000$.
The transmission of the monochromator beamline is close to $10 \%$.
The magnetic chicane delays the electron electron bunch accordingly,
so that the photon beam passing through the monochromator system
recombines with the same electron bunch. The chicane provides a
dispersion strength of about $2$ mm in order to match the optical
delay and also smears out the SASE microbunching generated in the
first $4$ cells of the undulator. The final design is under active
investigations, and is subject to many changes \cite{COCC0}. In
\cite{OSOF} we studied the performance of the above-described scheme
for the European XFEL upgrade. For bio-imaging beamline we adopt the
same design as in \cite{FENG,FENG2}. The layout of the bypass and of
the monochromator optics is schematically shown in Fig. \ref{biof4},
Fig. \ref{biof5} and Fig. \ref{biof6}.

For hard X-ray self-seeding, a monochromator usually consists of
crystals in the Bragg geometry. A conventional 4-crystal, fixed exit
monochromator introduces optical delay of, at least, a few
millimeters, which has to be compensated with the introduction of an
electron bypass longer than one undulator module. To avoid this
difficulty, a simpler self-seeding scheme was proposed in
\cite{OURY5b}, which uses the transmitted X-ray beam from the single
crystal to seed the same electron bunch, Fig. \ref{biof1}. Here we
propose to use a diamond crystal with a thickness of $0.1$ mm. For a
symmetric C(400) Bragg reflection, it will be possible to cover the
photon energy range from $8$ keV to $13$ keV.

One of the main technical problems to provide bio-imaging
capabilities in $3$ keV - $5$ keV photon energy range is the
extension of the self-seeding grating monochromator design from $2$
keV up to $5$ keV. Here we propose a method to get around this
obstacle. Our solution is based in essence on the fresh bunch
technique \cite{BZVI} and exploits the above described conservative
design of self-seeding setup based on a grating monochromator.  The
hardware requirement is minimal, and in order to implement a fresh
bunch technique it is sufficient to install an additional magnetic
chicane at a special position behind the soft X-ray self-seeding
setup. The function of this second chicane is both to smear out the
electron bunch microbunching, and to delay the electron bunch with
respect to the monochromatic soft X-ray pulse produced in the second
undulator. In this way, only half of the electron bunch is seeded,
and saturates in the third undulator. Finally, the second half of
the electron bunch, which remains unspoiled, is seeded by the third
harmonic of the monochromatic radiation pulse generated in the third
undulator, which is also monochromatic. The final delay of the
electron bunch with respect to the seed radiation pulse can be
obtained with a third, hard X-ray self-seeding magnetic chicane,
which in this mode of operation is simply used to provide magnetic
delay. The monochromatic third harmonic radiation pulse used as seed
for the unperturbed part of the electron bunch is in the GW power
level, and the combination of self-seeding and fresh bunch technique
is extremely insensitive to non-ideal effects. The final undulator,
composed by $29$ cells, is tuned to the third harmonic frequency,
and is simply used to amplify the X-ray pulse up to $1$ TW output
power level.

It should be noted that the delay of the radiation pulse with
respect to the electron bunch in the second and in the third chicane
has a different sign. The solution to this problem is to install a
mirror chicane within the second magnetic chicane, as shown in Fig.
\ref{biof2}. The function of the mirror chicane is to delay the
radiation in the range between $1$ keV and $1.7$ keV relatively to
the electron bunch. The glancing angle of the mirrors is as small as
$3$ mrad. At the undulator location, the transverse size of the
photon beam is smaller than $0.1$ mm, meaning that the mirror length
would be just about $5$ cm. The single-shot mode of operation will
relax the heat-loading issues. The mirror chicane can be built in
such a way to obtain a delay of the radiation pulse of about
$23~\mu$m. This is enough to compensate a bunch delay of about
$20~\mu$m from the magnetic chicane, and to provide any desired
shift in the range between $0~\mu$m and $3~\mu$m. Note that for the
European XFEL parameters, $1$ nm microbunching is washed out with a
weak dispersive strength corresponding to an $R_{56}$ in the order
of ten microns. The dispersive strength of the proposed magnetic
chicane is more than sufficient to this purpose. Thus, the
combination of magnetic chicane and mirror chicane removes the
electron microbunching produced in the second undulator and acts as
a tunable delay line within $0~\mu$m and $3~\mu$m with the required
choice of delay sign.

%Altogether, the proposed setup is composed of four undulators
%separated by three magnetic chicanes as shown in Fig. \ref{biof3}.
%These undulators will respectively consist of $4$, $3$, $4$ and $29$
%undulator cells, each of them $6$ m-long. Each magnetic chicane is
%compact enough to fit a single $5$ m-long undulator segment. The
%installation of these chicanes does not perturb the undulator
%focusing system.  The implementation the self-seeding scheme for
%soft and hard X-ray would exploit the first and third magnetic
%chicane, respectively, whereas the second chicane would enable the
%implementation of a fresh-bunch technique. Both self-seeding setups
%should be compact enough to fit one undulator module.

\subsection{\label{sub:verysoft} Generation of TW pulses in the $0.3$ keV - $1.7$  keV photon energy
range}

%7
\begin{figure}[tb]
\includegraphics[width=1.0\textwidth]{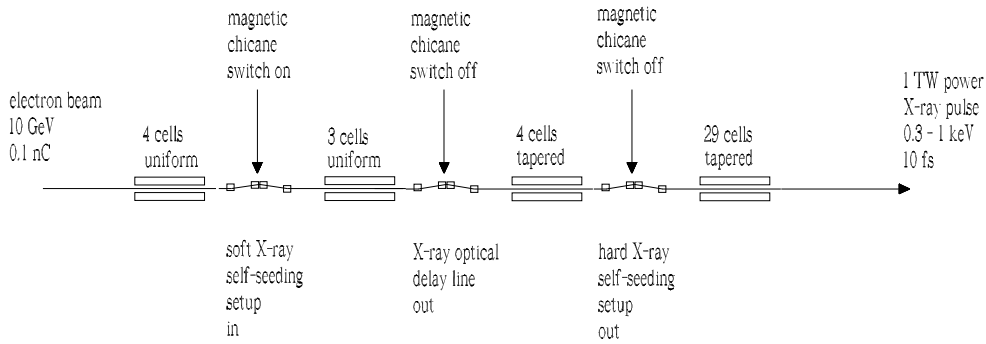}
\caption{Design of the undulator system for $1$ TW power mode of
operation in the soft X-ray photon energy range.  The method
exploits a combination of self-seeding scheme with grating
monochromator and an undulator tapering technique. } \label{biof7}
\end{figure}

The four-undulator configuration in Fig. \ref{biof3} can be
naturally taken advantage of at different photon energies ranging
from soft to hard X-rays. Fig. \ref{biof7} shows the basic setup for
the high-power mode of operation in the soft X-ray wavelength range.
The second and the third chicanes are not used for such regime, and
must be switched off. After the first undulator (4 cells-long) and
the grating monochromator, the output undulator follows. The  first
section of the output undulator (consisting of second and third
undulator) is composed by $7$ untapered cells , while tapering is
implemented in the fourth undulator. The monochromatic seed is
exponentially amplified by passing through the first untapered
section of the output undulator. This section is long enough to
allow for saturation, and yields an output power of about $100$ GW.
Such monochromatic FEL output is finally enhanced up to $1$ TW in
the second output-undulator section, by tapering the undulator
parameter over the last cells after saturation. Under the
constraints imposed by undulator and chicane parameters it is only
possible to operate at an electron beam energy of $10$ GeV. The
setup was optimized based on results of start-to-end simulations for
a nominal electron beam with 0.1 nC charge. Results were presented
in \cite{OSOF}, where we studied the performance of this scheme for
the SASE3 upgrade.

\subsection{Generation of TW pulses in the $3$ keV - $5$ keV photon energy
range}

%8
\begin{figure}[tb]
\includegraphics[width=1.0\textwidth]{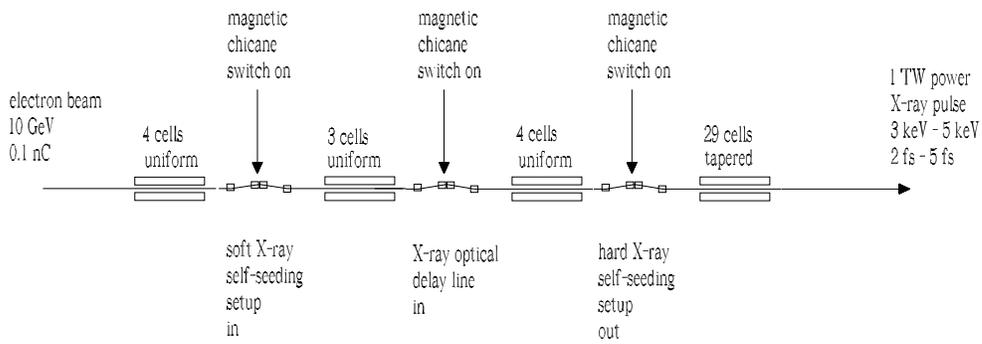}
\caption{Design of the undulator system for $1$ TW power mode of
operation in the $3$ keV - $5$ keV photon energy range. The method
exploits a combination of self-seeding scheme with grating
monochromator, fresh bunch and undulator tapering techniques. }
\label{biof8}
\end{figure}
%
%9
\begin{figure}[tb]
\includegraphics[width=1.0\textwidth]{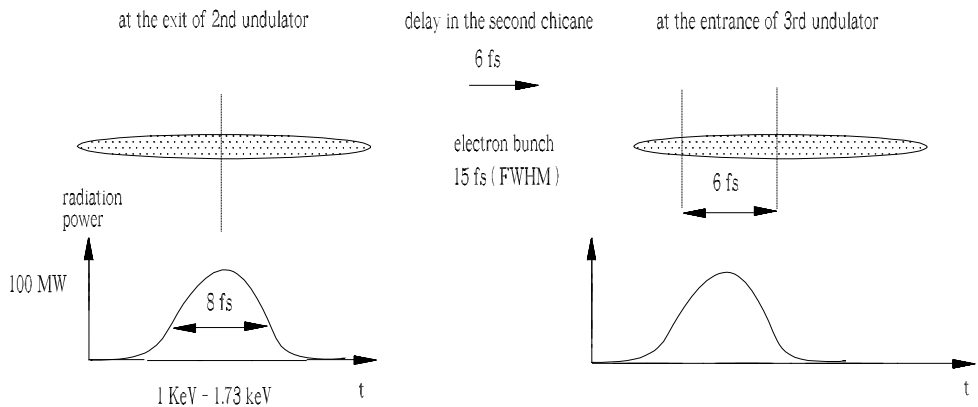}
\caption{Principle of the fresh bunch technique for the high power
mode of operation in the photon energy range between $3$ keV and $5$
keV. The second chicane smears out the electron microbunching and
delays the monochromatic soft X-ray pulse with respect to the
electron bunch of $6$ fs. In this way, half of of the electron bunch
is seeded and saturates in the third undulator.} \label{biof9}
\end{figure}
%
%10
\begin{figure}[tb]
\includegraphics[width=1.0\textwidth]{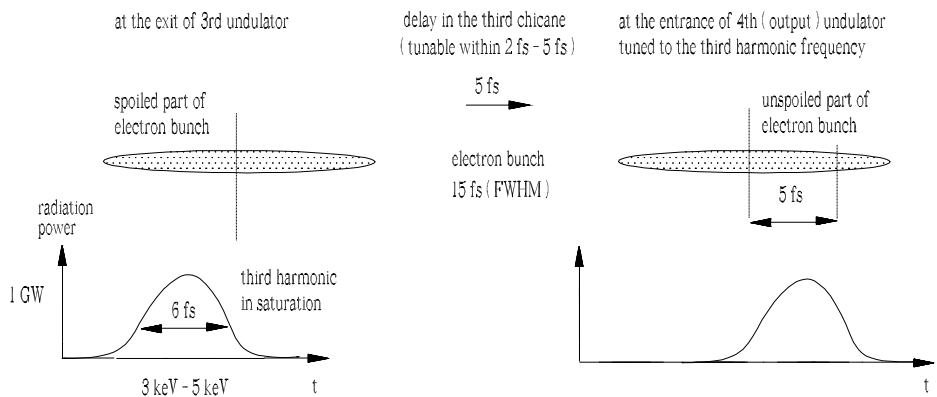}
\caption{Principle of the fresh bunch technique for the high power
mode of operation in the photon energy range between $3$ keV and $5$
keV. The third magnetic chicane smears out the electron
microbunching and delays the electron bunch with respect to the
radiation pulse. The unspoiled part of electron bunch is seeded by a
GW level monochromatic pulse at third harmonic frequency. Tunability
of the output pulse duration can be easily obtained by tuning the
magnetic delay of the third chicane.} \label{biof10}
\end{figure}

\begin{figure}[tb]
\includegraphics[width=1.0\textwidth]{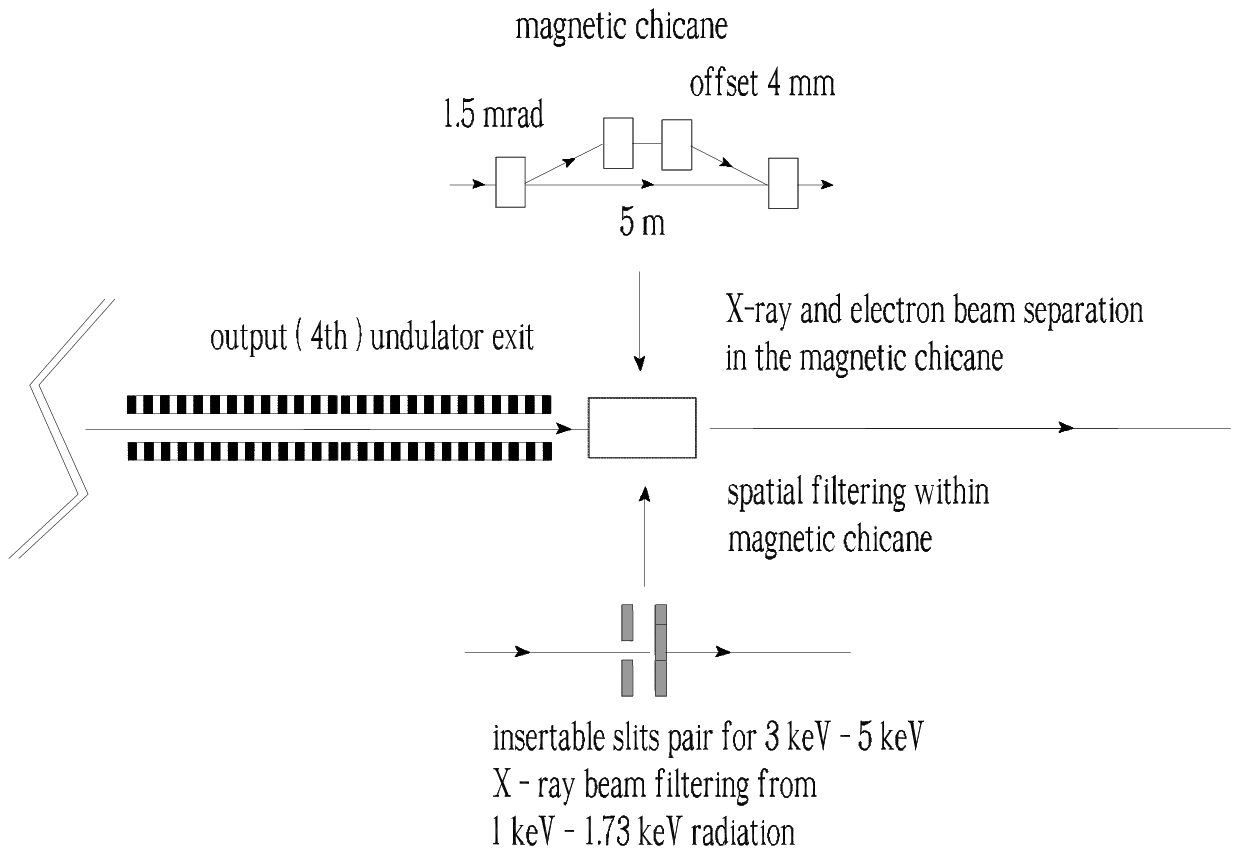}
\caption{Schematics of a background filtering setup downstream of
the extended SASE3 undulator. The scheme for spatial filtering will
make use of a short magnetic chicane immediately behind the exit of
the output undulator, so that the electron beam bypasses the slits.
Vertical and horizontal slits will positioned at $70$ m (Phase one)
or $200$ m (Phase two) downstream of the third undulator, where the
background radiation (between $1$ keV and $1.73$ keV) is
characterized by a spot size ten times larger than that of the main
X-ray beam (between $3$ keV and  $5$ keV). } \label{biof10b}
\end{figure}
Fig. \ref{biof8} shows the basic setup for high power mode of
operation in the most preferable photon energy range for single
biomolecule imaging. All three chicanes are used for such regime,
and must be switched on. The third chicane is used as a magnetic
delay only, and the crystal must be removed from the light path. We
propose to perform monochromatization at photon energies ranging
between $1$ keV and $1.7$ keV  with the help of a grating
monochromator, and to amplify the seed in the second undulator up to
the power level of $0.2$ GW. The second chicane smears out the
electron microbunching and delays the monochromatic soft X-ray pulse
of $2~\mu$m with respect to the electron bunch. In this way, half of
the electron bunch is seeded and saturates in the third undulator up
to $40$ GW, Fig. 9. At saturation, the electron beam generates
considerable monochromatic radiation at the third harmonic in the GW
power level. The third magnetic chicane smears out the electron
microbunching and delays the electron bunch with respect to the
radiation of $2~\mu$m. Thus, the unspoiled part of the electron
bunch is seeded by the GW-level monochromatic pulse at the third
harmonic frequency, Fig. 10. The fourth, 29 cells-long undulator is
tuned to the third harmonic frequency (between $3$ keV and $5$ keV),
and is used to amplify the radiation pulse up to $1$ TW.  The
additional advantage of the proposed setup for bio-imaging is the
tunability of the output pulse duration, which is obtained by tuning
the magnetic delay of the third chicane. Simulations show that the
X-ray pulse duration can be tuned from $2$ fs to $5$ fs. The
production of such pulses is of great importance when it comes to
single biomolecule imaging experiments.

The soft X-ray background can be easily eliminated by using a
spatial window positioned downstream of the fourth undulator exit,
Fig. \ref{biof10b}. Since the soft X-ray radiation has an angular
divergence of about $0.02$ mrad FWHM, and the slits are positioned
more than $100$ m downstream of the third undulator, the background
has much larger spot size compared with the $3$ keV - $5$ keV
radiation spot size, which is about 0.1 mm at the exit of the fourth
undulator. Therefore, the background radiation power can be
diminished of more than two orders of magnitude without any
perturbations of the main pulse.

Under the constraints imposed by the soft X-ray self-seeding setup,
it is only possible to operate at an electron beam energy of $10$
GeV. The setup was optimized based on results of start-to-end
simulations for a nominal electron bunch with a charge of $0.1$ nC.
Results are presented in Section 4. The proposed undulator setup
uses the electron beam coming from the SASE1 undulator. We assume
that SASE1 operates at the photon energy of $12$ keV, and that the
FEL process is switched off for one single dedicated electron bunch
within each macropulse train. A method to control the FEL
amplification process is based on betatron switcher described in
\cite{SWIT1,SWIT2}. Due to quantum energy fluctuations in the SASE1
undulator, and to wakefields in the SASE1 undulator pipe, the energy
spread and the energy chirp of the electron bunch at the entrance of
the bio-imaging beamline significantly increases compared with the
same parameters at the entrance of the SASE1 undulator. The
dispersion strength of the first chicane has been taken account from
the viewpoint of the electron beam dynamics, because it disturbs the
electron beam distribution. The other two chicanes have tenfold
smaller dispersion strength compared with first one. The electron
beam was tracked through the first chicane using the code Elegant
\cite{ELEG}. The electron beam distortions complicate the simulation
procedure. However, simulations show (see Section 4) that the
proposed setup is not significantly affected by perturbations of the
electron phase space distribution, and yields about the same
performance as in the case for an electron beam without chicane
transformation (see below).

Finally, the design of the grating monochromator for the soft X-ray
self-seeding scheme is under active investigation. For example, a
more compact grating monochromator design has appeared very recently
\cite{COCC0}. This novel design is based on the use of a toroidal
grating, and adopts a constant entrance-angle mode of operation. The
resolution and the photon energy range are the same, but the delay
of the photons is about three times smaller. Therefore, the
perturbations of the electron beam distributions generated in this
way would be negligible. %Simulation results for this unperturbed
%electron beam case are presented in Appendix.

\subsection{Generation of TW pulses in 8 keV - 13 keV energy range}

%11
\begin{figure}[tb]
\includegraphics[width=1.0\textwidth]{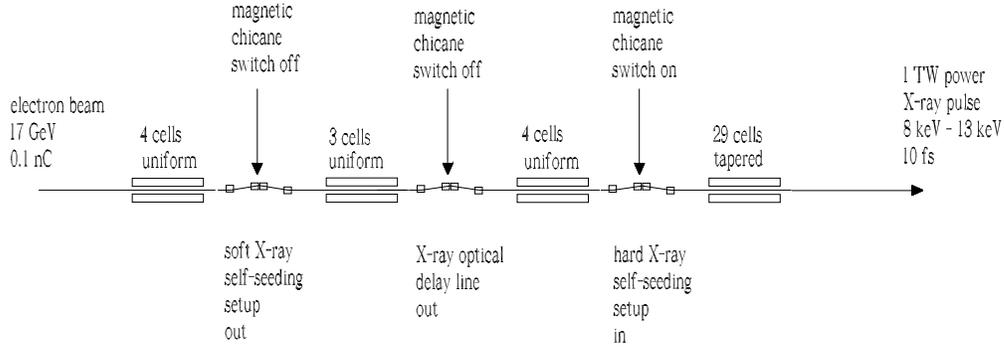}
\caption{Design of the undulator system for $1$ TW power mode of
operation in the $8$ keV - $13$ keV photon energy range. The method
exploits a self-seeding scheme with crystal monochromator. }
\label{biof11}
\end{figure}
Fig. \ref{biof11} shows the basic setup for the high-power mode of
operation in the hard X-ray photon energy range.  The first and the
second chicane are not used in this regime, and are switched off.
After the first three undulators and the single-crystal
monochromator, a fourth output-undulator follows. Under the
constraints imposed by the undulator parameters it is possible to
operate at two nominal electron beam energies of 14 GeV and 17.5
GeV. The setup was optimized based on results of the start-to-end
simulations for an electron beam energy of 17.5 GeV and a nominal
electron beam with 0.1 nC charge. Results are presented in section
4. The output undulator is long enough to reach  $1$ TW power. The
duration of the output pulses is of about $10$ fs. In this mode of
operation  there is no possibility to tune the pulse duration
without changing the electron beam distribution. If tunability of
the pulse duration is requested in this energy range, this is most
easily achieved by providing additional delay with a magnetic
chicane installed behind the hard X-ray self-seeding setup.

\subsection{Location and expansion plans}

\begin{figure}[tb]
\includegraphics[width=1.0\textwidth]{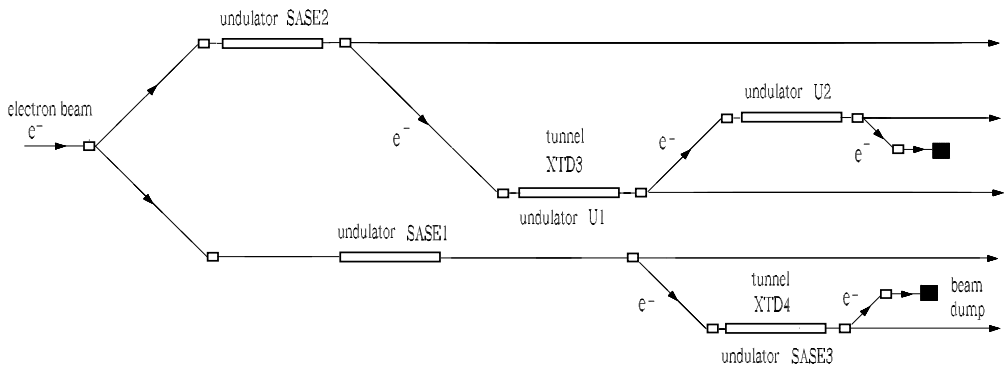}
\caption{Original design of the European XFEL facility.}
\label{biof12}
\end{figure}

\begin{figure}[tb]
\includegraphics[width=1.0\textwidth]{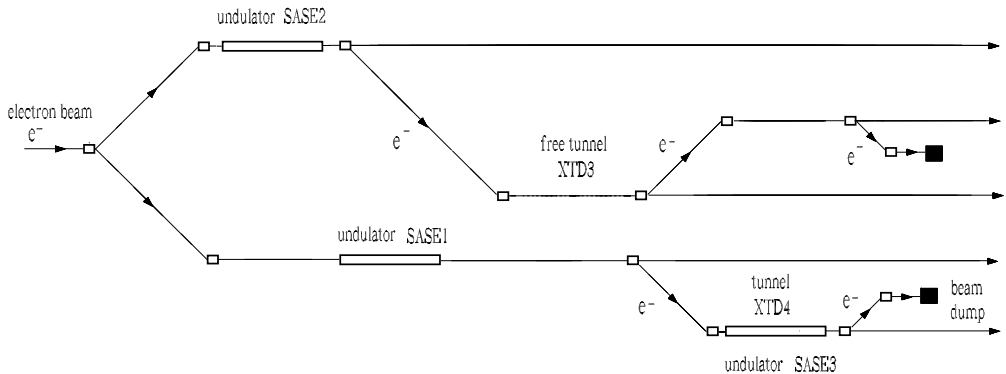}
\caption{Current design of the European XFEL facility.}
\label{biof13}
\end{figure}

\begin{figure}[tb]
\includegraphics[width=1.0\textwidth]{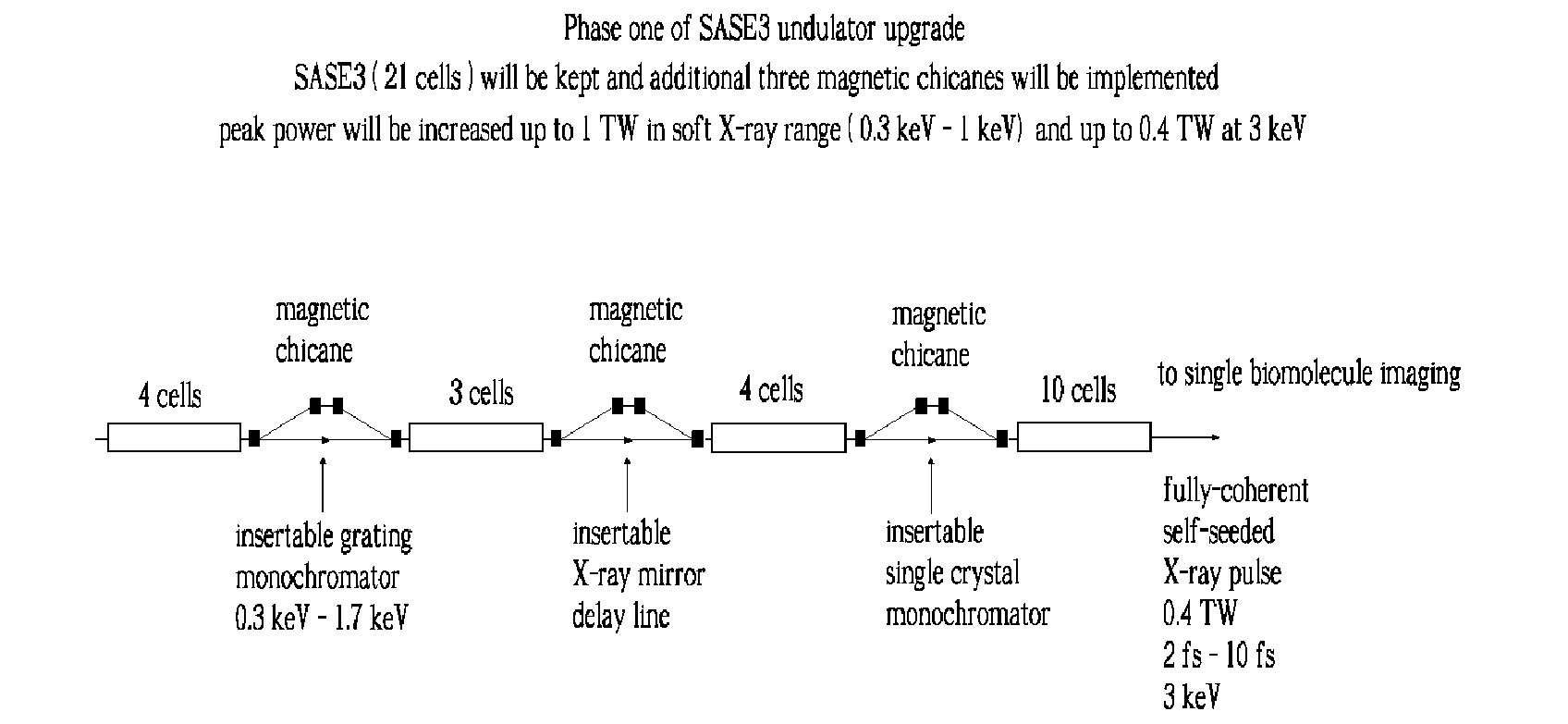}
\caption{Phase 1 of the SASE3 upgrade. Schematic layout of the SASE3
beamline after installation of three additional magnetic chicanes.
Phase 1 experiments will aim to demonstrate the effectiveness of the
combination of self-seeding, fresh bunch and undulator tapering
techniques in the photon energy range between $3$ keV and $5$ keV,
and to serve as a test bench before the installation of additional
undulators (Phase 2). The Phase 1 upgrade will allow for single
biomolecule imaging in the preferable photon energy range between
$3$ keV and $5$ keV.} \label{biof14}
\end{figure}

\begin{figure}[tb]
\includegraphics[width=1.0\textwidth]{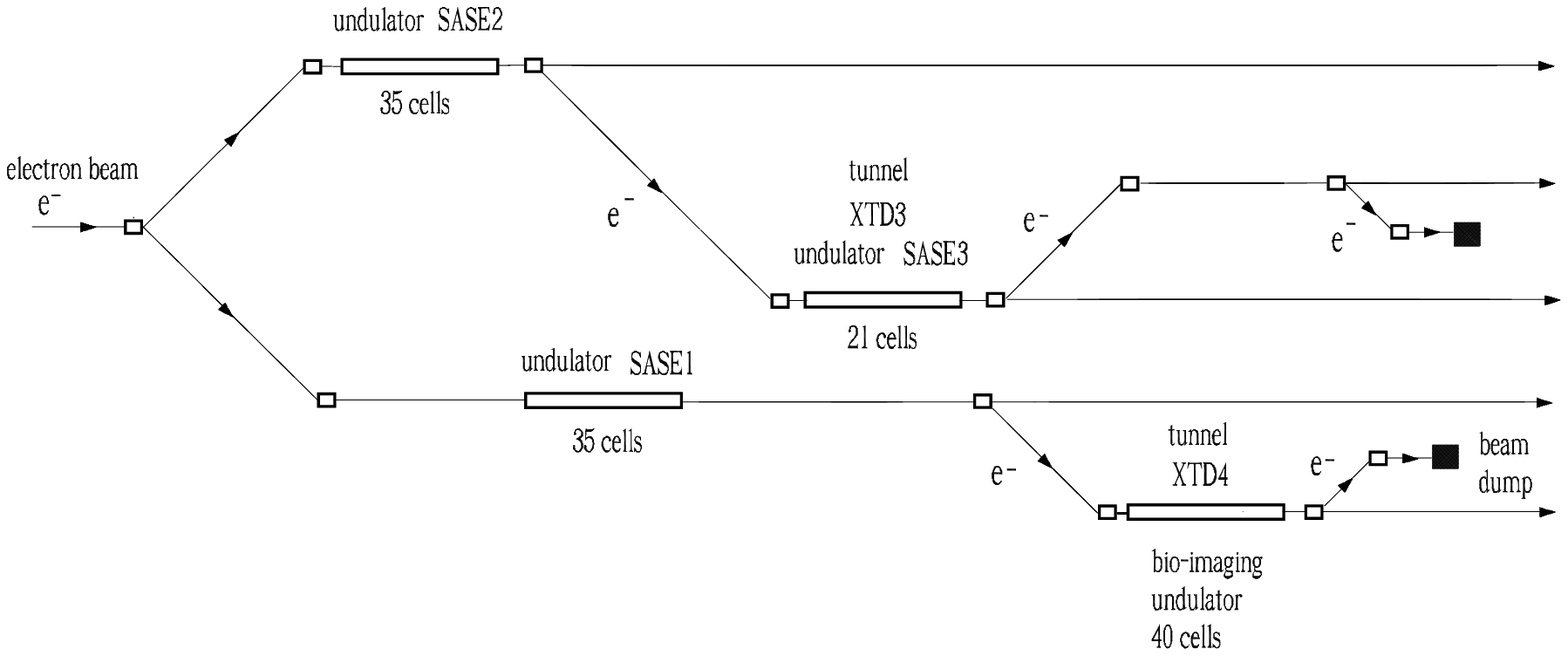}
\caption{Schematic of the proposed extension of the European XFEL
facility.} \label{biof15}
\end{figure}

The original design of the European XFEL \cite{tdr-2006} was
optimized to produce XFEL radiation at $0.1$ nm, simultaneously at
two undulator lines, SASE1 and SASE2. Additionally, the design
included one FEL line in the soft X-ray range, SASE3, and two
undulator lines for spontaneous synchrotron radiation, U1 and U2,
Fig. \ref{biof12}. The soft X-ray SASE3 beamline used the spent
electron beam from SASE1, and the U1 and U2 beamlines used the spent
beam from SASE2. In fact, although the electron beam performance is
degraded by the FEL process, the beam can still be used in
afterburner mode in the SASE3 undulator, which will be equipped with
a $126$ m-long undulator system, for a total of $21$ cells.

After a first design report, the layout of the European XFEL
changed. In the last years after the achievement of the LCLS, and
the subsequent growth of interest in XFEL radiation by the
scientific community, it became clear that the experiments with XFEL
radiation, rather than with spontaneous synchrotron radiation, had
to be prioritized. In the new design, the two beamlines behind SASE2
are now free for future XFEL undulators installations, Fig.
\ref{biof13}.

Recently it was also realized that the amplification process in the
XFEL undulators can be effectively controlled by betatron FEL
switchers \cite{SWIT1,SWIT2}. The SASE3 undulator was then optimized
for generating soft X-rays. However, due to the possibility of
switching the FEL process in SASE1, it is possible to produce high
power SASE3 radiation in a very wide photon energy range between
$0.3$ keV and $13$ keV.  The SASE3 beamline is now expected to
provide excellent performance, and to take advantage of its location
in the XTD4 tunnel, which is close to the experimental hall and has
sufficient free space behind the undulator for future expansion
($140$ m). After this section, the electron beam will be separated
from the photon beam and will be bent down to an electron beam dump,
Fig. \ref{biof12}.  In the photon energy range between $3$ keV and
$13$ keV, the SASE3 beamline is now expected to provide even better
conditions for users than SASE1 and SASE2.

In this article we propose to build the bio-imaging beamline in the
XTD4 tunnel as an upgrade of SASE3.

The first installation phase of the bio-imaging beamline described
in this report includes the installation of the full SASE3 undulator
system, constituted by $21$ cells, together with three magnetic
chicanes, after the 4th, the 7th and the 11th cell, Fig.
\ref{biof14}. These three chicanes will be equipped with
self-seeding setups in the soft X-ray (first chicane), in the hard
X-ray (third chicane), and with an X-ray mirror delay line (second
chicane). They constitute the most important elements of proposed
bioimaging beamline design. The start of commissioning and operation
of the entire beamline is defined by the completion of these
elements. Our results (see section 4) suggest that $400$ GW output
power in the wavelength range between $3$ keV and $5$ keV is
possible from the last tapered undulator part, consisting of $10$
cells. The cost for soft X-ray self-seeding setup, fresh bunch
technique setup and hard X-ray self-seeding setup can be estimated
as $2$ M EUR, $1$ M EUR and $1$ M EUR respectively. This
installation phase will play a most important role in the
construction of the new beamline.

After the experimental demonstration of self-seeding and fresh bunch
techniques feasibility, a second installation phase will follow,
including the extension of the line with additional $19$ undulator
cells for achieving 1 TW peak power in the entire photon energy
range. The  cost for one cell can be estimated as $0.5$ M EUR
\footnote{The cost of one undulator segment and intersection setup
can be estimated as  $0.3$ M EUR and $0.13$ M EUR, respectively.}.
This amounts to about $10$ M EUR for the undulator extension.

In a possible third stage, a new SASE3 undulator composed by $21$
cells can be installed in the free XTD3 tunnel, which is shorter
than the XTD4 tunnel but sufficiently long for such installation,
Fig. \ref{biof15}.

The bio-imaging beamline will support experiments carried out over a
rather wide photon energy range. It is therefore proposed that the
photon beam transport of the new beamline includes two lines. Line A
uses $0.5$m-long mirrors operating at a grazing angle of $2$ mrad.
This line is dedicated to the transport of X-ray radiation in the
photon energy range from $3$ keV up to $13$ keV. This would be
complementary to the designed SASE3 line, Line B, that is now
optimized in the soft X-ray range between $0.3$ keV and $3$ keV. The
distance from the 40-cells-long undulator exit to the first mirror
system will be only of about $100$m\footnote{This is in contrast
with SASE1 and SASE2 beamlines, where an opening angle of $0.003$
mrad at $3$ keV FEL radiation leads to unacceptable mirror length of
2 m due to long distance of about 500 m between the source and
mirror system. For these beamlines there is no possibility to use
identical configuration of mirrors within the photon energy range
from 3 keV to 13 keV. }.

%4.

\section{\label{sec:FEL} FEL studies}

\begin{figure}[tb]
\includegraphics[width=0.5\textwidth]{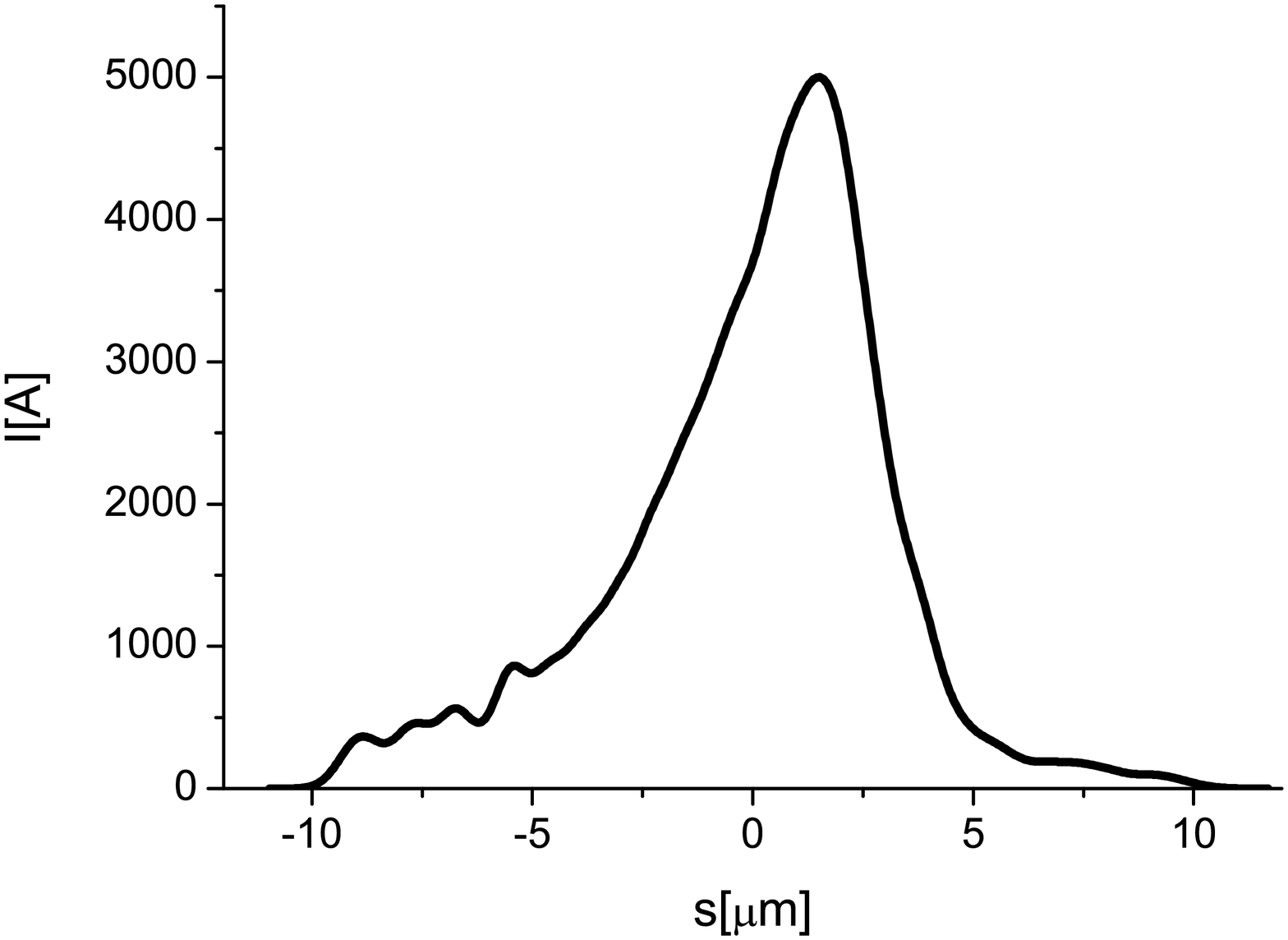}
\includegraphics[width=0.5\textwidth]{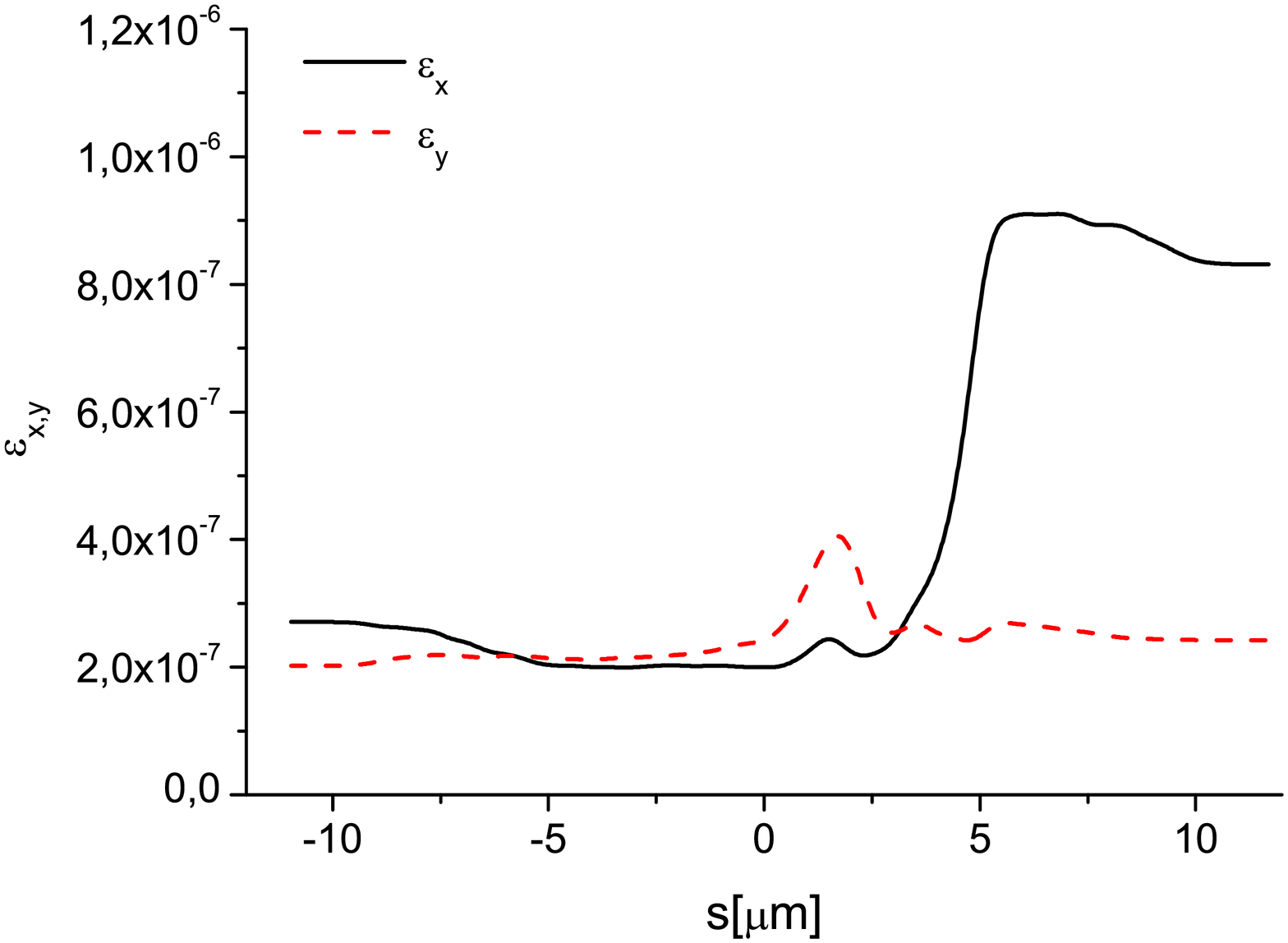}
\includegraphics[width=0.5\textwidth]{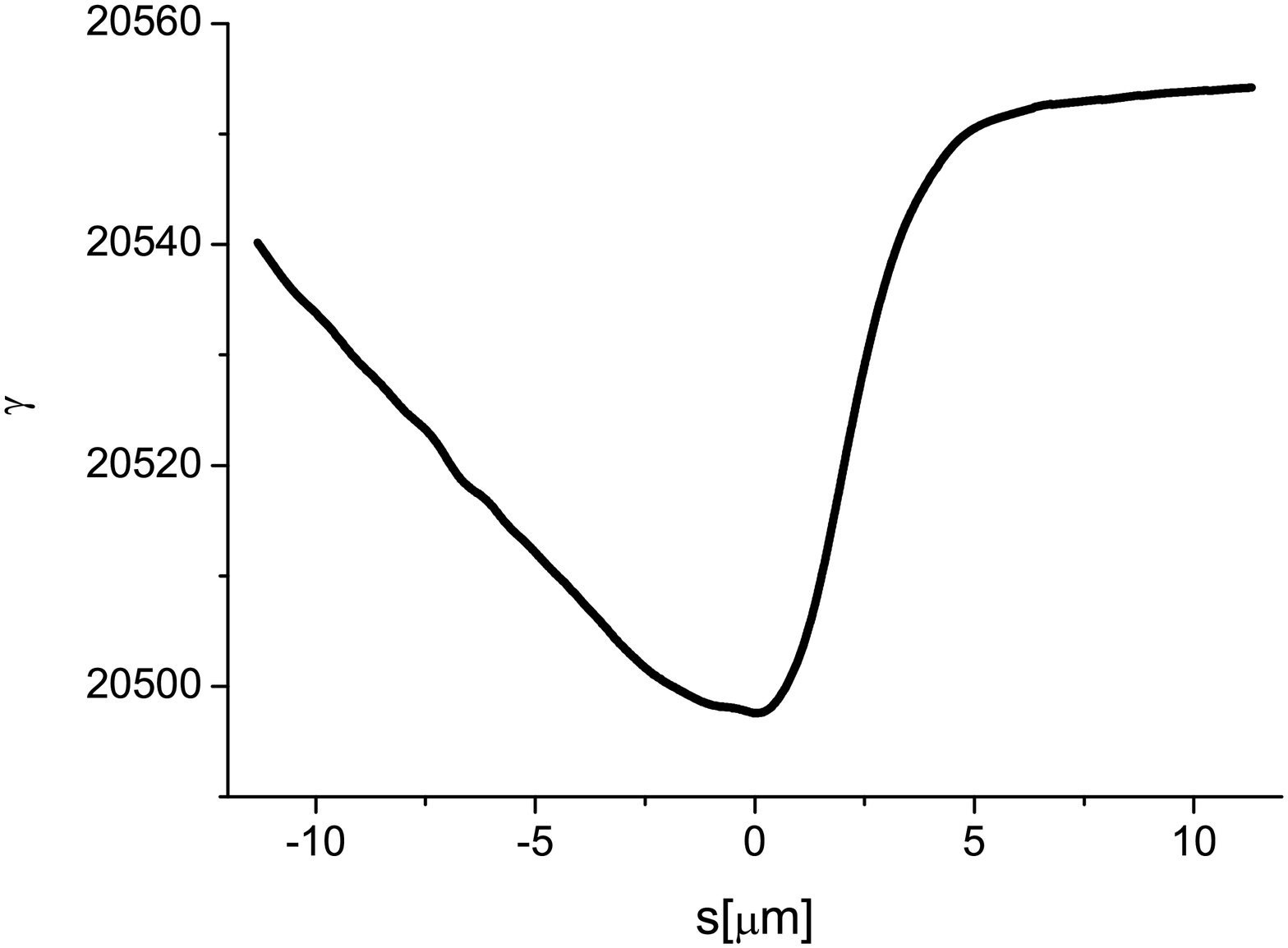}
\includegraphics[width=0.5\textwidth]{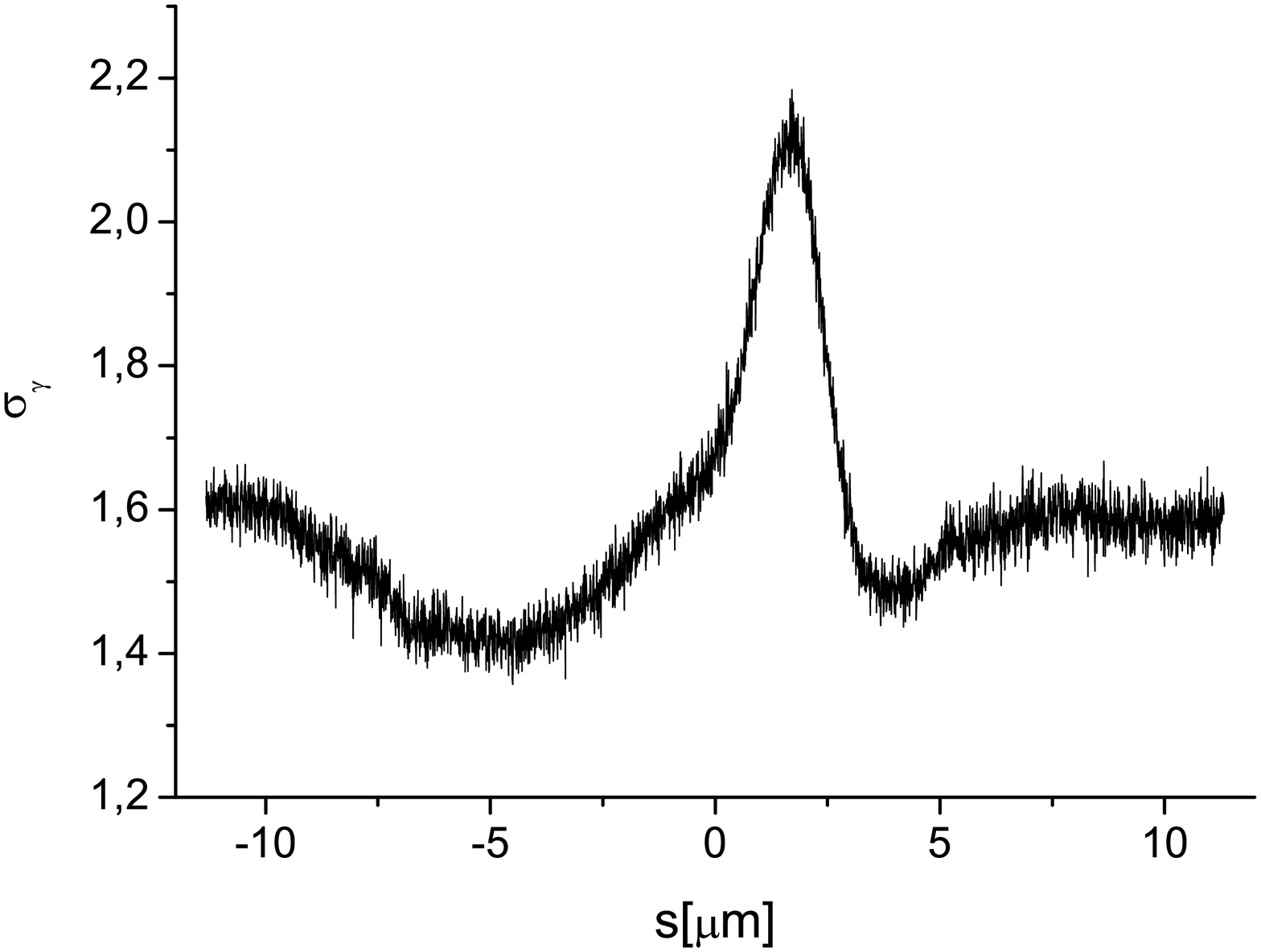}
\begin{center}
\includegraphics[width=0.5\textwidth]{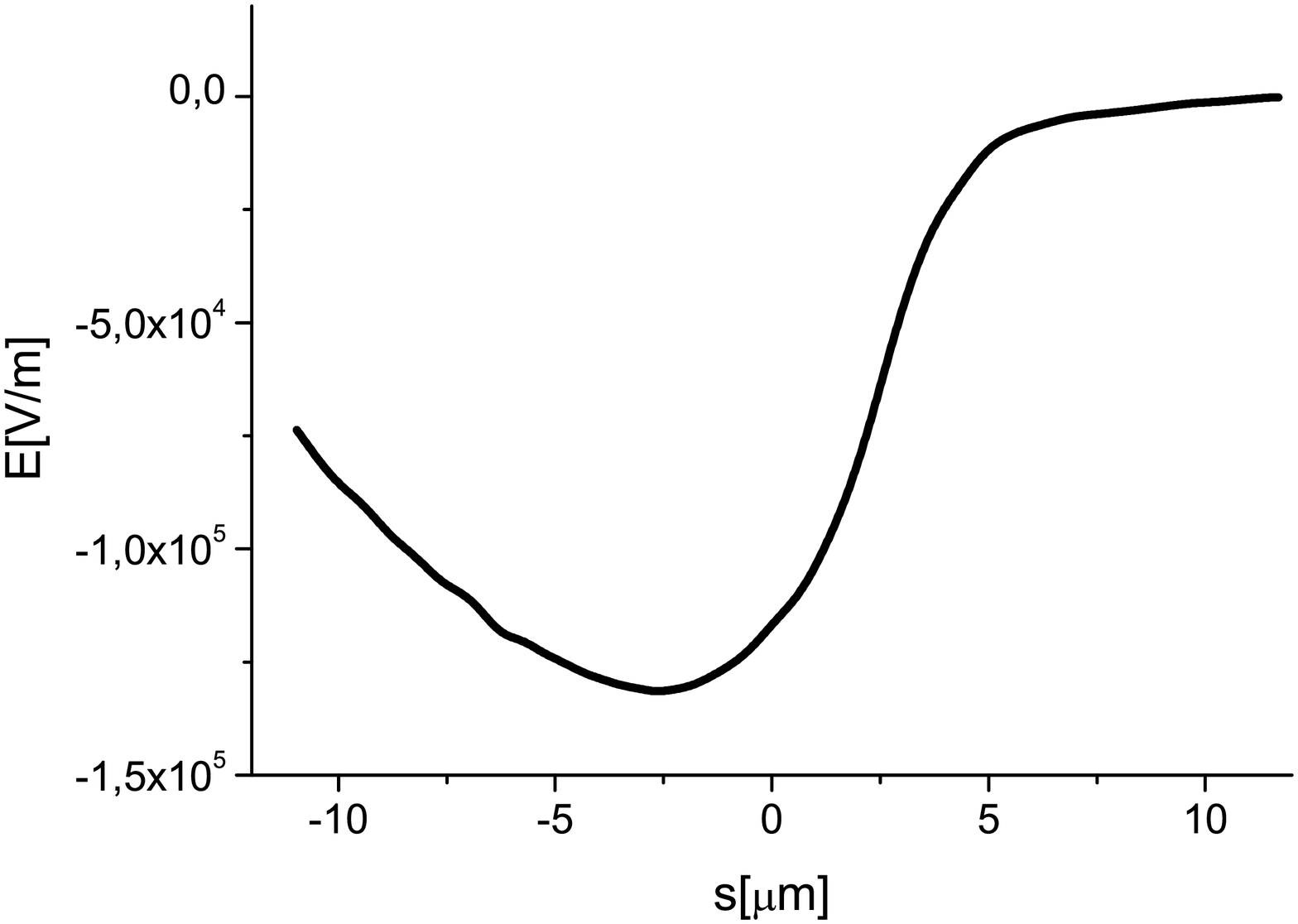}
\end{center}
\caption{Results from electron beam start-to-end simulations at the
entrance of SASE3 \cite{S2ER} for the soft X-ray case. (First Row,
Left) Current profile. (First Row, Right) Normalized emittance as a
function of the position inside the electron beam. (Second Row,
Left) Energy profile along the beam. (Second Row, Right) Electron
beam energy spread profile. (Bottom row) Resistive wakefields in the
SASE3 undulator \cite{S2ER}. These results are used as starting
point for our investigations in the soft X-ray regime, Section
\ref{subsso} and \ref{subso}.} \label{biof16}
\end{figure}

\begin{figure}[tb]
\includegraphics[width=0.5\textwidth]{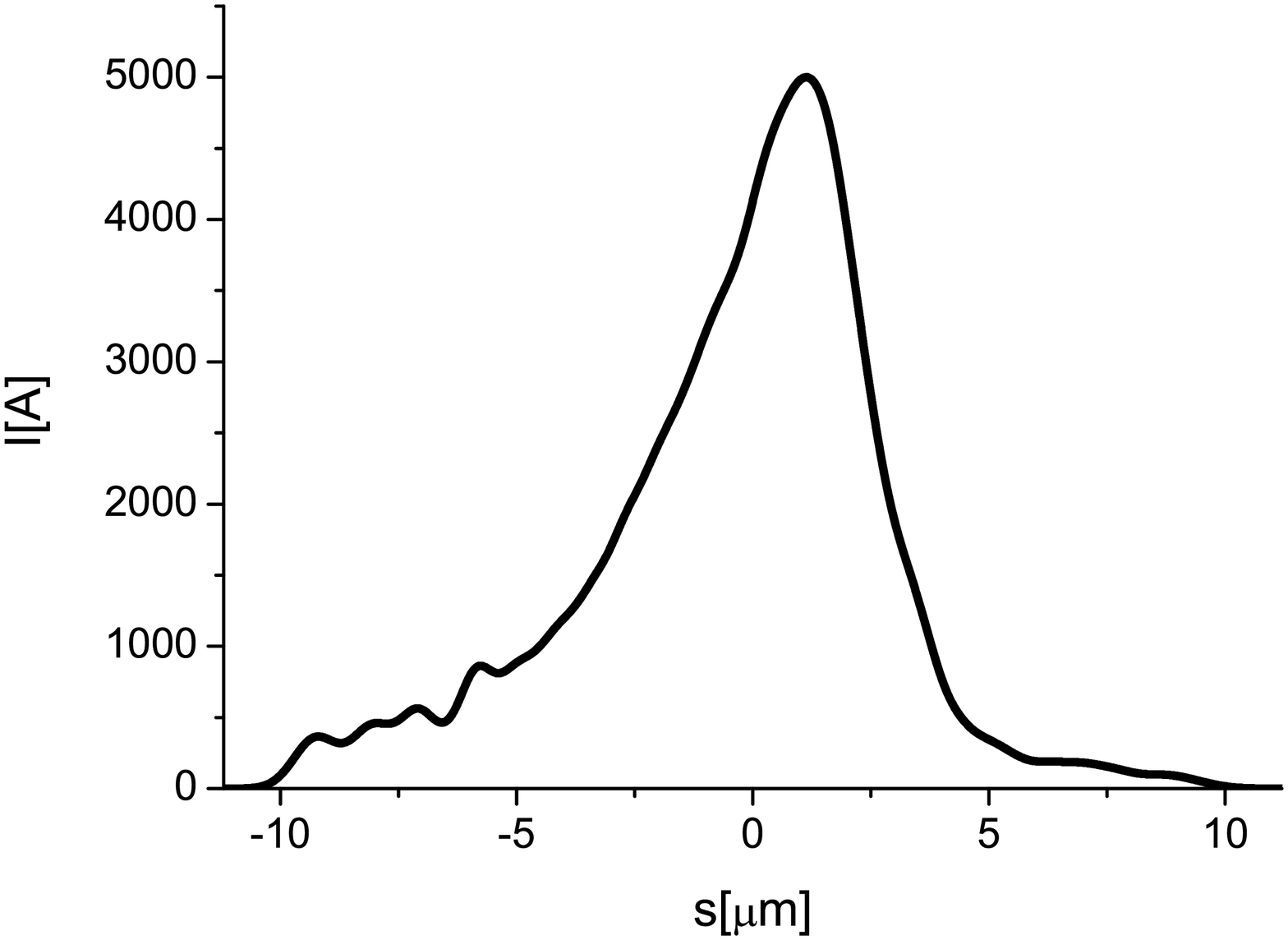}
\includegraphics[width=0.5\textwidth]{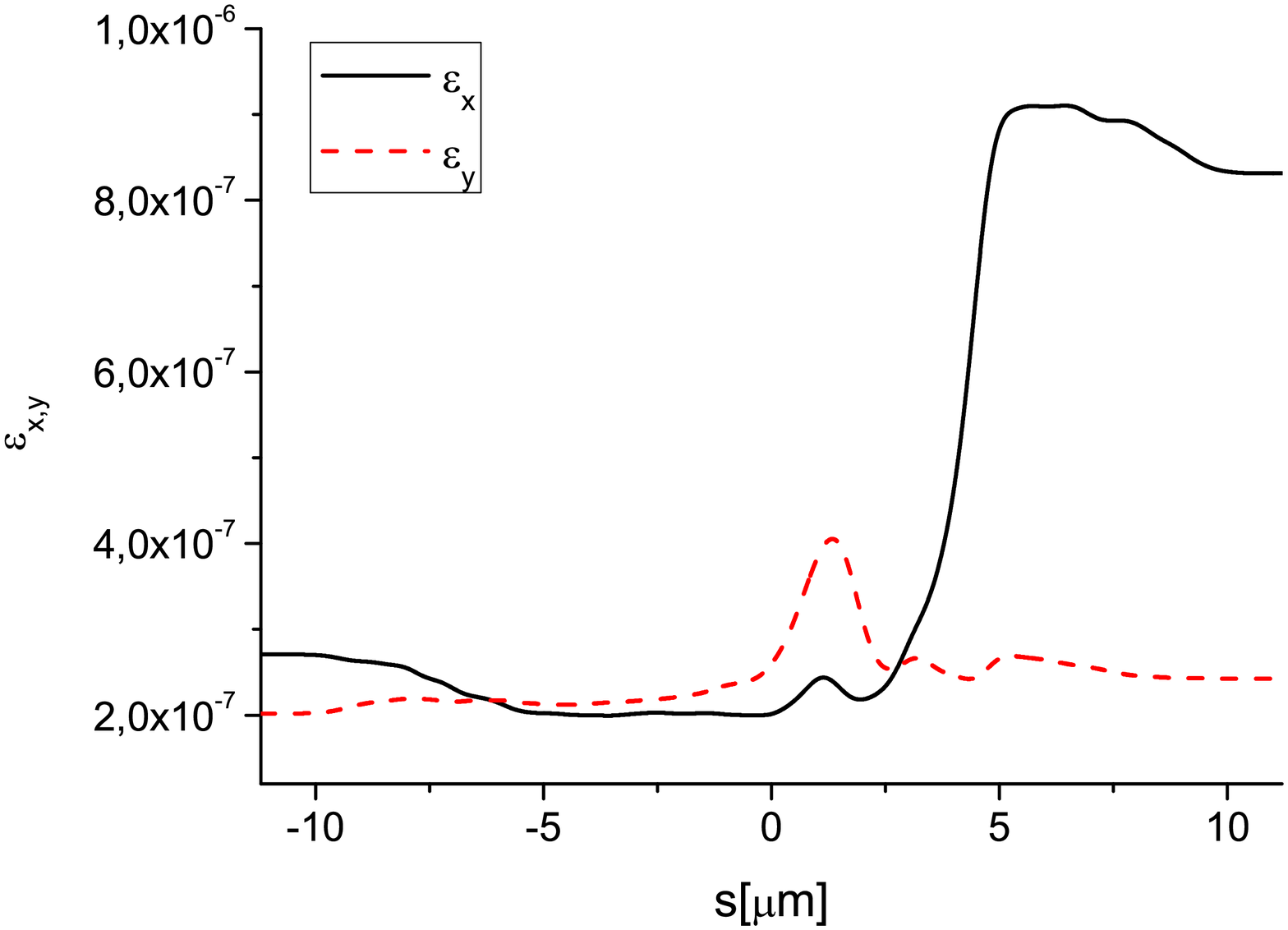}
\includegraphics[width=0.5\textwidth]{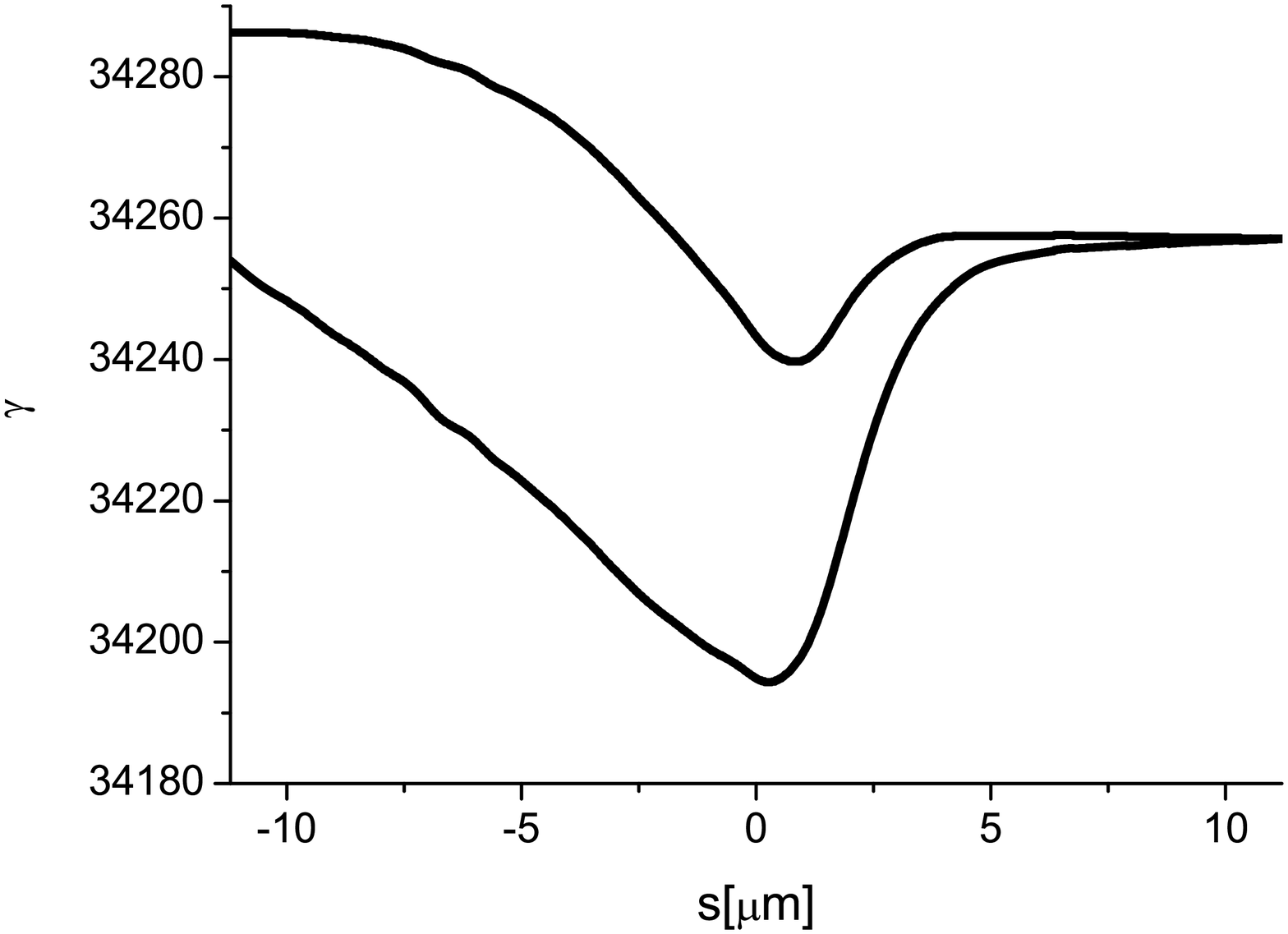}
\includegraphics[width=0.5\textwidth]{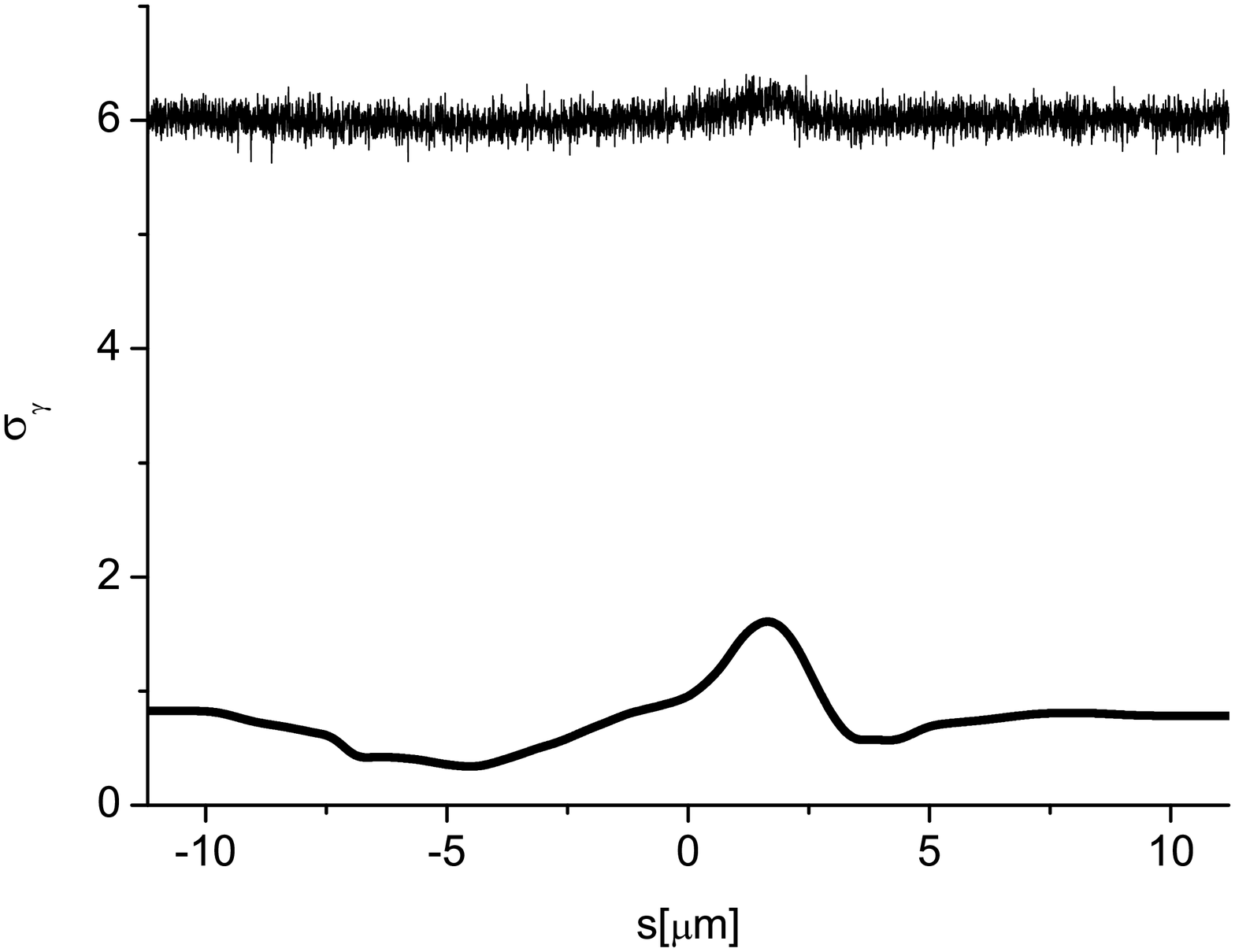}
\begin{center}
\includegraphics[width=0.5\textwidth]{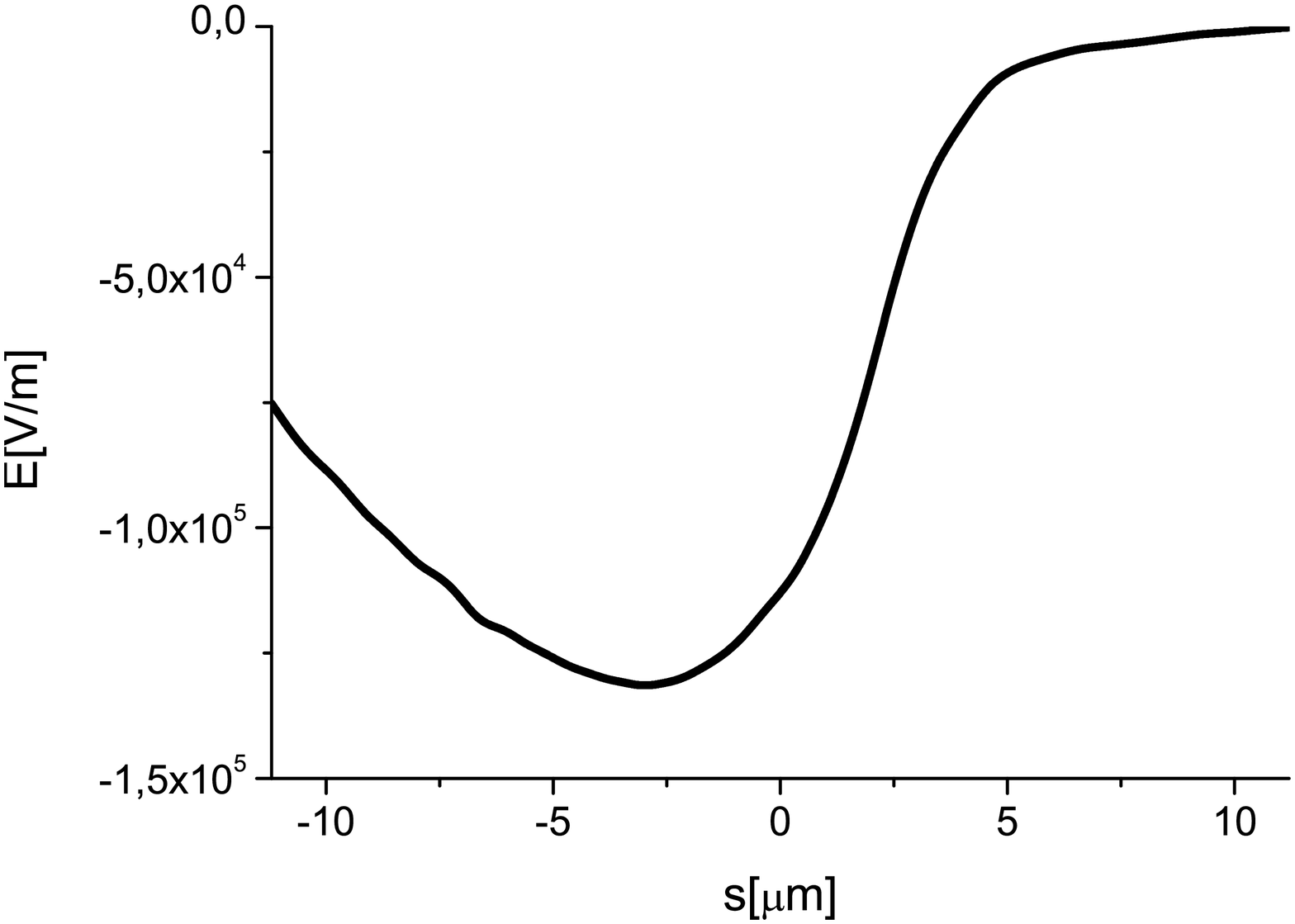}
\end{center}
\caption{Results from electron beam start-to-end simulations at the
entrance of SASE3 \cite{S2ER} for the hard X-ray case. (First Row,
Left) Current profile. (First Row, Right) Normalized emittance as a
function of the position inside the electron beam. (Second Row,
Left) Energy profile along the beam, lower curve. The effects of
resistive wakefields along SASE1 are illustrated by the comparison
with the upper curve, referring to the entrance of SASE1 (Second
Row, Right) Electron beam energy spread profile, upper curve. The
effects of quantum diffusion along SASE1 are illustrated by the
comparison with the lower curve, referring to the entrance of SASE1.
(Bottom row) Resistive wakefields in the SASE3 undulator
\cite{S2ER}. These results are used as starting point for our
investigations in the soft X-ray regime, Section \ref{subhard}.}
\label{biof16b}
\end{figure}

With reference to Fig. \ref{biof7}, Fig. \ref{biof8} and Fig.
\ref{biof11} we performed feasibility studies pertaining the three
energy ranges considered in this articles. These studies were
performed with the help of the FEL code GENESIS 1.3 \cite{GENE}
running on a parallel machine. Simulations are based on a
statistical analysis consisting of $100$ runs.

\begin{table}
\caption{Undulator parameters}

\begin{small}\begin{tabular}{ l c c}
\hline & ~ Units &  ~ \\ \hline
Undulator period      & mm                  & 68     \\
Periods per cell      & -                   & 73   \\
K parameter (rms)     & -                   & 1.5-7\\
Total number of cells & -                   & 40    \\
Intersection length   & m                   & 1.1   \\
Photon energy         & keV                 & 0.3-13 \\
\hline
\end{tabular}\end{small}
\label{tt1}
\end{table}

In order to specify the electron beam parameters for the simulations
we use results from electron beam start-to-end simulations at the
entrance of SASE3 \cite{S2ER}. In particular, Fig. \ref{biof16}
refers to the soft X-ray case, where the energy of the electron beam
is $10.5$ GeV, while Fig. \ref{biof16b} refers to the hard X-ray
case, where the energy of the electron beam is $17.5$ GeV. It should
be remarked that the start-to-end simulations in \cite{S2ER}
actually give back the beam parameters at the entrance of SASE1.
Therefore, resistive wakefields up to the SASE3 entrance modify the
electron beam energy. Moreover, here we assume that the lasing in
SASE1 is inhibited with the help of the betatron switcher technique
\cite{SWIT1,SWIT2}, but that the undulator gap is not opened.
Therefore, energy spread due to quantum fluctuations in SASE1 must
be accounted for. We illustrated the change in both energy and
energy spread due to these processes in Fig. \ref{biof16b}, where
two curves are present in the pictures in the second row. In the
following simulations we will use the current profile the normalized
emittance the energy profile the electron beam energy spread and the
resistive wakefields in the SASE3 undulator from Fig. \ref{biof16}
(Section \ref{subsso} and \ref{subso}) and from Fig. \ref{biof16b}
(Section \ref{subhard}). The undulator parameters are presented in
Table \ref{tt1}.

\subsection{\label{subsso} Soft X-ray photon energy range below $1$ keV}

Production of soft X-rays with photon energies below $1$ keV is
enable by configuring the setup as described in Fig. \ref{biof7},
and discussed in Section \ref{sub:verysoft}. A feasibility study for
this case has been already carried out in \cite{OSOF}, to which we
refer the reader for further details and simulation results.

\subsection{\label{subso}Photon energy range between $3$ keV and $5$ keV}

We now turn to analyze the case described in Fig. \ref{biof8}, which
pertains the energy range between $3$ keV and $5$ keV.

\begin{figure}[tb]
\includegraphics[width=0.5\textwidth]{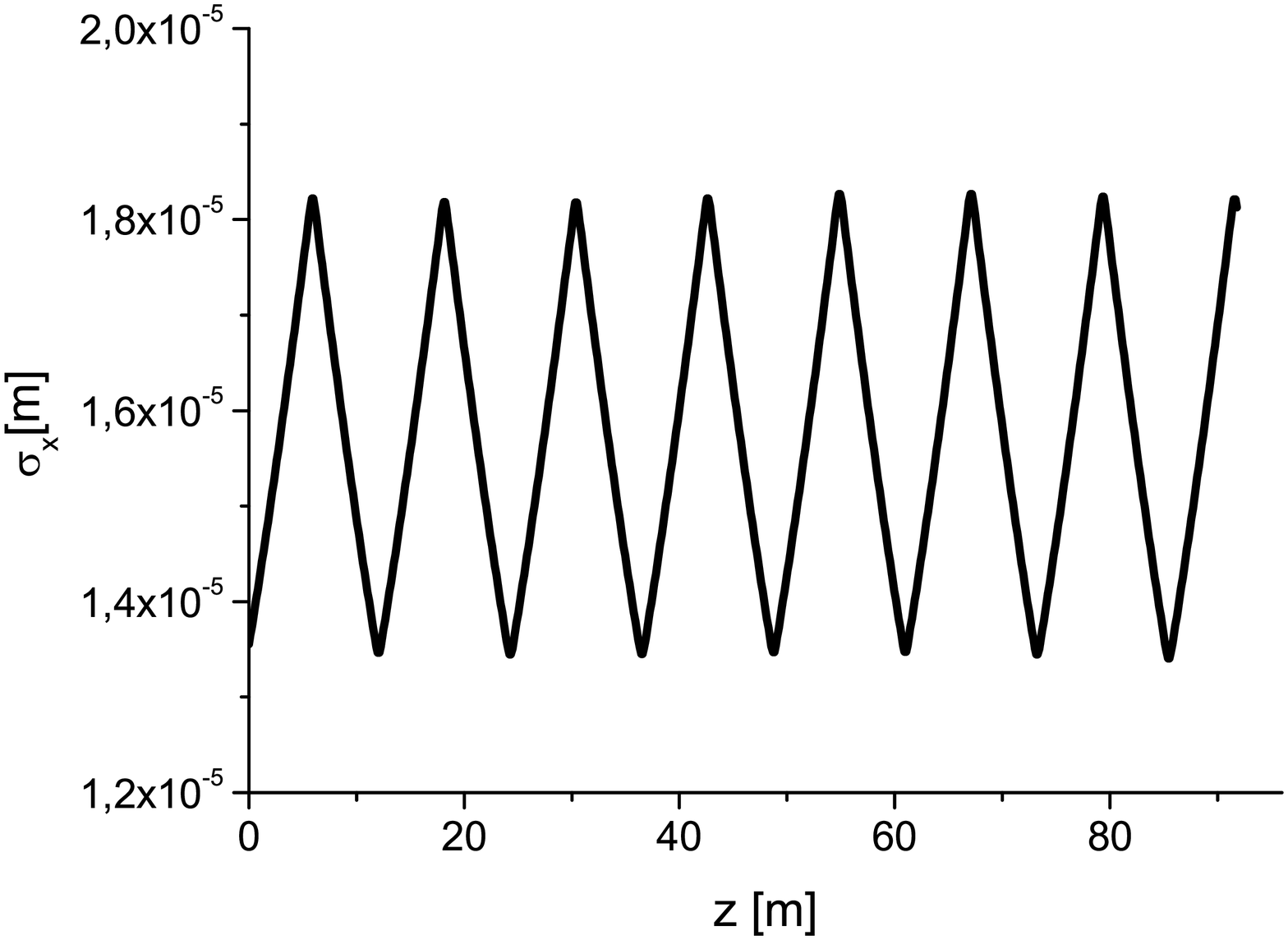}
\includegraphics[width=0.5\textwidth]{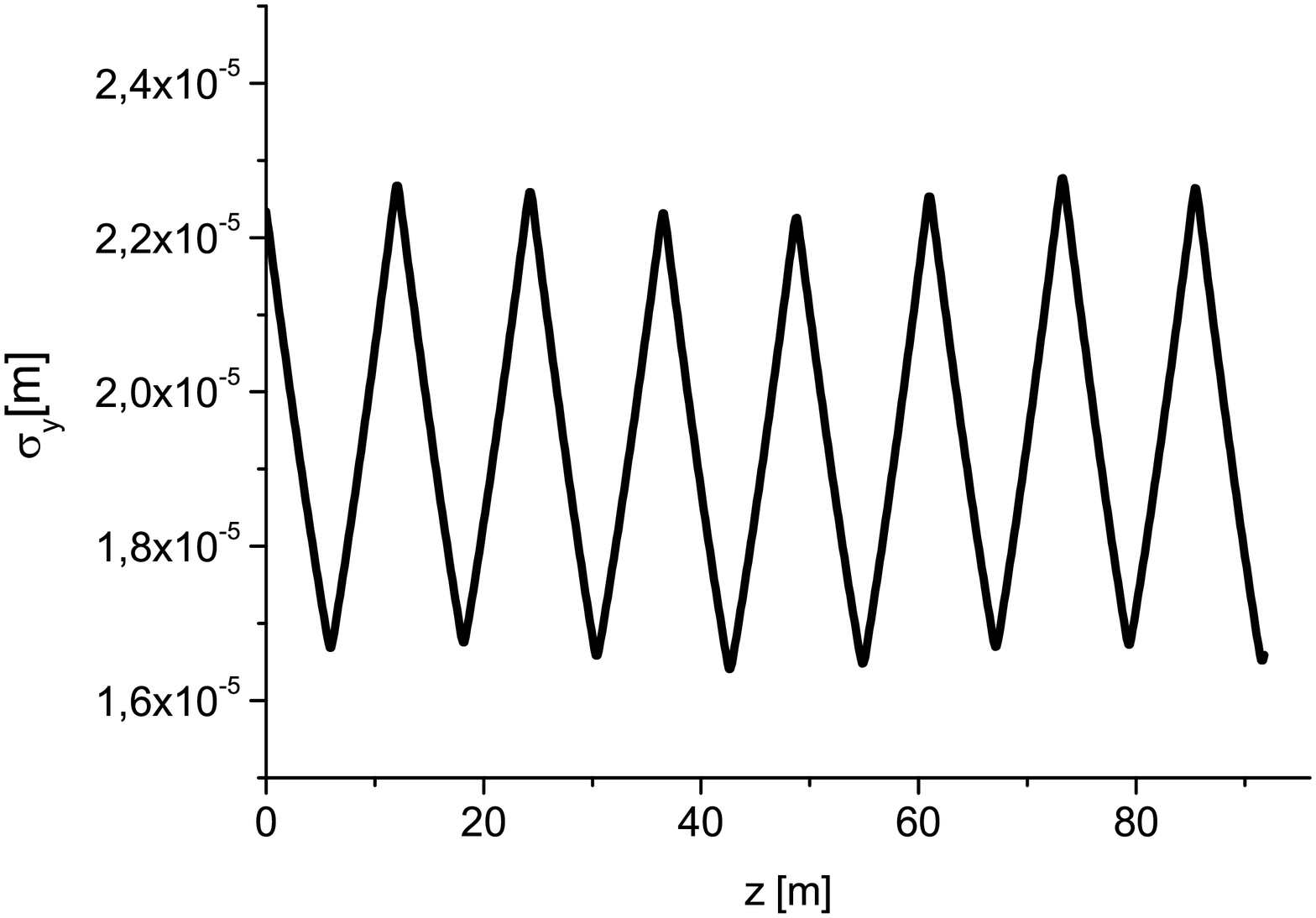}
\caption{Evolution of the horizontal (left plot) and vertical (right
plot) dimensions of the electron bunch as a function of the distance
inside the undulator at $10.5$ GeV. The plots refer to the
longitudinal position inside the bunch corresponding to the maximum
current value.} \label{sigma}
\end{figure}
The expected beam parameters at the entrance of the SASE3 undulator,
and the resistive wake inside the undulator are shown in Fig.
\ref{biof16}. The evolution of the transverse electron bunch
dimensions are plotted in Fig. \ref{sigma}.

\begin{figure}[tb]
\includegraphics[width=0.5\textwidth]{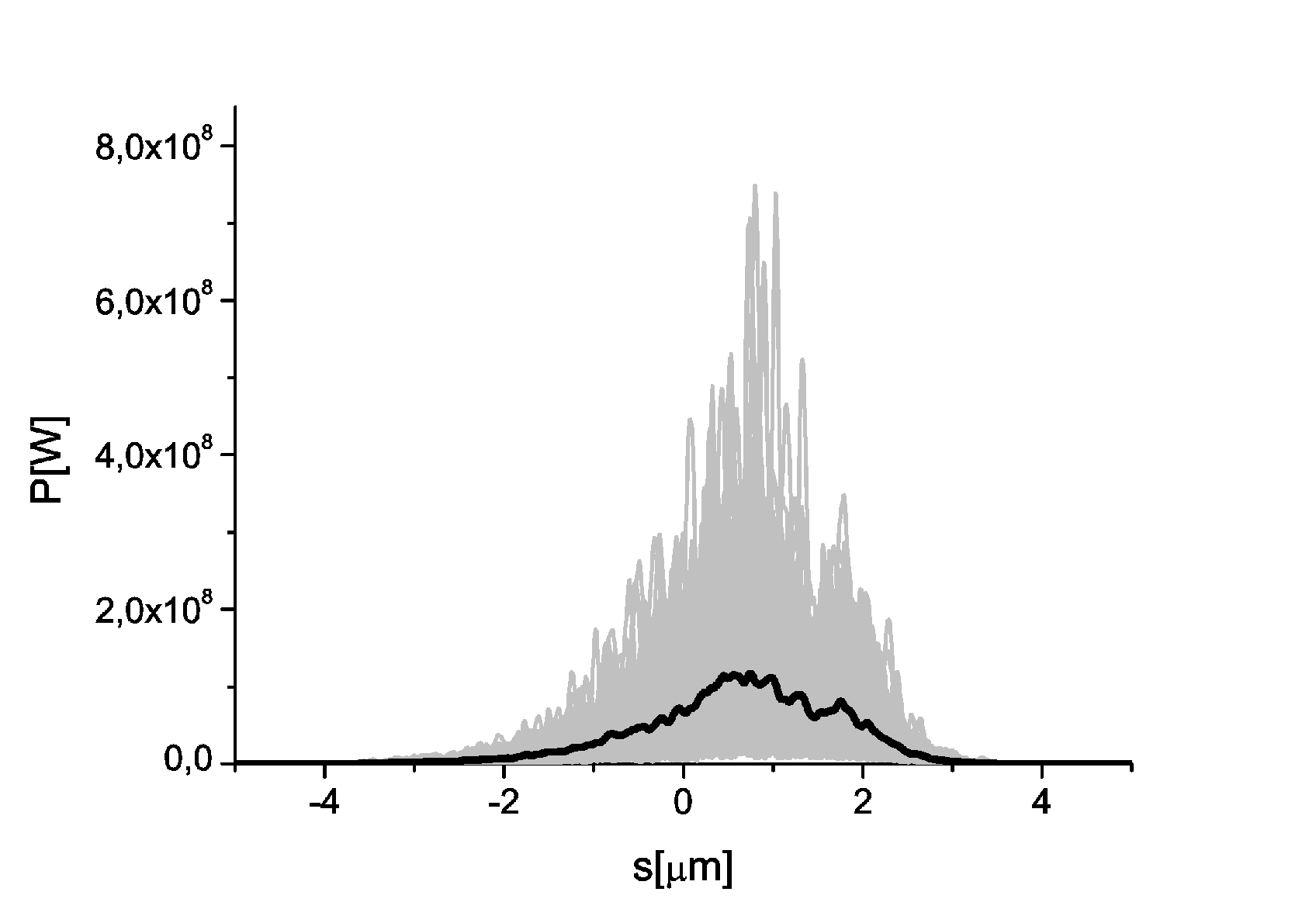}
\includegraphics[width=0.5\textwidth]{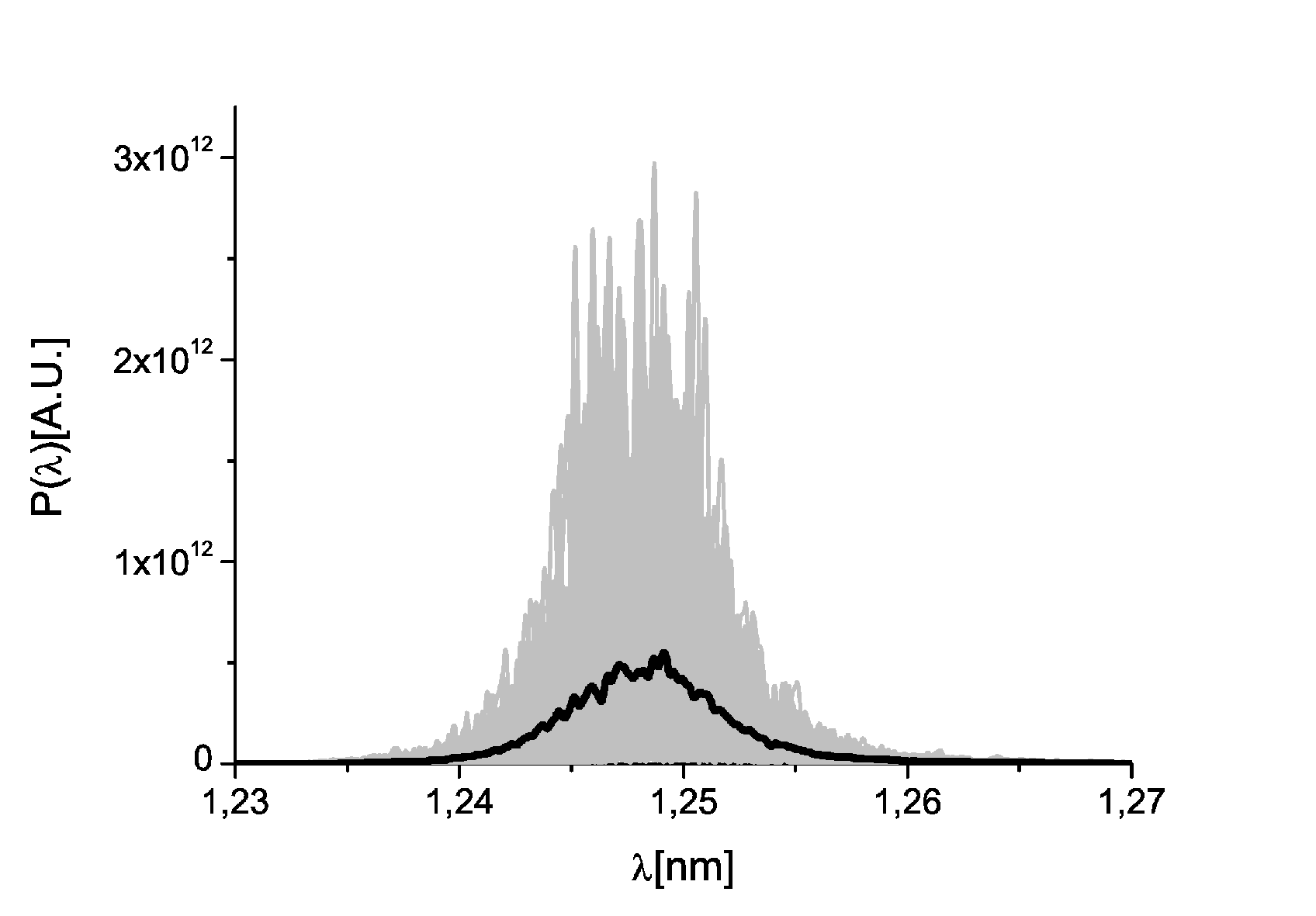}
\caption{Power and spectrum before the first magnetic chicane. Grey
lines refer to single shot realizations, the black line refers to
the average over a hundred realizations.} \label{biof17}
\end{figure}
We begin our investigation by simulating the SASE power and spectrum
after the first $4$ undulator cells, that is before the first
magnetic chicane. Results are shown in Fig. \ref{biof17}.

\begin{figure}[tb]
\includegraphics[width=0.5\textwidth]{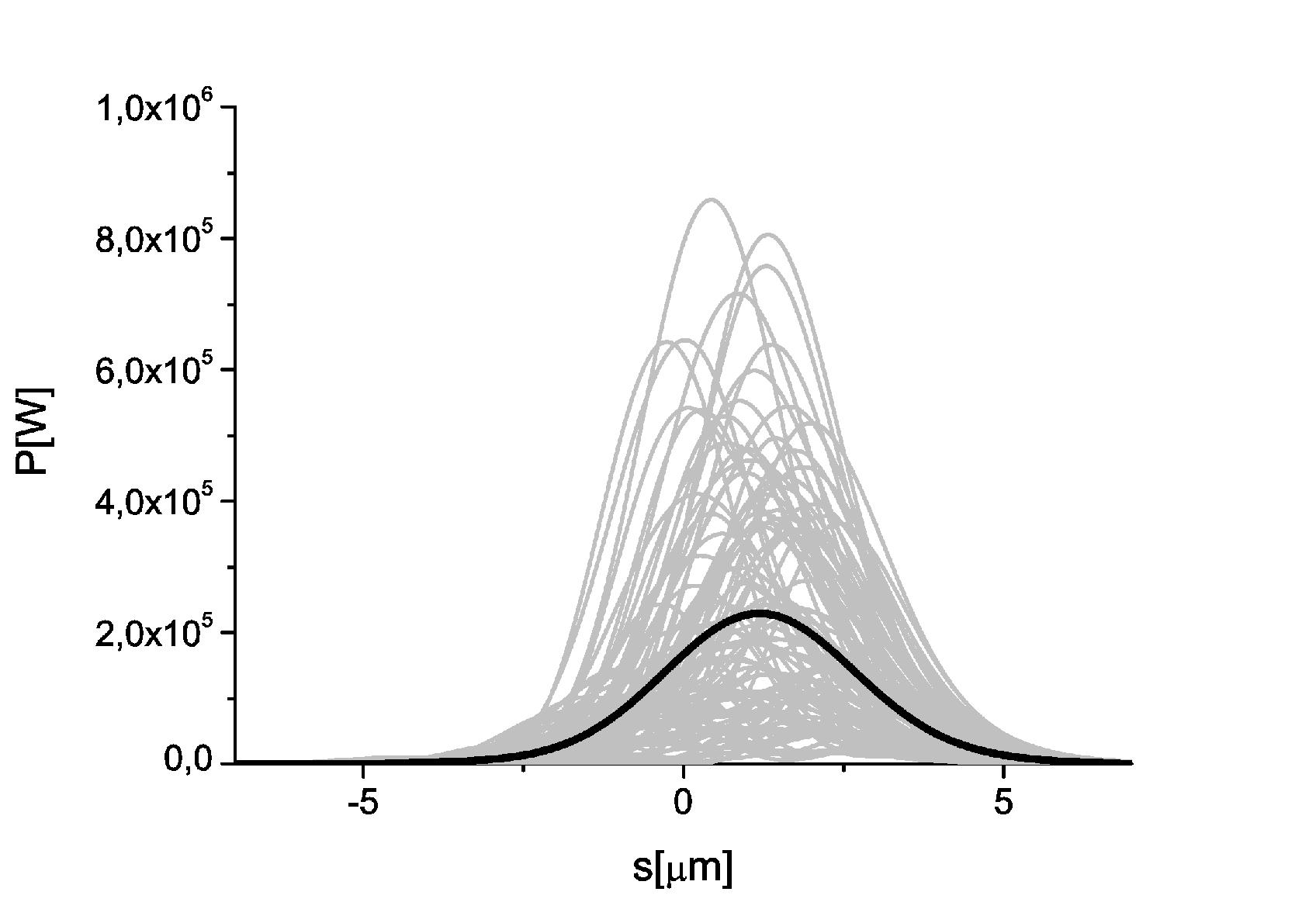}
\includegraphics[width=0.5\textwidth]{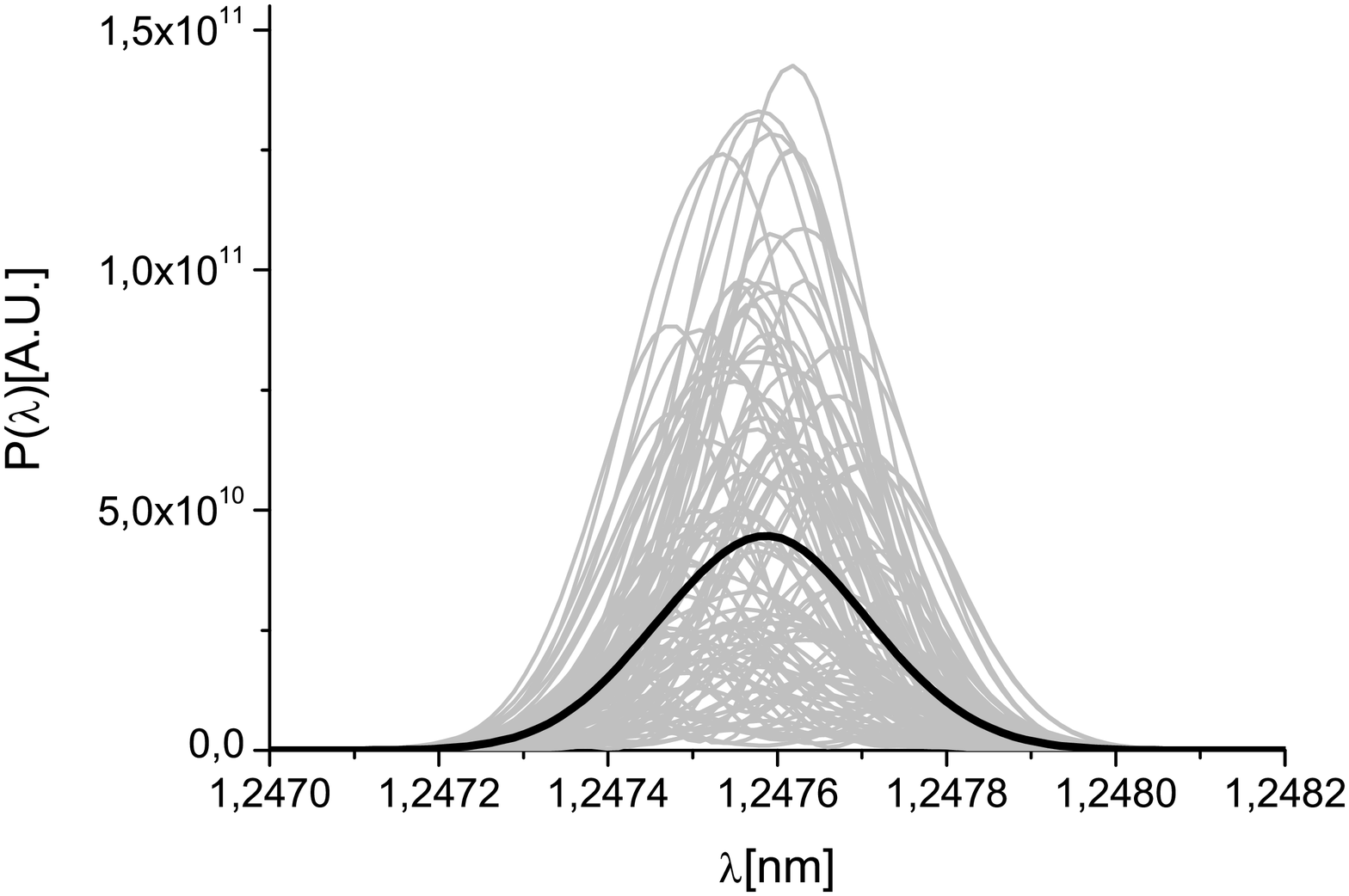}
\caption{Power and spectrum after the first magnetic chicane and
soft X-ray monochromator. Grey lines refer to single shot
realizations, the black line refers to the average over a hundred
realizations.} \label{biof18}
\end{figure}
The magnetic chicane is switched on, an the soft X-ray monochromator
is inserted. Assuming a monochromator efficiency of $10\%$, a
Gaussian line, and a resolving power of $5000$ we can filter the
incoming radiation pulse in Fig. \ref{biof17} accordingly, to obtain
the power and spectrum in Fig. \ref{biof18}. This power and spectrum
are used for seeding.

\begin{figure}[tb]
\includegraphics[width=0.5\textwidth]{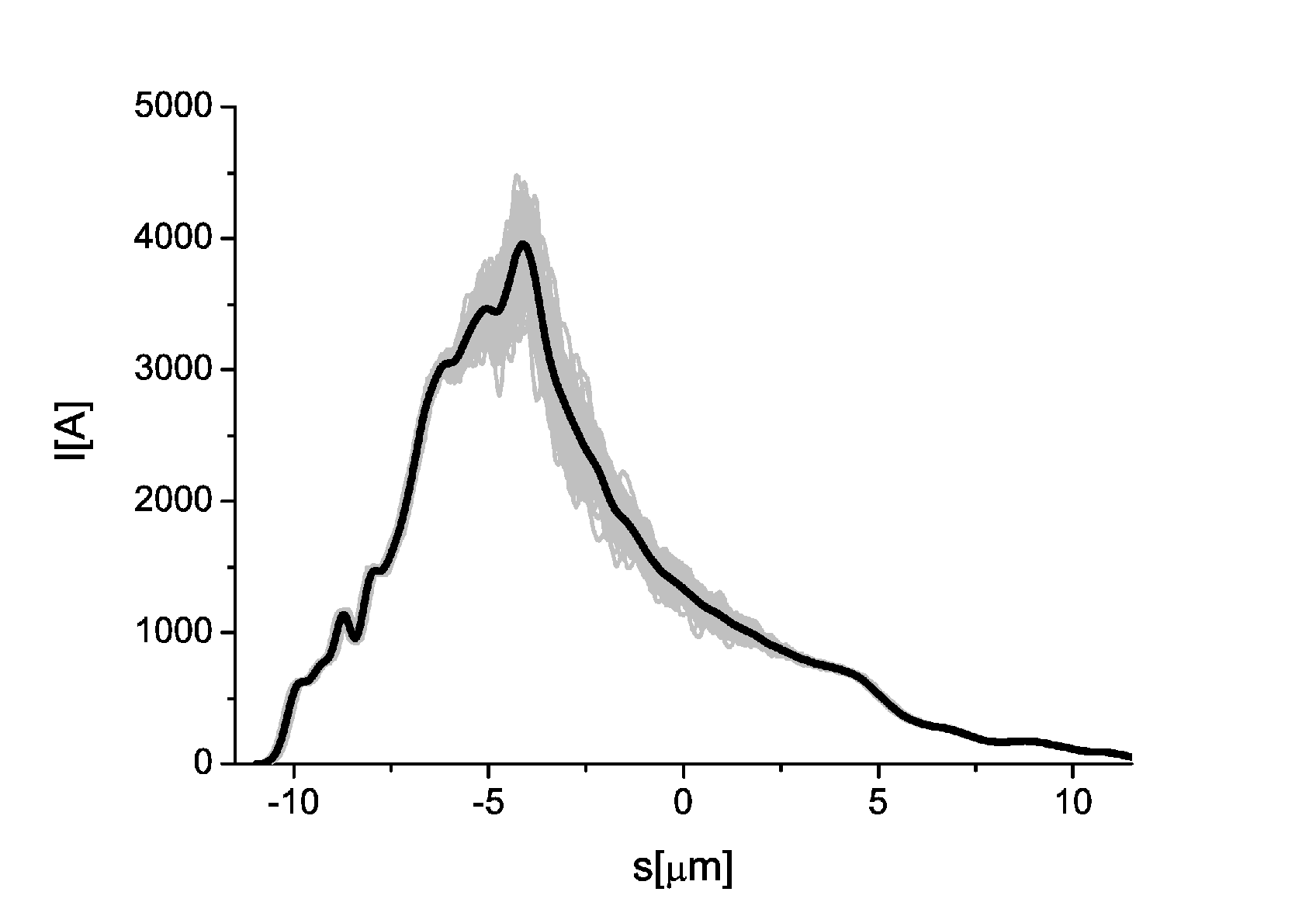}
\includegraphics[width=0.5\textwidth]{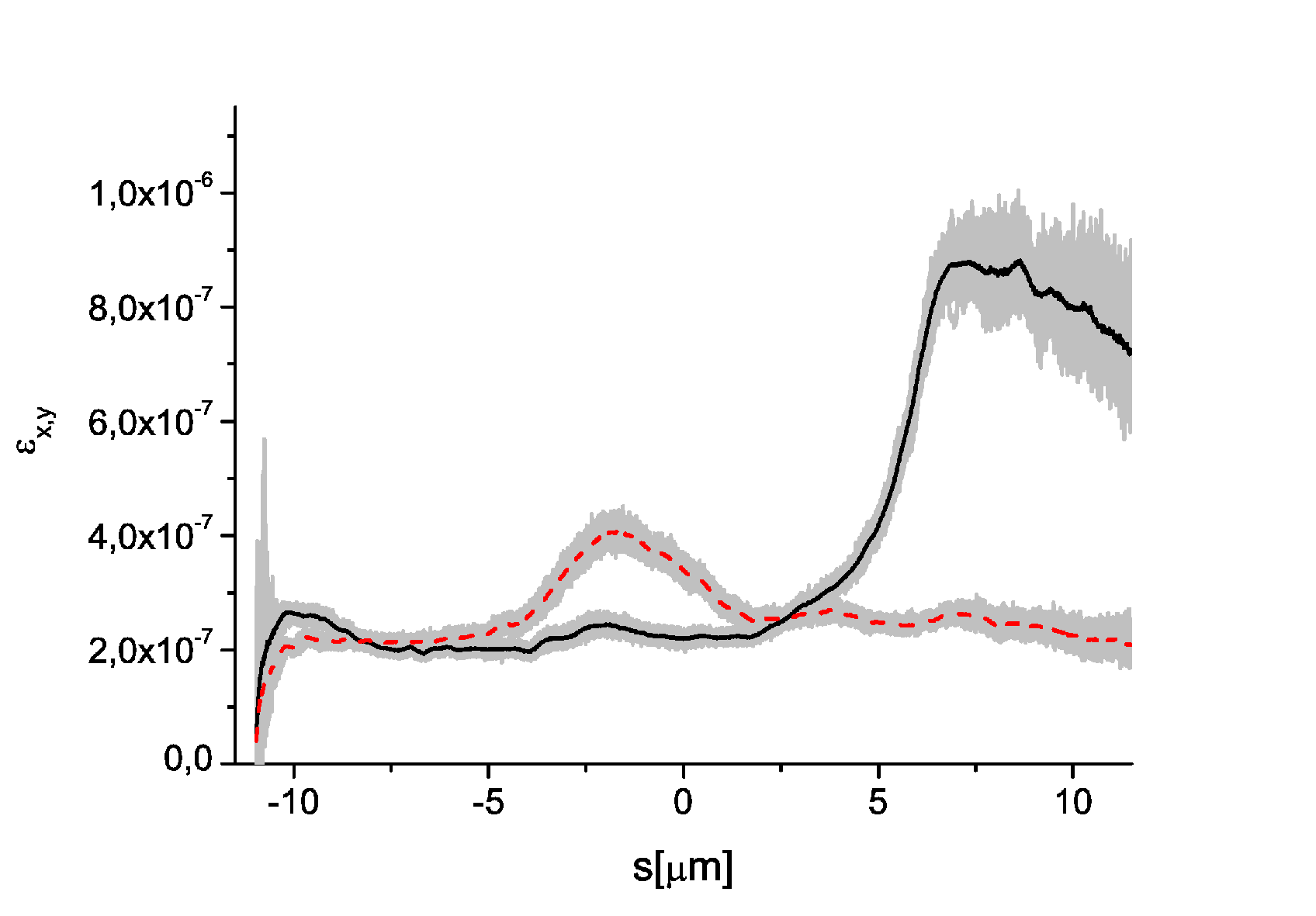}
\includegraphics[width=0.5\textwidth]{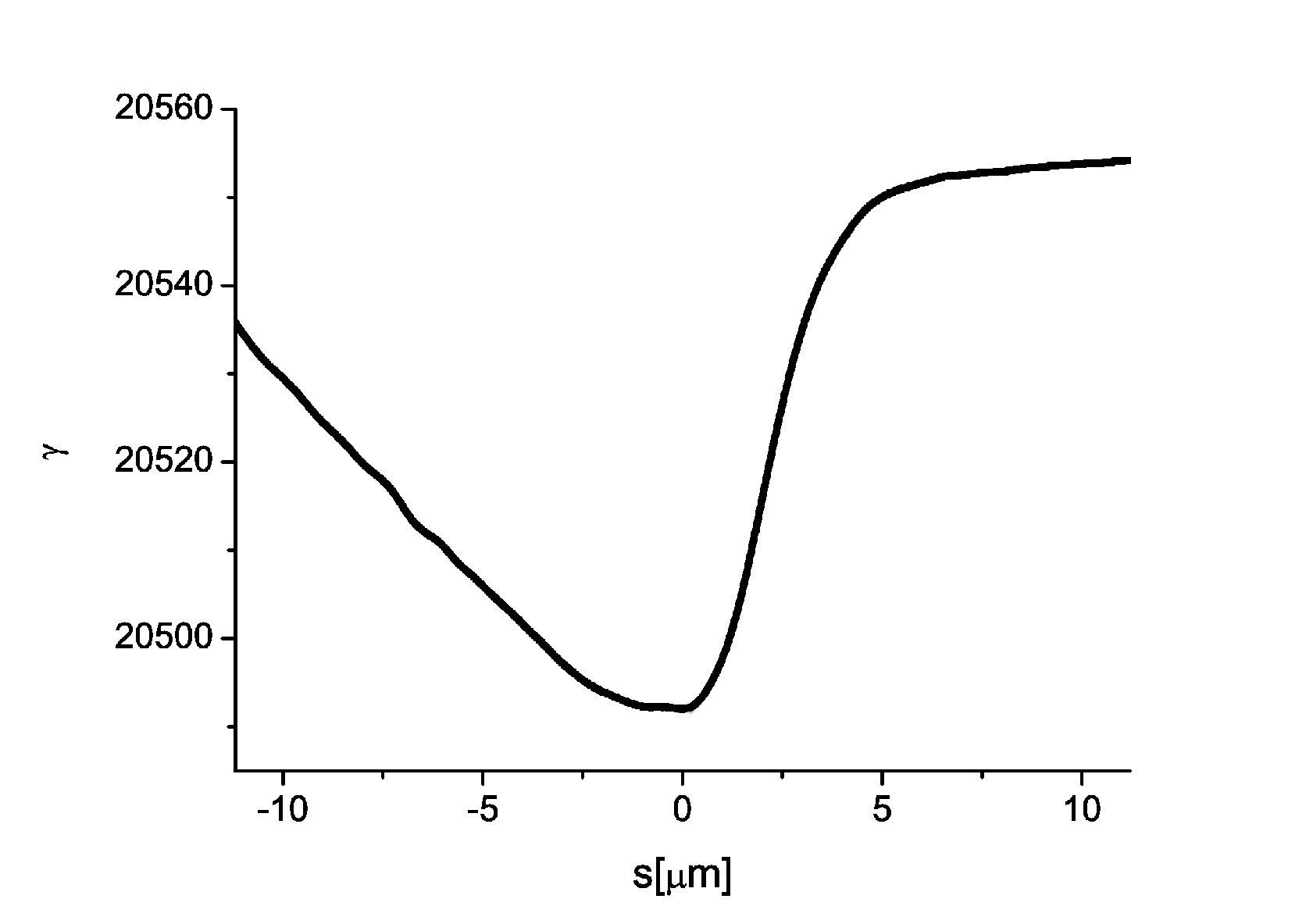}
\includegraphics[width=0.5\textwidth]{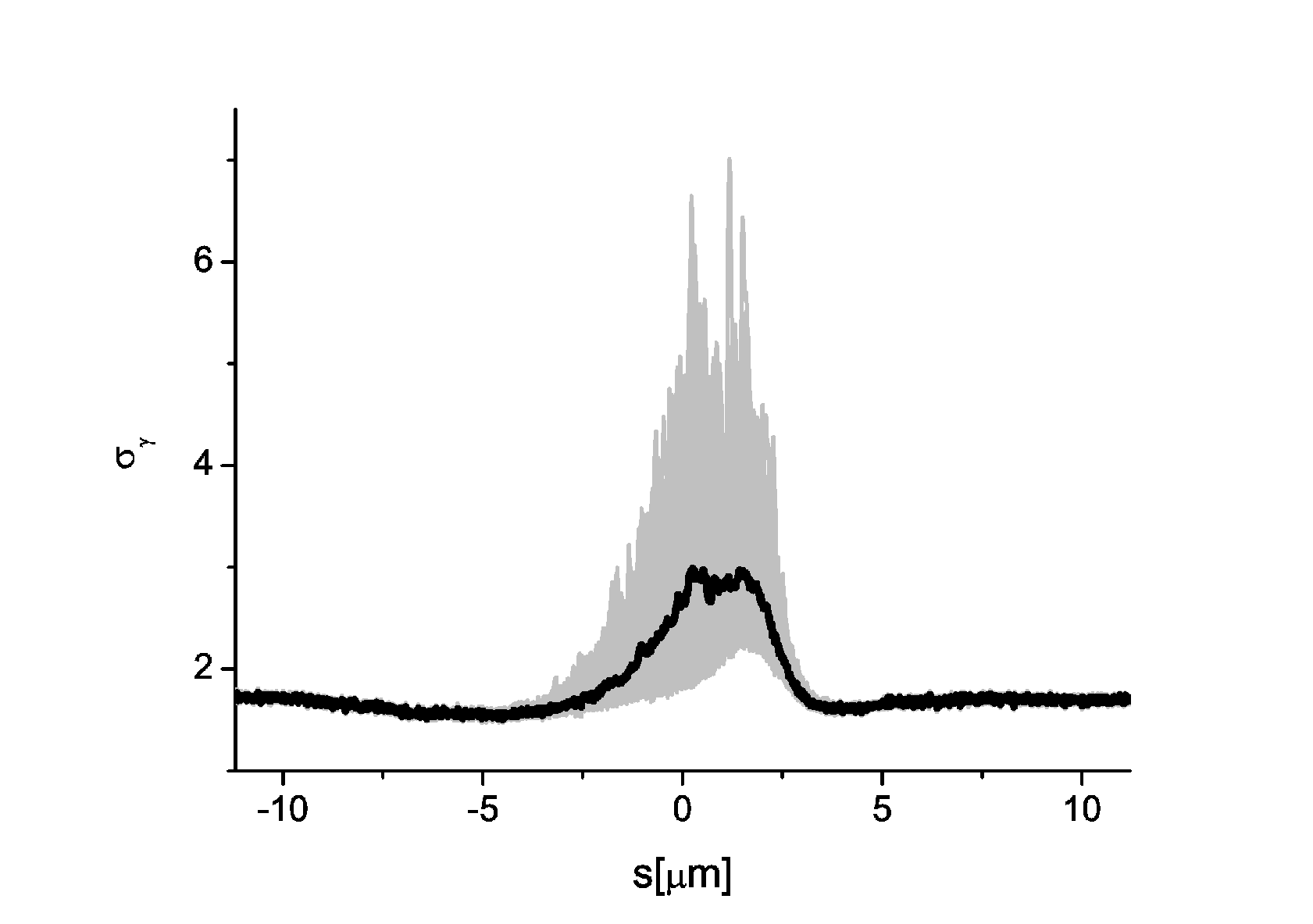}
\begin{center}
\includegraphics[width=0.5\textwidth]{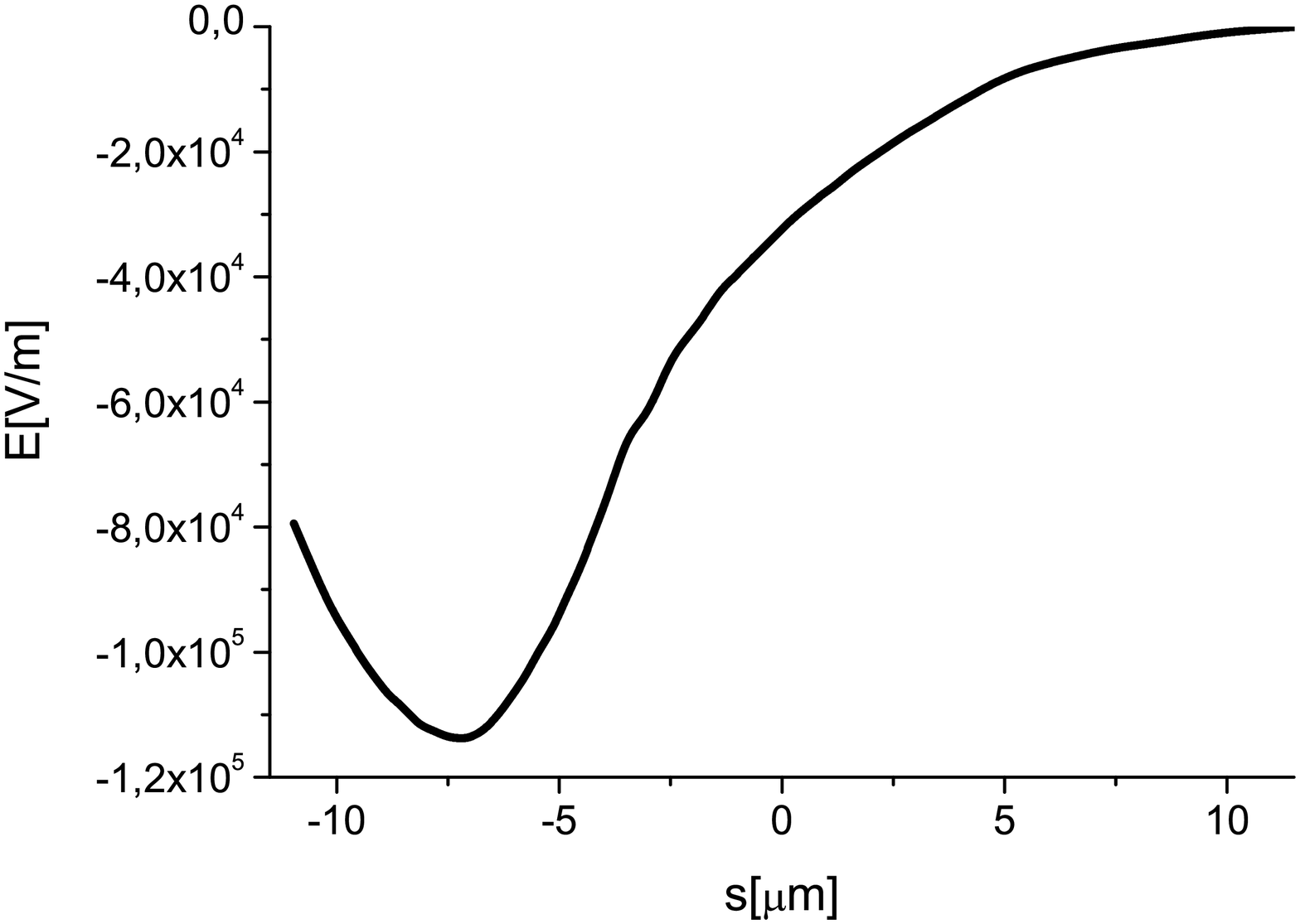}
\end{center}
\caption{Electron beam characteristics after the second magnetic
chicane. (First Row, Left) Current profile. (First Row, Right)
Normalized emittance as a function of the position inside the
electron beam. (Second Row, Left) Energy profile along the beam.
(Second Row, Right) Electron beam energy spread profile. (Bottom
row) Resistive wakefields in the SASE3 undulator \cite{S2ER}.}
\label{biof19}
\end{figure}
Since we now deal with a conventional grating monochromator, the
photon pulse is delayed with respect to the electron pule. In our
study case we assume a relatively large delay of about $3$ ps. In
order to compensate for such delay, one needs a chicane with a
relatively large dispersion strength $R_{56} \sim 2$ mm. This
assumption is conservative, and as the design efforts of the LCLS
crew continue, a better monochromator design with shorter photon
delay is likely to be designed, which will have positive effects on
the design of the chicane. However, assuming for the time being
$R_{56} = 2$ mm, we cannot neglect, in principle, the effects of the
chicane dispersion on the electron bunch properties. We accounted
for them with the help of the code Elegant \cite{ELEG}, which was
used to propagate the electron beam distribution through the
chicane. Results are shown in Fig. \ref{biof19}.

\begin{figure}[tb]
\includegraphics[width=0.5\textwidth]{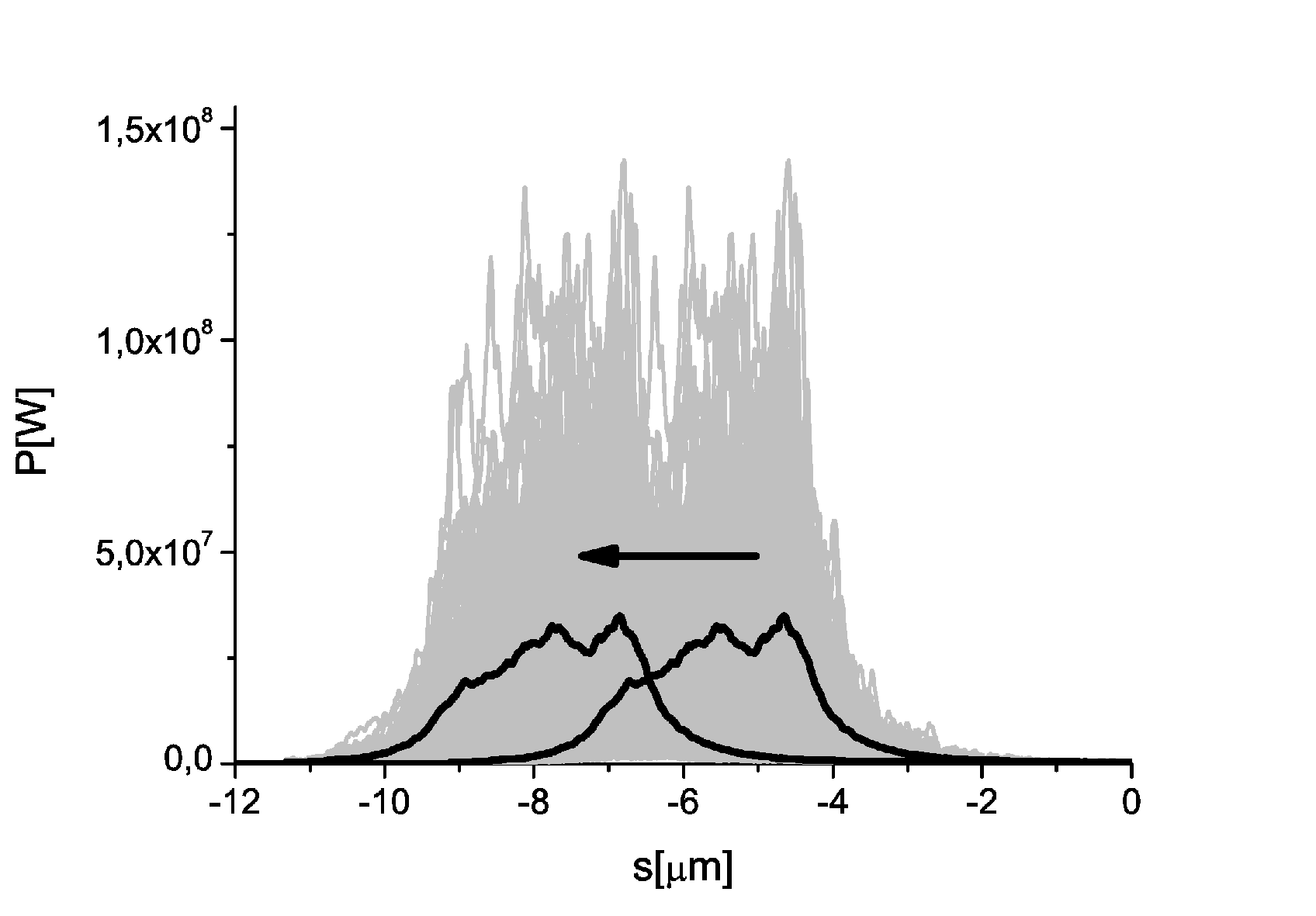}
\includegraphics[width=0.5\textwidth]{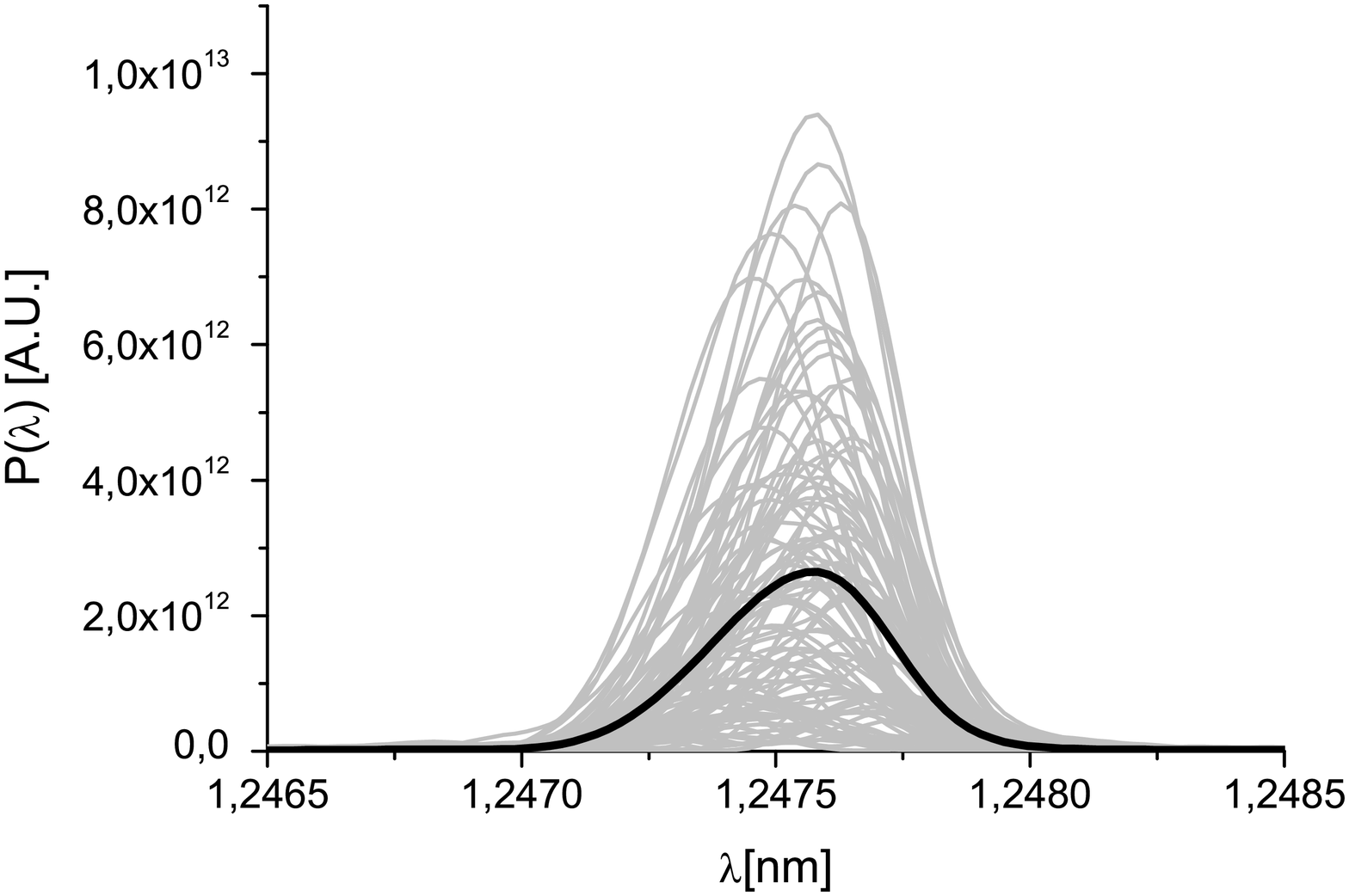}
\caption{Power and spectrum at the fundamental harmonic after the
second chicane equipped with the X-ray optical delay line, delaying
the radiation pulse with respect to the electron bunch. Grey lines
refer to single shot realizations, the black line refers to the
average over a hundred realizations.} \label{biof20}
\end{figure}
Since the $R_{56}$ is large enough to wash out the electron beam
microbunching, we assume a fresh bunch, at the entrance of the
following undulator part constituted by $3$ undulator cells. This
means that the results in Fig. \ref{biof19} are taken to generate a
new beam file to be fed into GENESIS. The electron bunch is now
seeded with the monochromatized radiation pulse in Fig.
\ref{biof18}, so that the seed is amplified in the $3$ undulator
cells following the chicane. After that, the electron beam is sent
through the second chicane, while the radiation pulse goes through
the X-ray optical delay line described in Fig. \ref{biof2}, where
the radiation pulse is delayed of about $6$ fs with respect to the
electron beam, as shown in Fig. \ref{biof9}. The power and spectrum
of the radiation pulse after the optical delay line are shown in
Fig. \ref{biof20}, where the effect of the optical delay is
illustrated.

\begin{figure}[tb]
\includegraphics[width=0.5\textwidth]{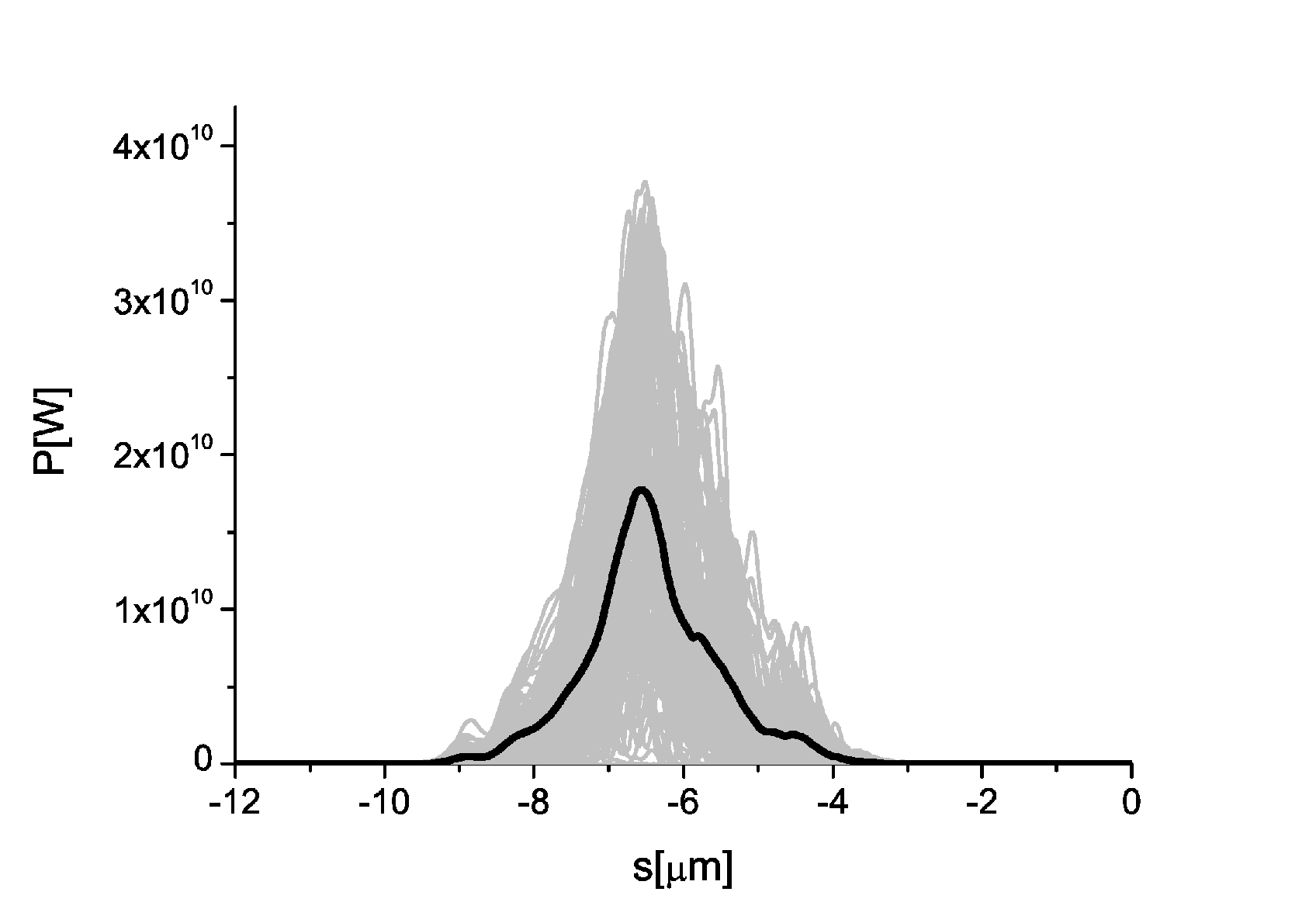}
\includegraphics[width=0.5\textwidth]{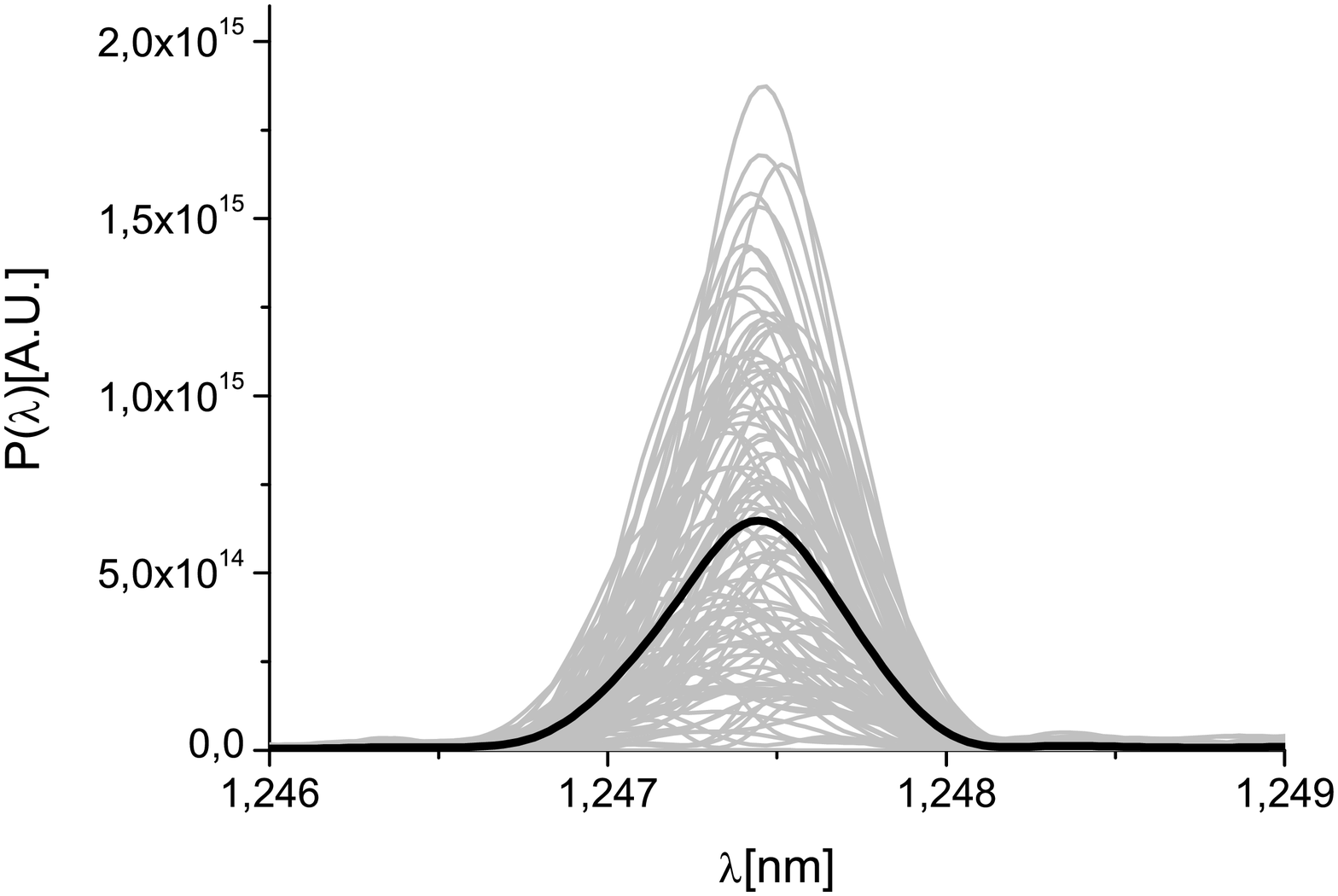}
\caption{Power and spectrum at the fundamental harmonic at the exit
of the third undulator and before the third magnetic chicane. Grey
lines refer to single shot realizations, the black line refers to
the average over a hundred realizations.} \label{biof20}
\end{figure}
Besides allowing for the installation of the optical delay line,
which delays the radiation pulse of about half of the electron bunch
size, the second chicane also smears out the microbunching in the
electron bunch. As a result, at the entrance of the third undulator
part the electron bunch can be considered as unmodulated, and half
of it is seeded with the radiation pulse. The seeded half of the
electron bunch amplifies the seed in the third undulator part,
composed by four cells. After that, electrons and radiation are
separated once more going through the third chicane. The hard X-ray
self-seeding crystal is out, and the chicane simply acts as a delay
line for the electron beam, which also smears out the microbunching.
Power and spectrum following the third chicane are shown in Fig.
\ref{biof20}.

\begin{figure}[tb]
\includegraphics[width=0.5\textwidth]{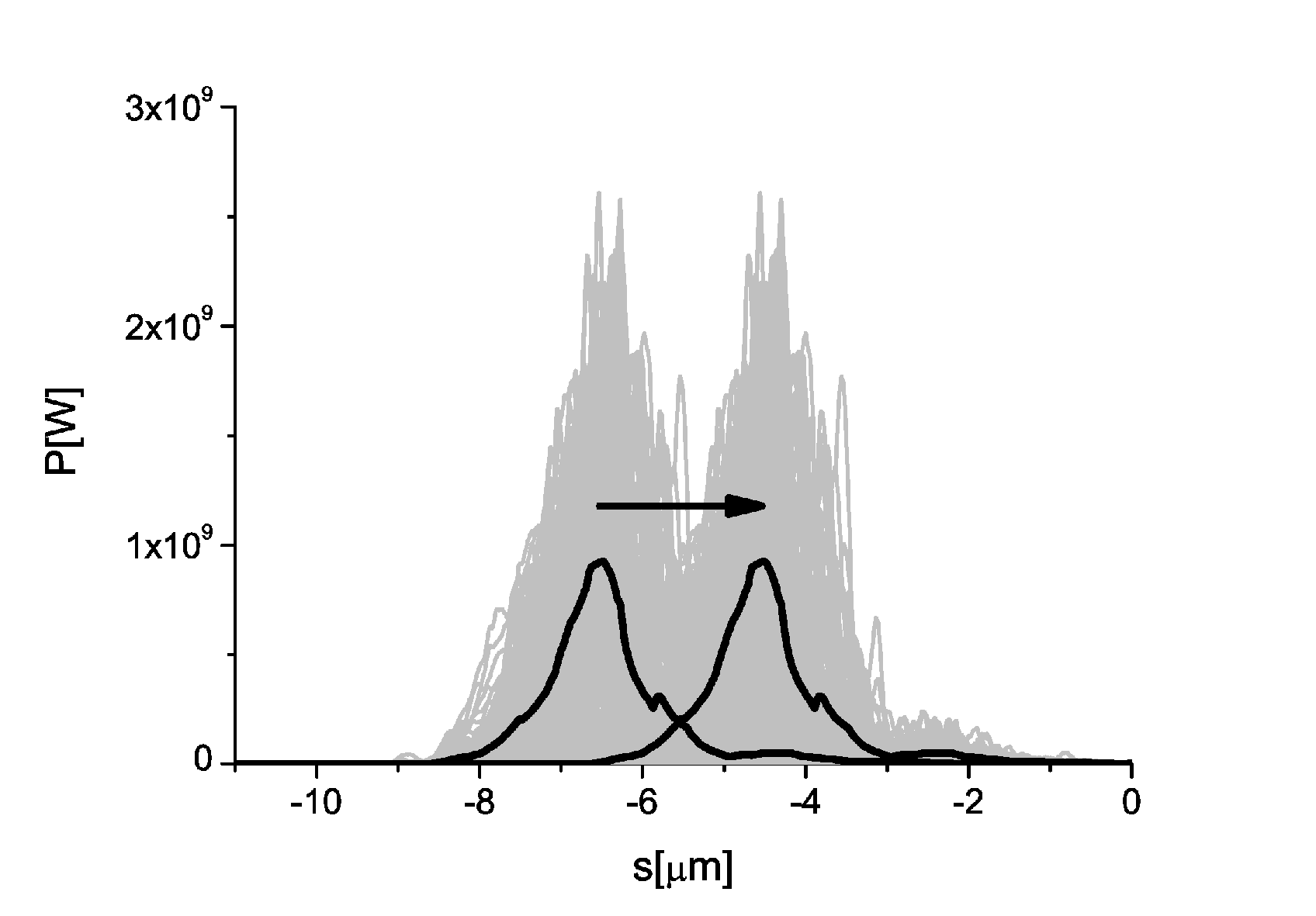}
\includegraphics[width=0.5\textwidth]{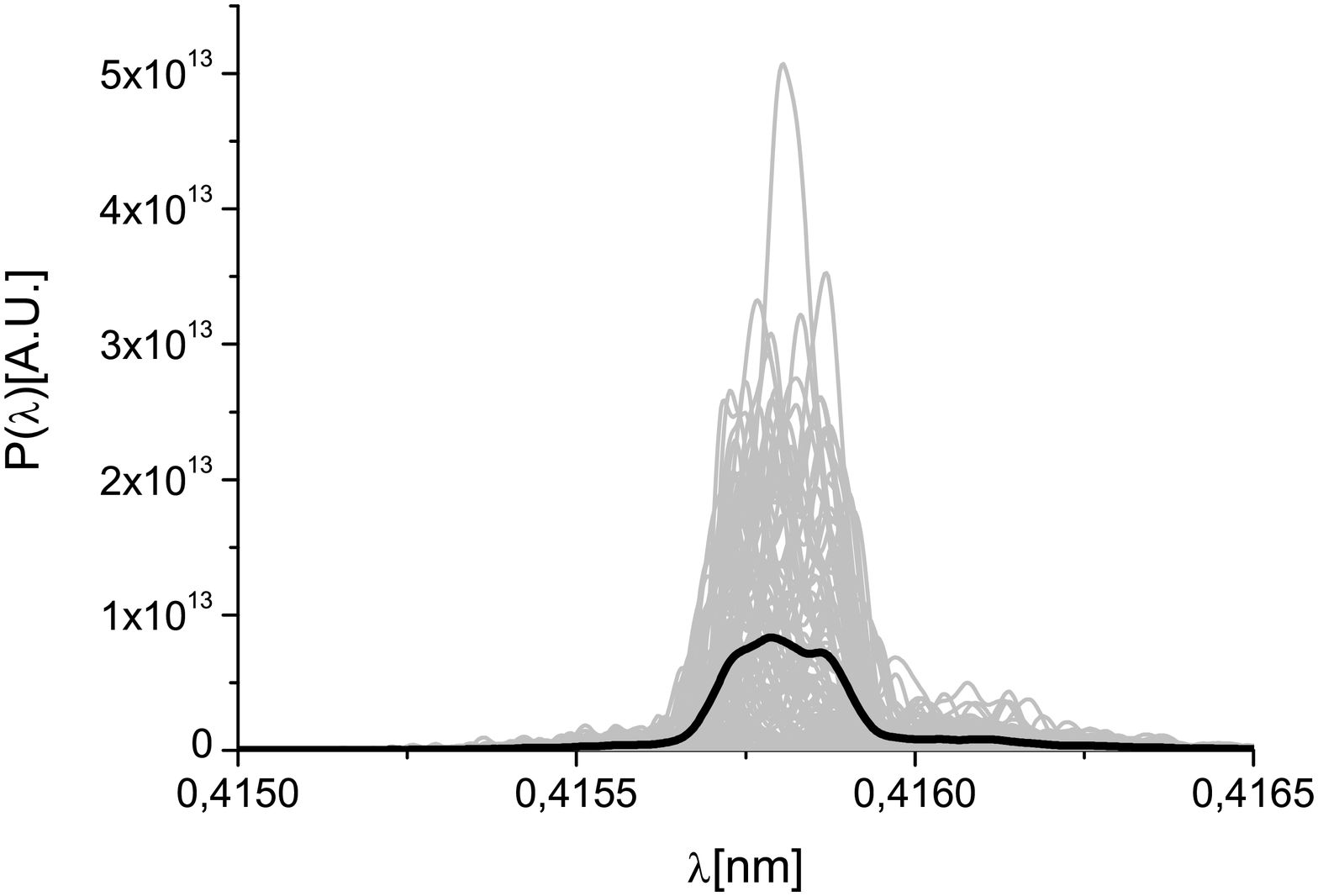}
\caption{Power and spectrum at the third harmonic after the third
magnetic chicane. Grey lines refer to single shot realizations, the
black line refers to the average over a hundred realizations.}
\label{biof21}
\end{figure}
The seeded part of the electron bunch is now spent, and its quality
has deteriorated too much for further lasing. However, by tuning the
third chicane in the proper way, one can superimpose the radiation
beam onto that part of the electron bunch that has not been seeded
in the third undulator part. This is fresh, and can lase again in
the fourth undulator part. The fourth undulator part is not tuned at
the fundamental harmonic, but rather at the third harmonic. This
allows to reach the photon energy range between $3$ keV and $5$ keV.
In fact, the power at the third harmonic, shown in Fig. \ref{biof21}
together with the relevant spectrum, is sufficient to act as a seed
in the last part of the undulator. Fig. \ref{biof21} also shows the
effect of the magnetic chicane, which delays the electron bunch
relative to the radiation pulse in order to allow for the seeding of
the fresh part of the bunch.

\begin{figure}[tb]
\begin{center}
\includegraphics[width=0.5\textwidth]{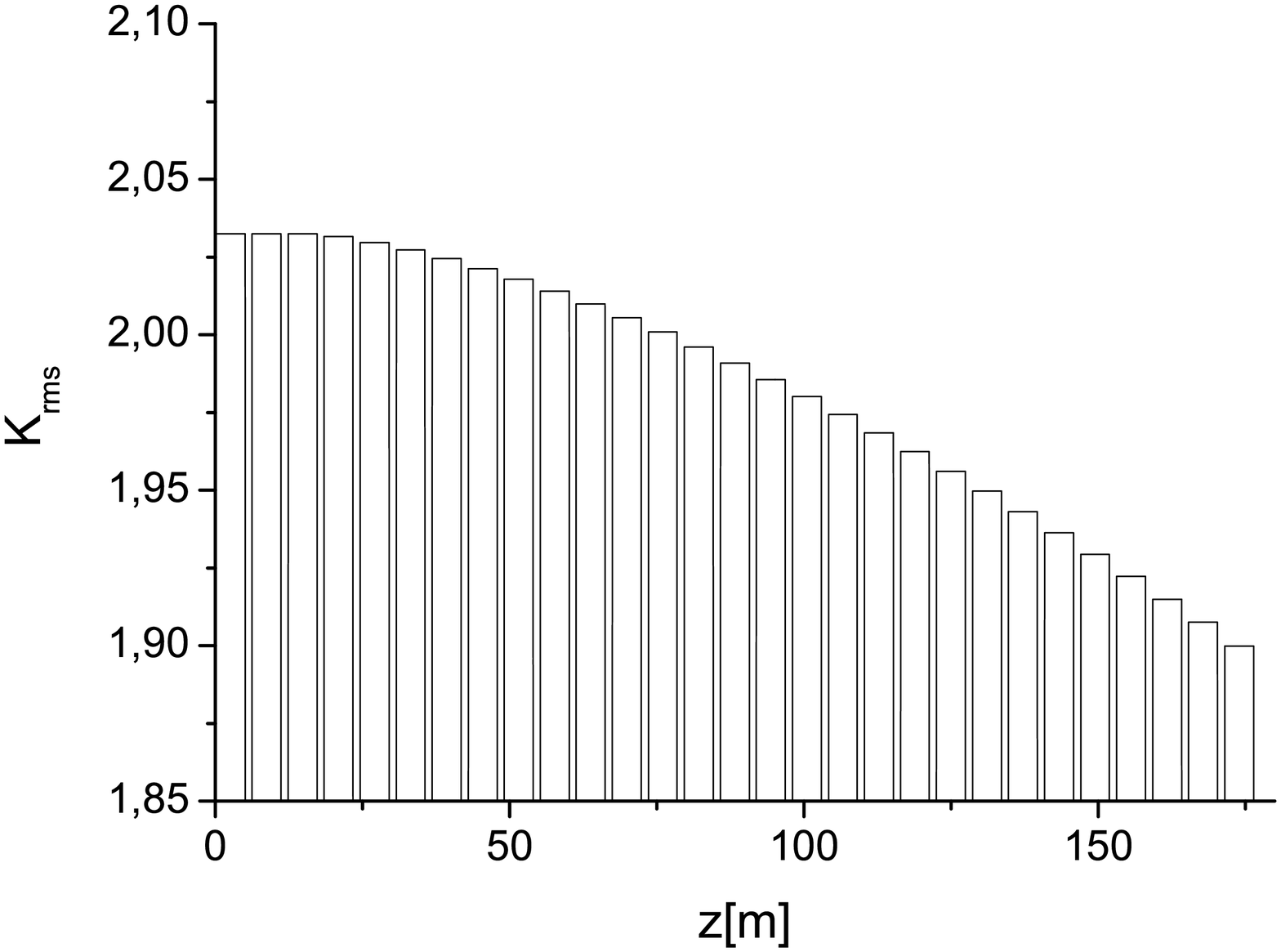}
\end{center}
\caption{Tapering law for the case $\lambda = 0.4$ nm.}
\label{biof22}
\end{figure}
The fourth and last part of the undulator is composed by $29$ cells,
and is partly tapered, post-saturation, to allow for increasing the
region where electrons and radiation interact properly to the
advantage of the radiation pulse. Tapering is implemented by
changing the $K$ parameter of the undulator segment by segment
according to Fig. \ref{biof22}. The tapering law used in this work
has been implemented on an empirical basis.

\begin{figure}[tb]
\includegraphics[width=0.5\textwidth]{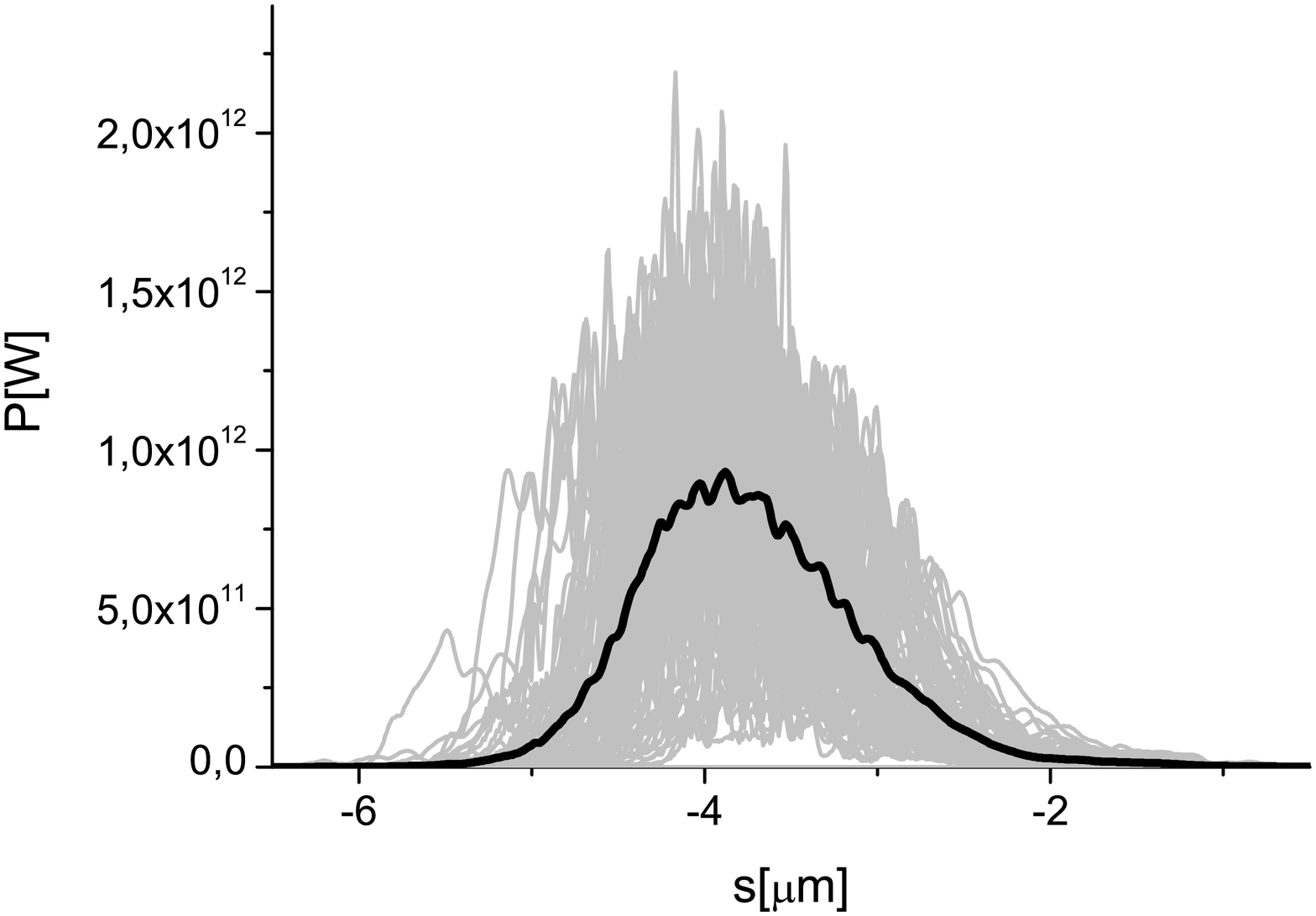}
\includegraphics[width=0.5\textwidth]{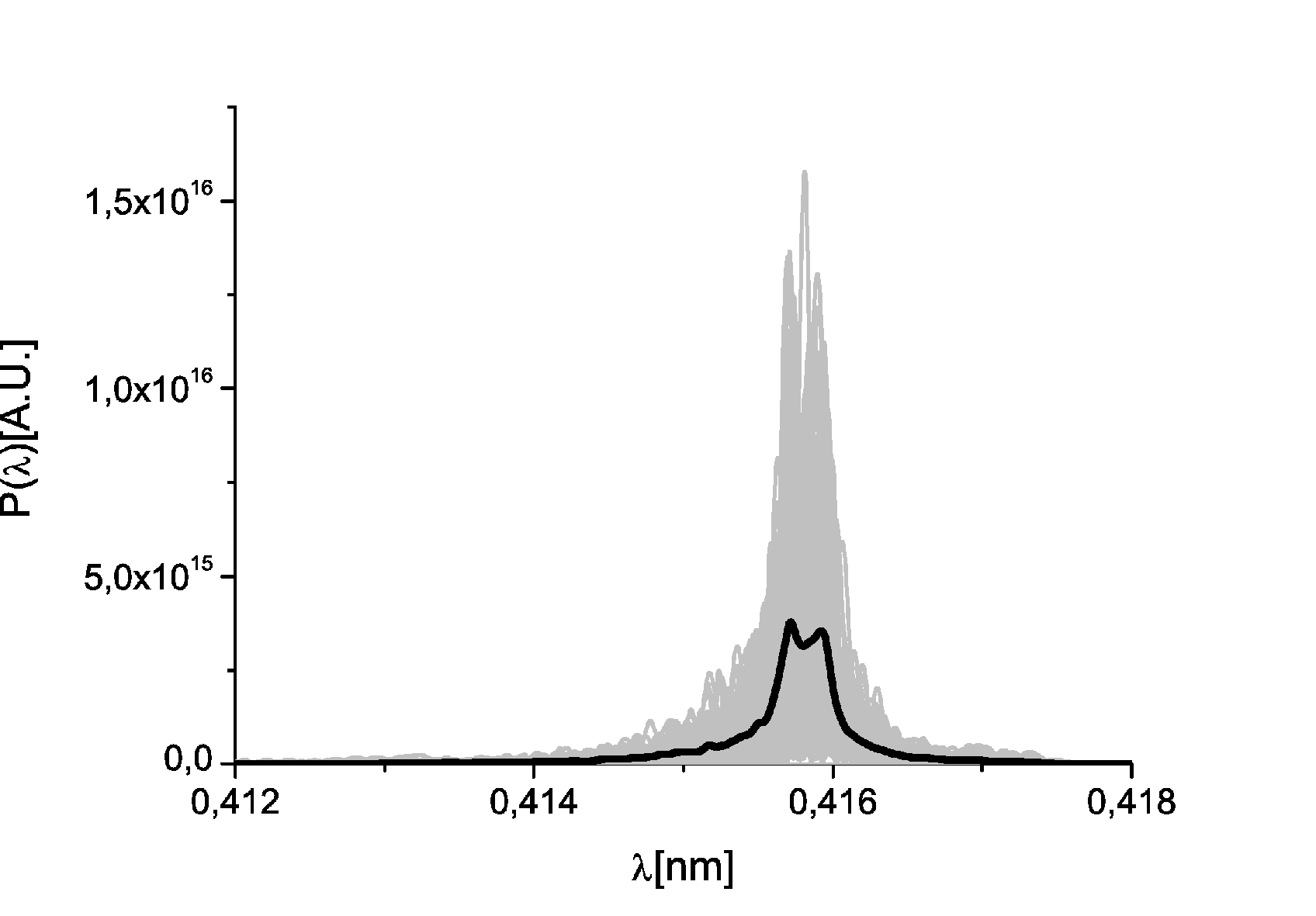}
\caption{Final output. Power and spectrum at the third harmonic
after tapering. Grey lines refer to single shot realizations, the
black line refers to the average over a hundred realizations.}
\label{biof23}
\end{figure}
The use of tapering together with monochromatic radiation is
particularly effective, since the electron beam does not experience
brisk changes of the ponderomotive potential during the slippage
process. The final output is presented in Fig. \ref{biof23} in terms
of power and spectrum. As one can see, simulations indicate an
output power of about $1$ TW.

\begin{figure}[tb]
\includegraphics[width=0.5\textwidth]{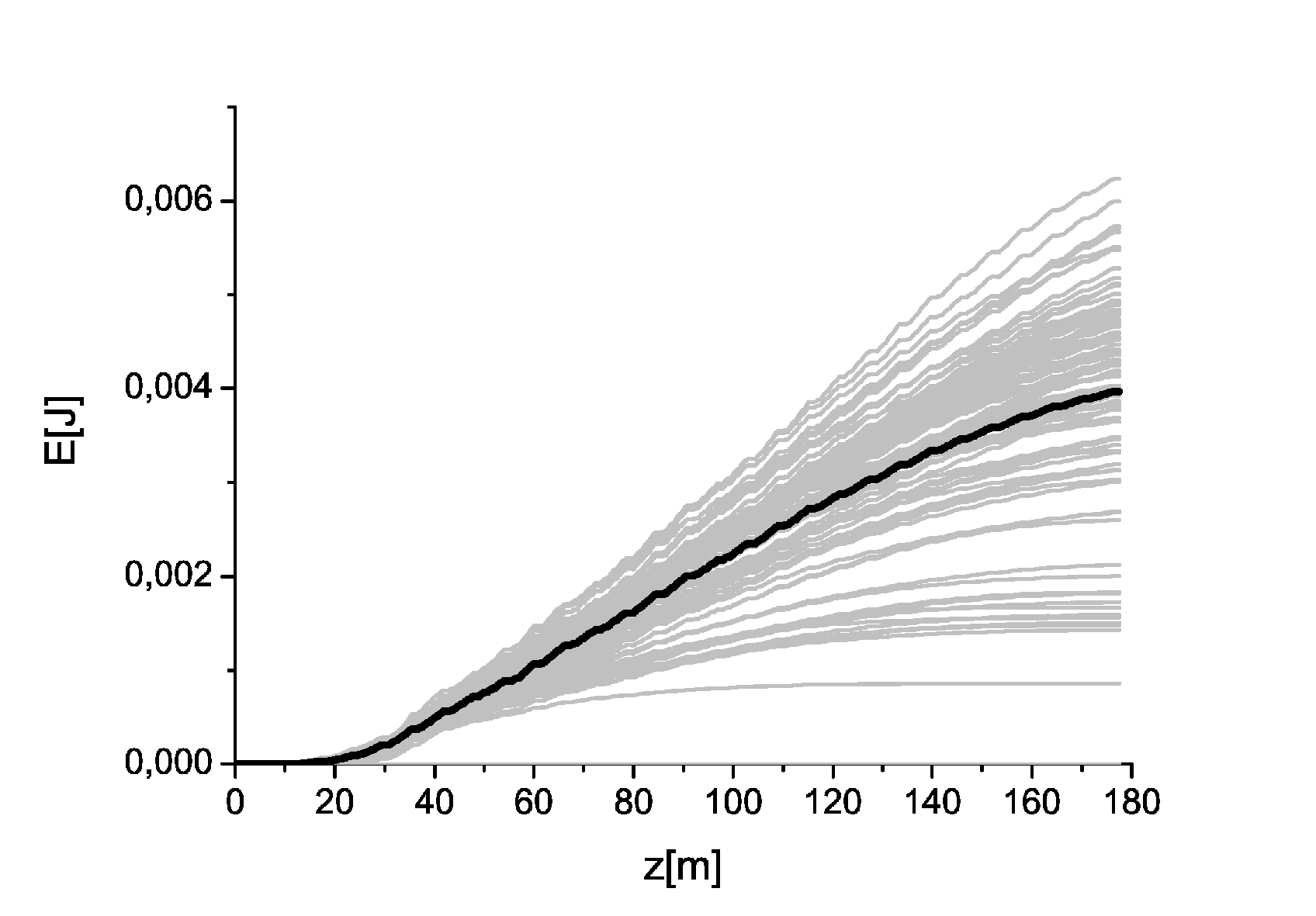}
\includegraphics[width=0.5\textwidth]{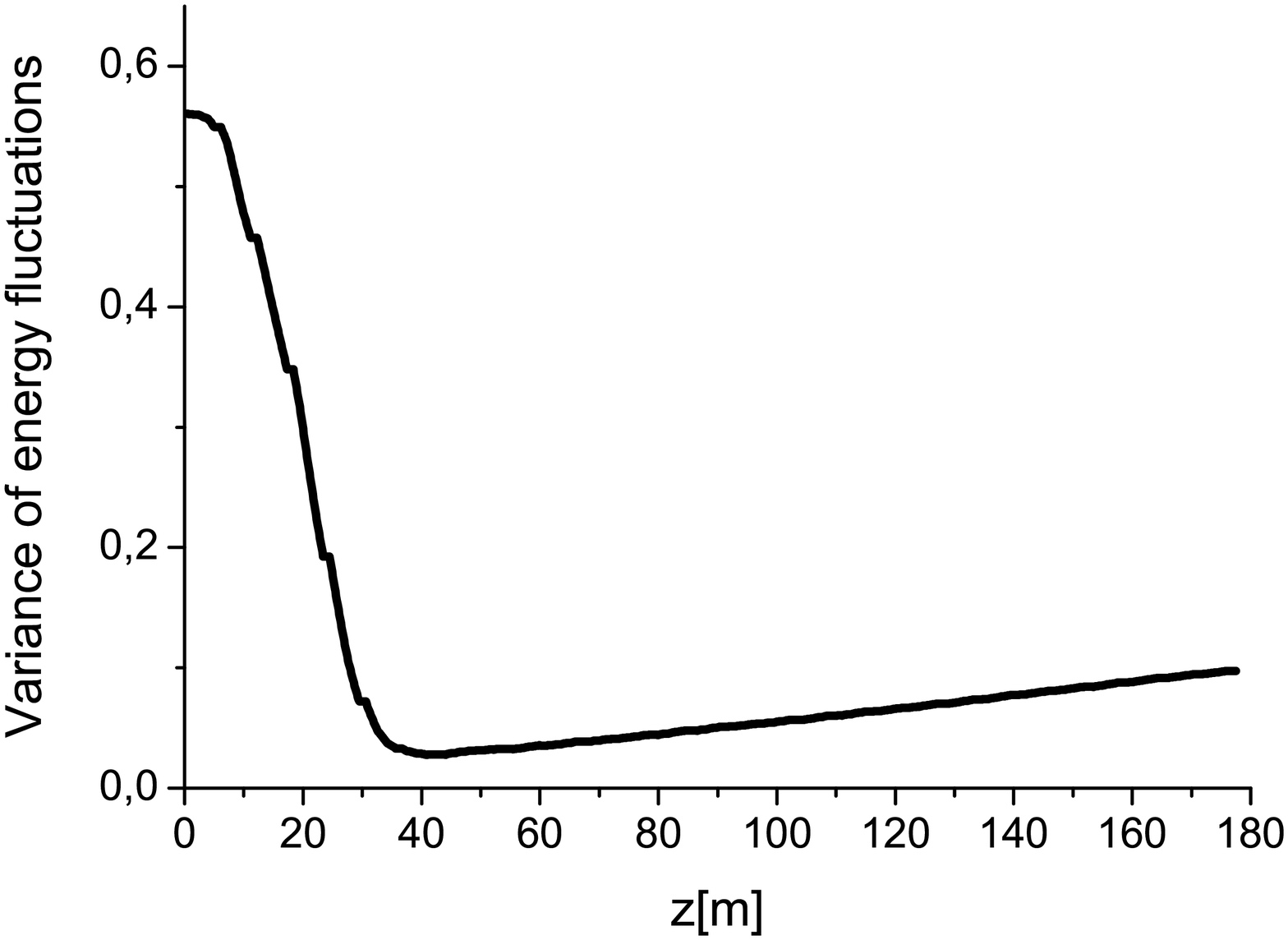}
\caption{Final output. Energy and energy variance of output pulses
for the case $\lambda = 0.4$ nm. In the left plot, grey lines refer
to single shot realizations, the black line refers to the average
over a hundred realizations.} \label{biof24}
\end{figure}

\begin{figure}[tb]
\includegraphics[width=0.5\textwidth]{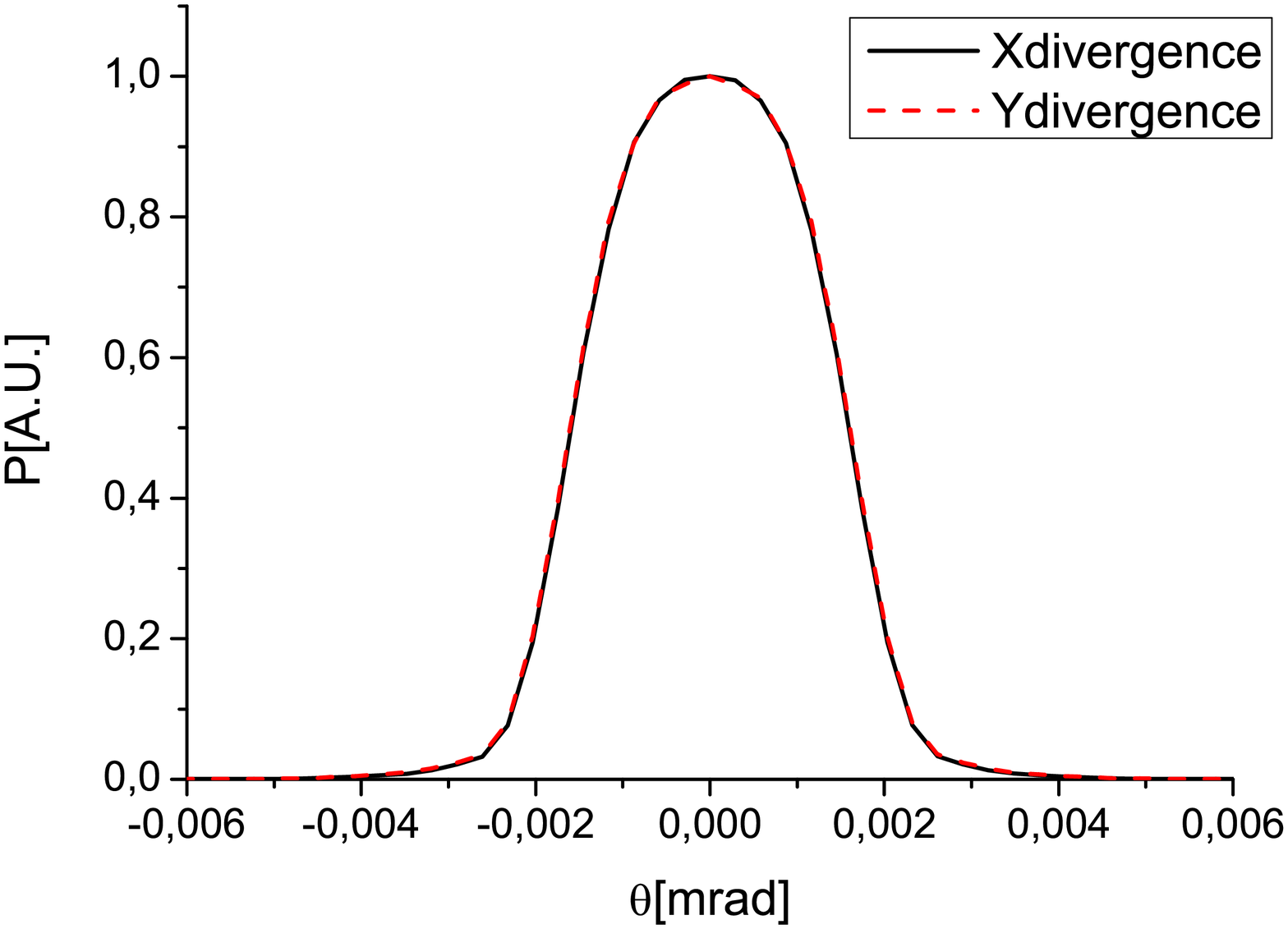}
\includegraphics[width=0.5\textwidth]{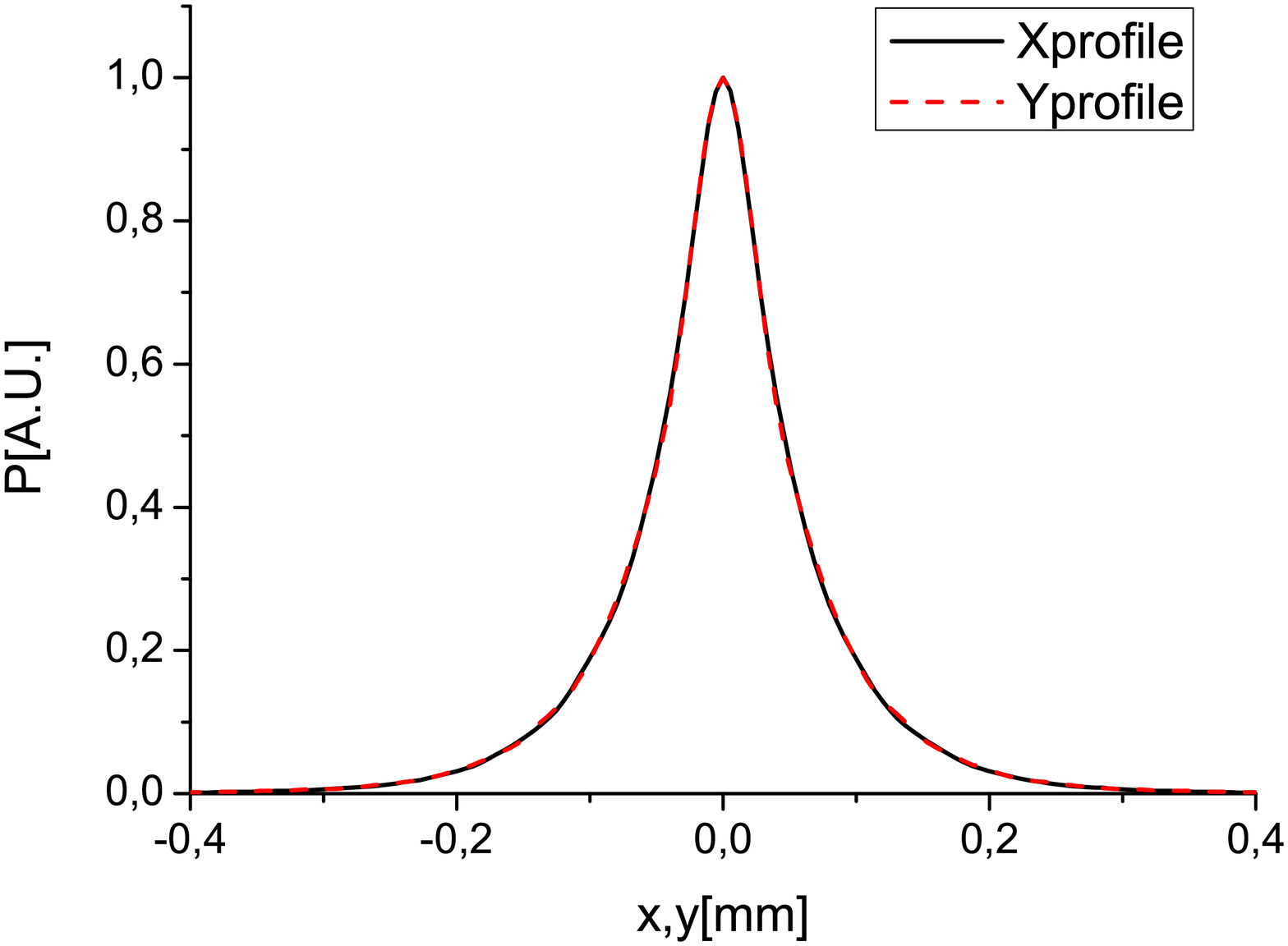}
\caption{Final output. X-ray radiation pulse energy distribution per
unit surface and angular distribution of the X-ray pulse energy at
the exit of output undulator for the case $\lambda = 0.4$ nm.}
\label{biof25}
\end{figure}
The energy of the radiation pulse and the energy variance are shown
in Fig. \ref{biof24} as a function of the position along the
undulator. The divergence and the size of the radiation pulse at the
exit of the final undulator are shown, instead, in Fig.
\ref{biof25}. In order to calculate the size, an average of the
transverse intensity profiles is taken. In order to calculate the
divergence, the spatial Fourier transform of the field is
calculated.

\begin{figure}[tb]
\includegraphics[width=0.5\textwidth]{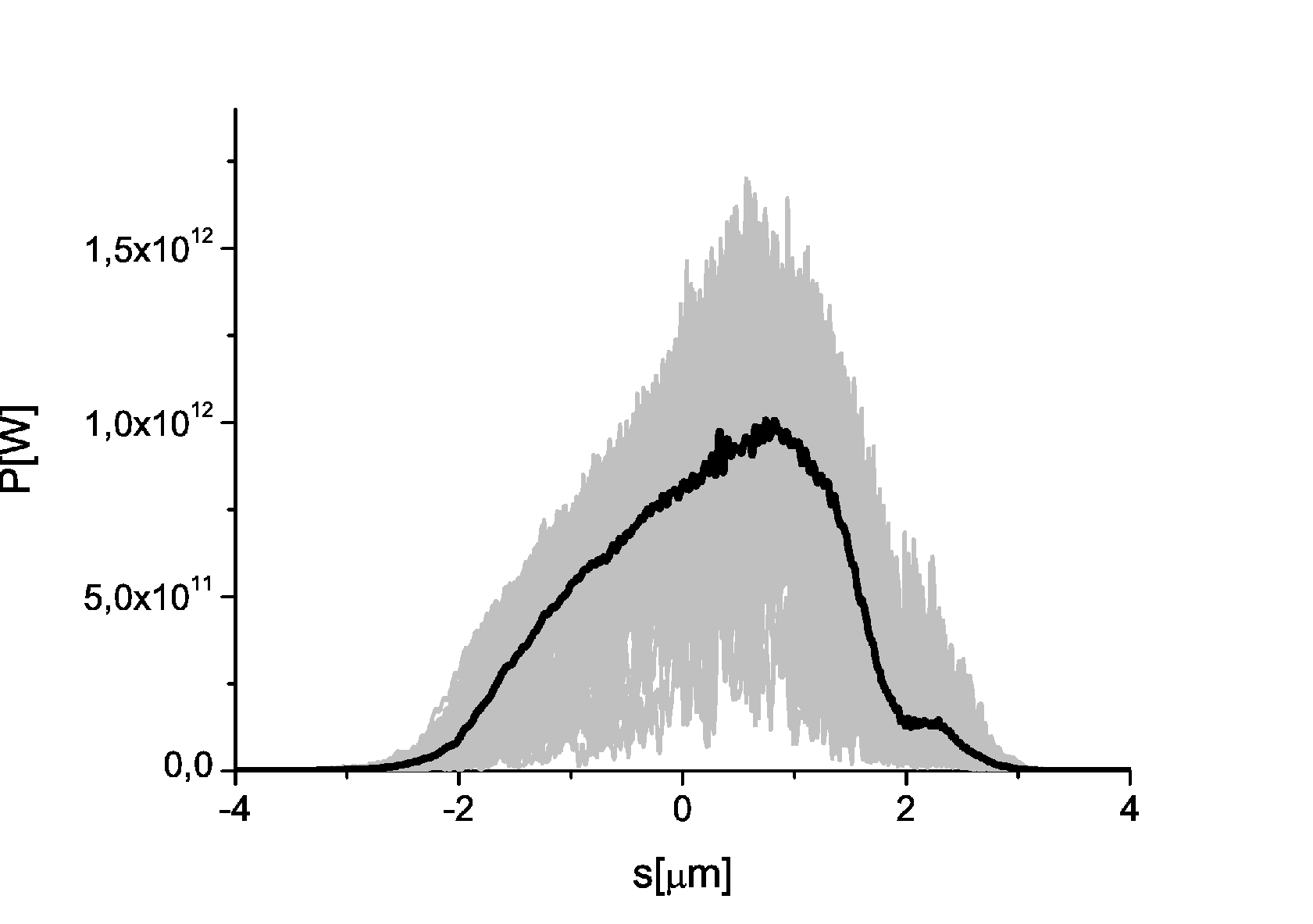}
\includegraphics[width=0.5\textwidth]{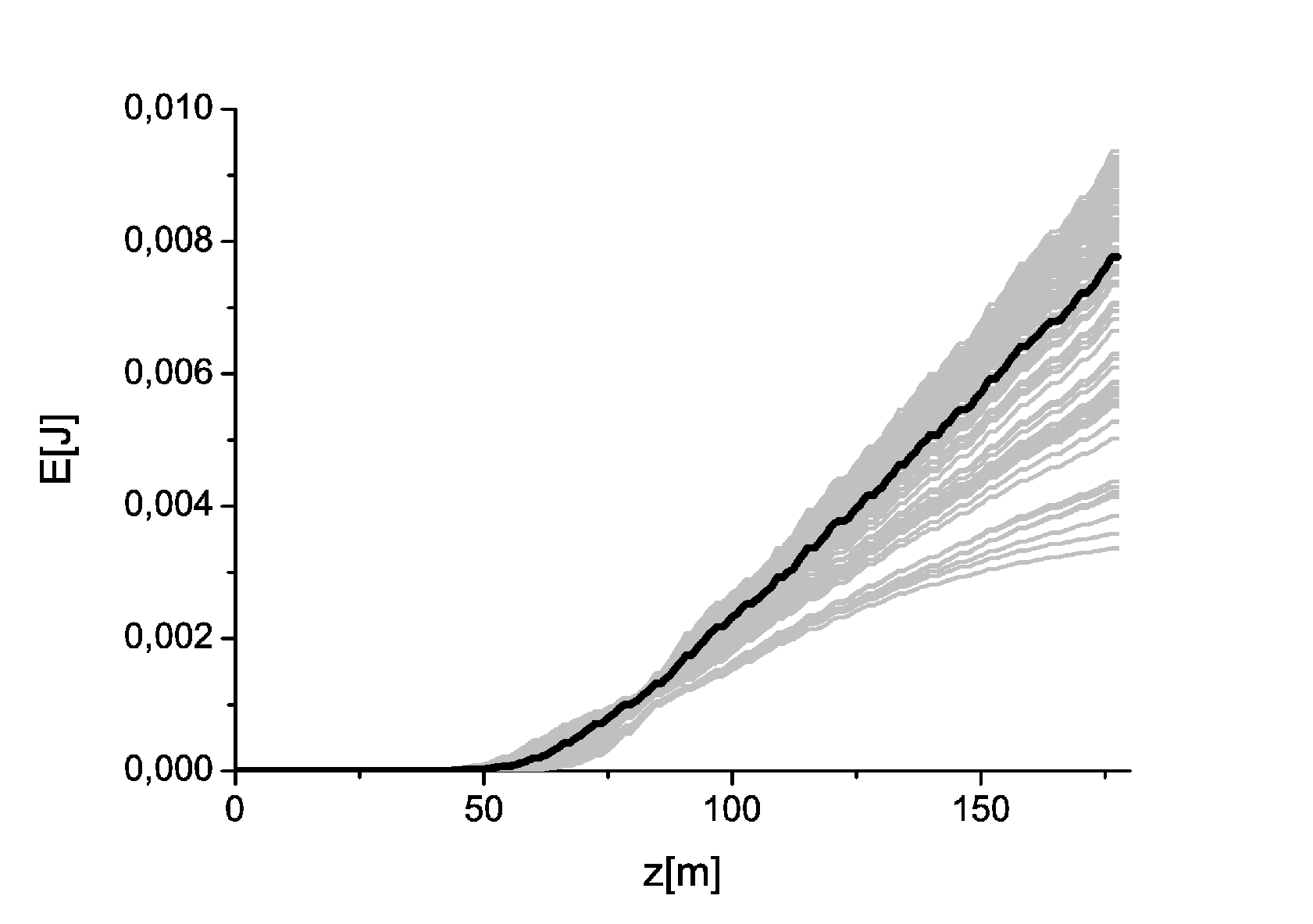}
\caption{Output radiation pulse obtained from simulations using
unperturbed electron bunch parameters (small momentum compaction
factor $R_{56}$ of first chicane) for the case $\lambda = 0.4$ nm.
They show (left) the temporal distribution of the output power and
(right) the energy per pulse as a function of the undulator length.
Note that the change in power and energy with respect to the  case
of perturbed electron bunch (where $R_{56} = 2$ mm), presented in
Fig. \ref{biof23} and Fig. \ref{biof24}, is relatively small. Grey
lines refer to single shot realizations, the black line refers to
the average over a hundred realizations.} \label{biof26}
\end{figure}
Finally, it is interesting to discuss the effect of the relatively
large value of the dispersion in the first chicane $R_{56} = 2$ mm.
As one can see by comparing Fig. \ref{biof16} and Fig. \ref{biof19}
the electron bunch characteristics changes in a seemingly
non-negligible way. However, using unperturbed electron bunch
parameters, that is assuming a negligible $R_{56}$ an thus using
Fig. \ref{biof16} for generating the beam file after the soft X-ray
self-seeding setup, results for the final power and energy in the
pulse do not appreciably change. This can  be seen by comparing Fig.
\ref{biof23} with Fig. \ref{biof26}, which is calculated in the
limit for a negligible $R_{56}$ in the first chicane. Small
differences in the behaviors of power and energy are due to slightly
different tuning of the simulation parameters.

\subsection{\label{subhard} Hard X-ray photon energy range above $8$ keV}

Finally, we consider the case described in Fig. \ref{biof9}, which
pertains the hard X-ray energy above $8$ keV.

\begin{figure}[tb]
\includegraphics[width=0.5\textwidth]{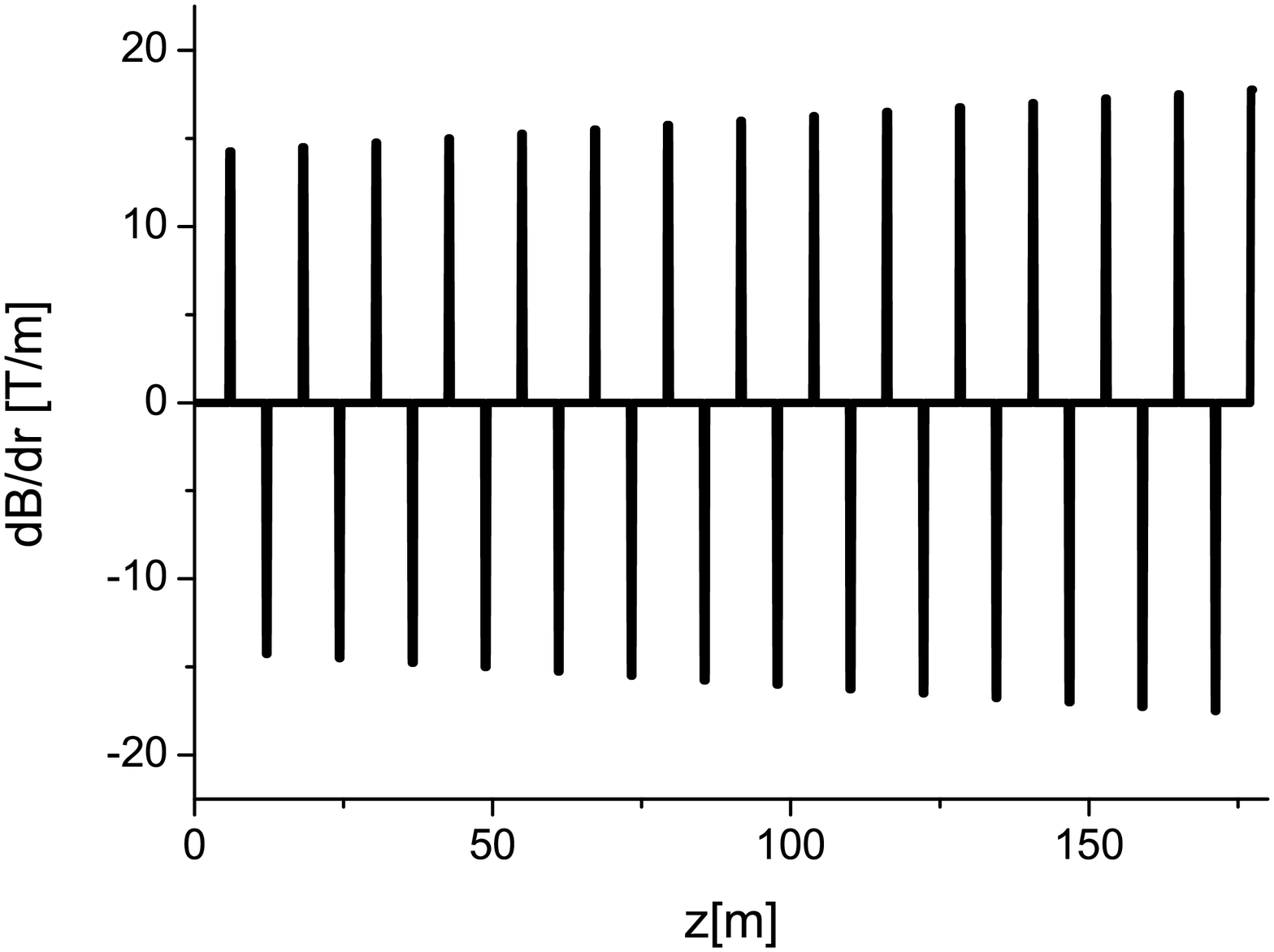}
\includegraphics[width=0.5\textwidth]{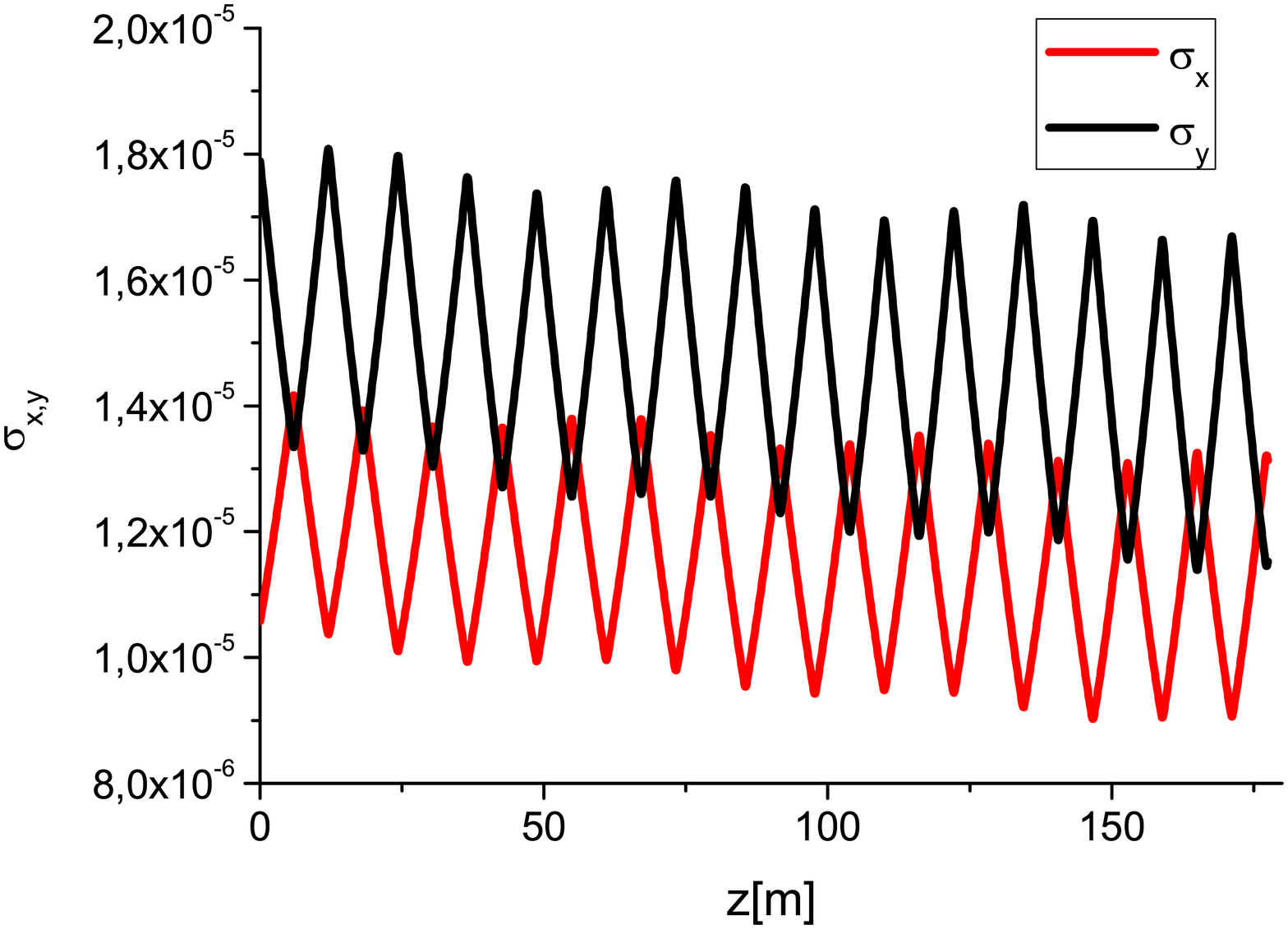}
\caption{Quadrupole strength (left) and electron beam size (right)
as a function of the position inside the undulator at $17.5$ GeV.
The electron beam size refers to the longitudinal position inside
the bunch corresponding to the maximum current value.}
\label{biofh7}
\end{figure}

The expected beam parameters at the entrance of the SASE3 undulator,
and the resistive wake inside the undulator are shown in Fig.
\ref{biof16b}. The evolution of the transverse electron bunch
dimensions are plotted in Fig. \ref{biofh7}, where the varying
quadrupole strength along the setup is also shown.

\begin{figure}[tb]
\includegraphics[width=0.5\textwidth]{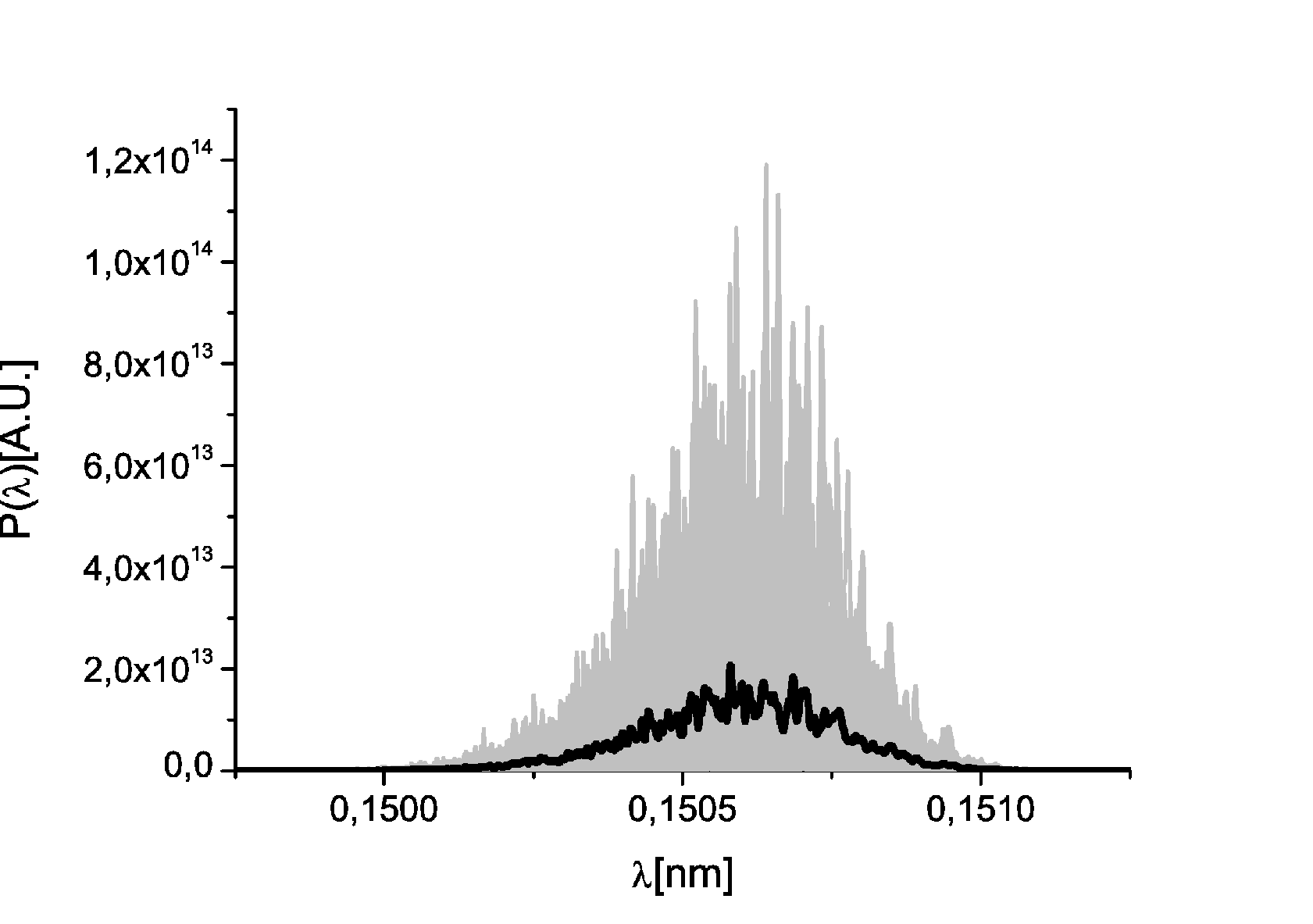}
\includegraphics[width=0.5\textwidth]{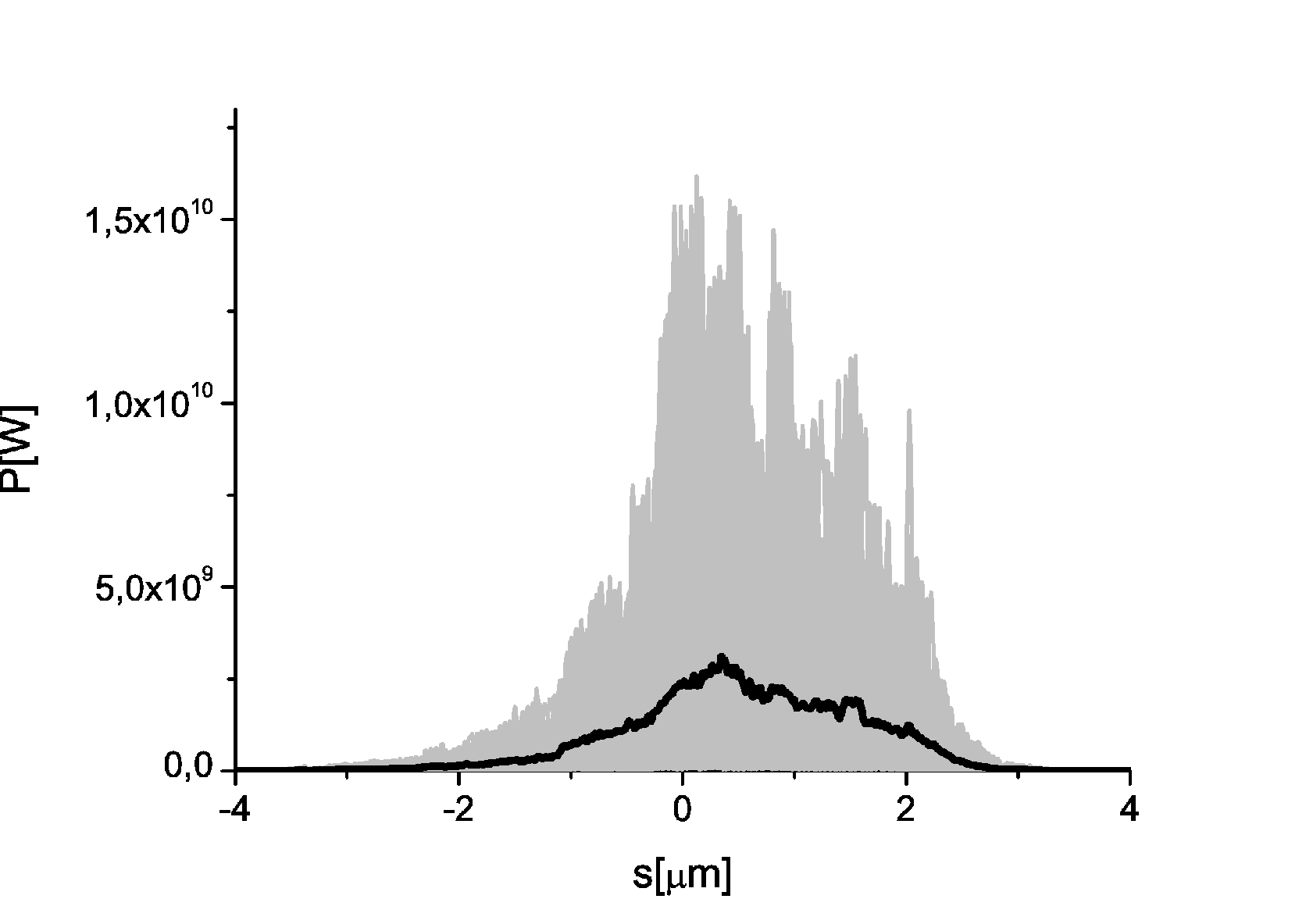}
\caption{Incident spectrum (left) and power (right) at the hard
X-ray self-seeding crystal setup.} \label{biofh9}
\end{figure}
The first two magnetic chicanes are now switched off, and both the
soft X-ray self-seeding setup and the X-ray optical delay line are
out. Therefore, the electron beam lases in SASE mode along the first
$11$ undulator cells before passing through the single-crystal
monochromator filter. The power and the spectrum of the incident
pulse are shown in Fig. \ref{biofh9}. The difference with respect to
a previously proposed hard X-ray self-seeding setup for the SASE1/2
lines (see \cite{OURY5}) is that in that case we were referring to a
cascade self-seeding setup, which considerably increases the
signal-to-noise ratio of the seeded signal compared to the SASE
background. This allowed for a considerably smaller incident power
level on the crystal (compare Fig. \ref{biofh9} with Fig. 13 of
reference \cite{OURY5}). In the present case the setup should run at
a repetition rate much smaller than in the case discussed in
\cite{OURY5}. As a result, the heat loading on the crystal will be
much less severe, and the requirement of a small incident power
level on the crystal can be relaxed.

\begin{figure}[tb]
\includegraphics[width=0.5\textwidth]{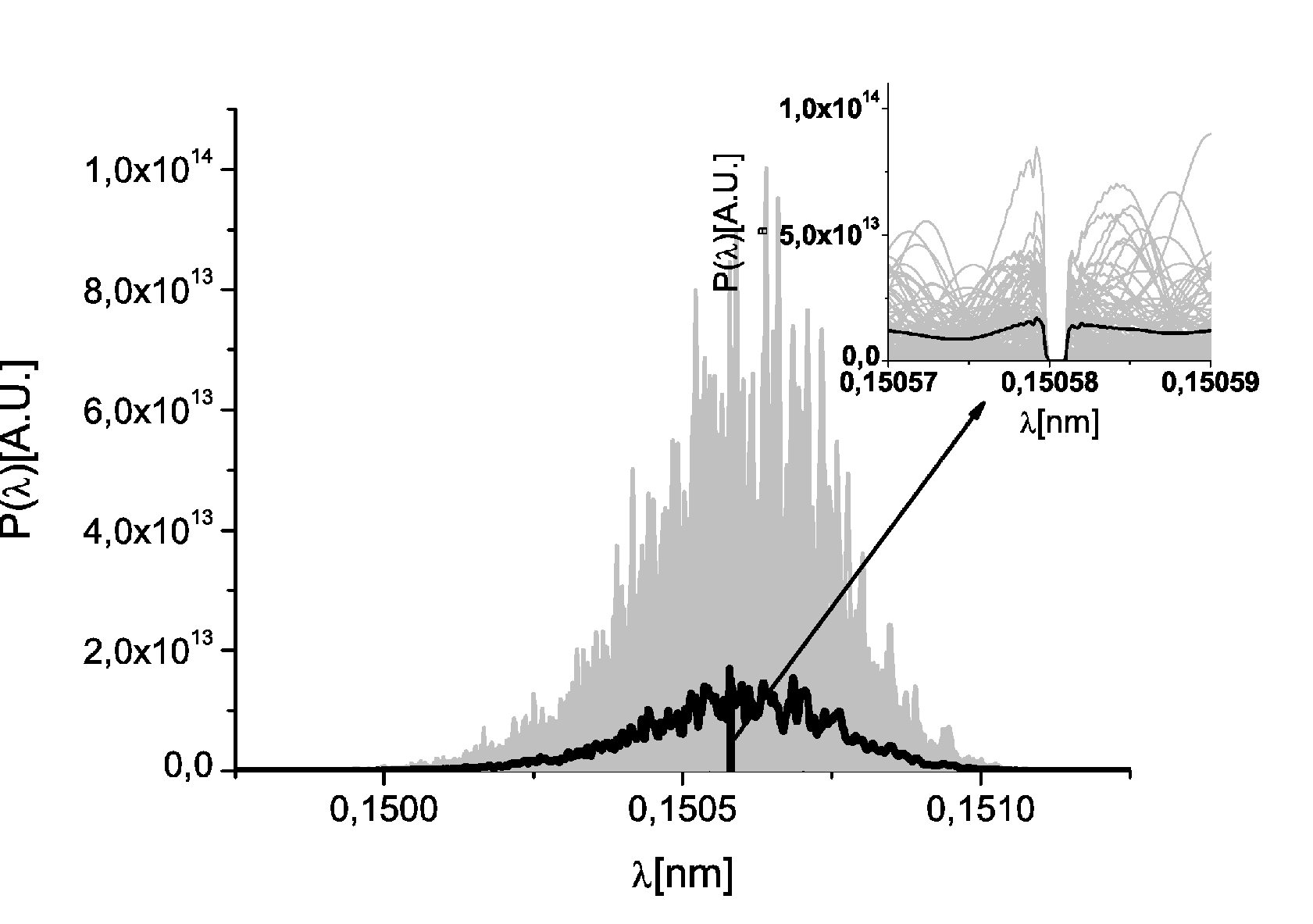}
\includegraphics[width=0.5\textwidth]{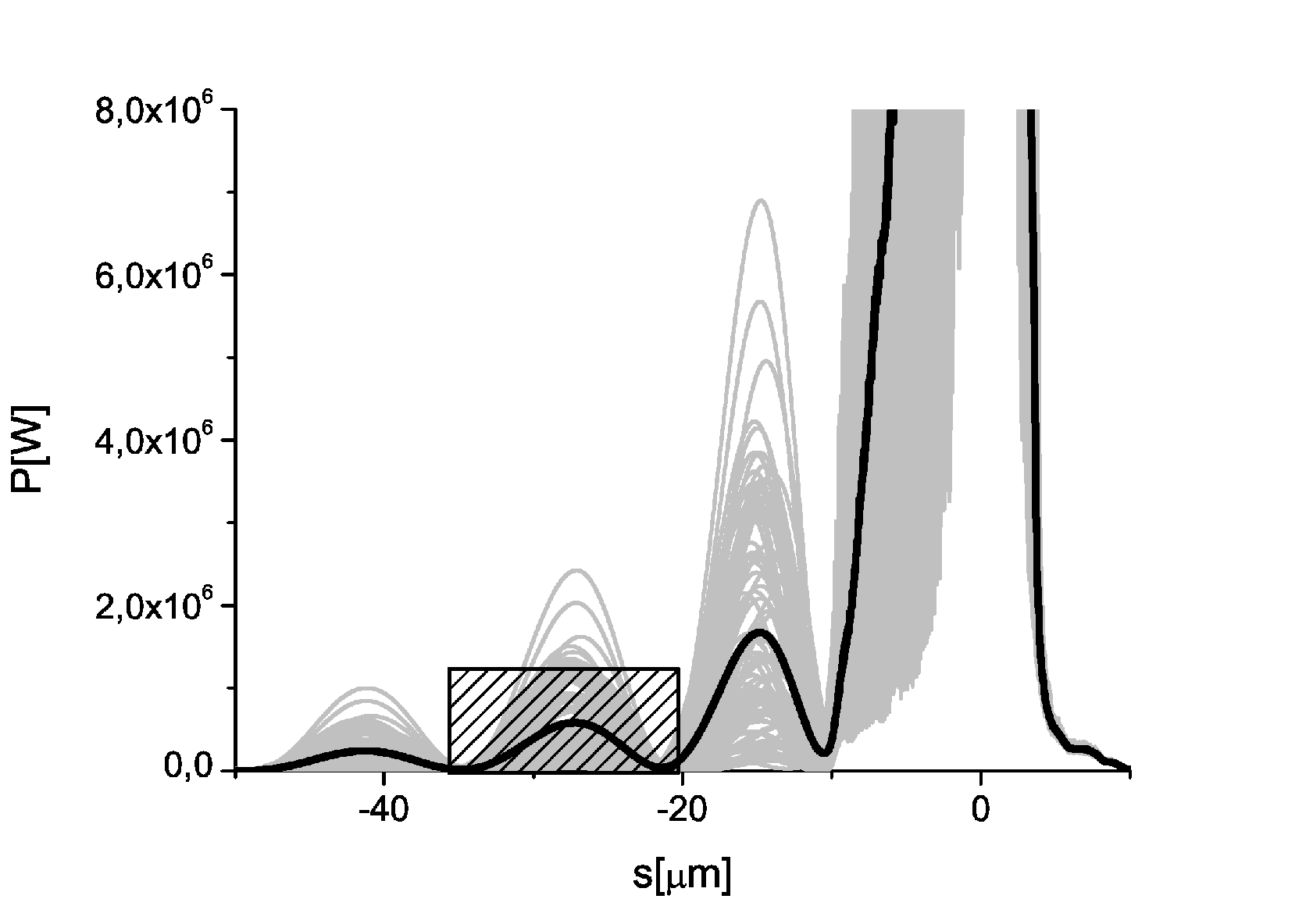}
\caption{Hard X-ray self-seeding mechanism in the frequency(left)
and in the time domain (right). In the frequency domain the diamond
crystal acts as a bandstop filter. Due to the presence of this
filter, a monochromatic tail follows the SASE pulse in the time
domain. The position of the electron bunch relative to such
monochromatic tail can be controlled after the chicane is
highlighted in the plot. } \label{biofh8}
\end{figure}
The effect of the filtering process is illustrated, both in the time
and in the frequency domain, in Fig. \ref{biofh8}. In the frequency
domain the filter acts as a bandstop filter, effectively drilling a
"hole" in the spectrum of the incident signal. In the time domain,
the outcoming pulse consists of the almost unperturbed incident
pulse transmitted through the crystal, followed by a monochromatic
tail of radiation, at much lower power level, which is due to the
presence of the crystal. The chicane is tuned in such a way that the
electron bunch is superimposed onto the monochromatic tail, which
now acts as seed.

\begin{figure}[tb]
\includegraphics[width=0.5\textwidth]{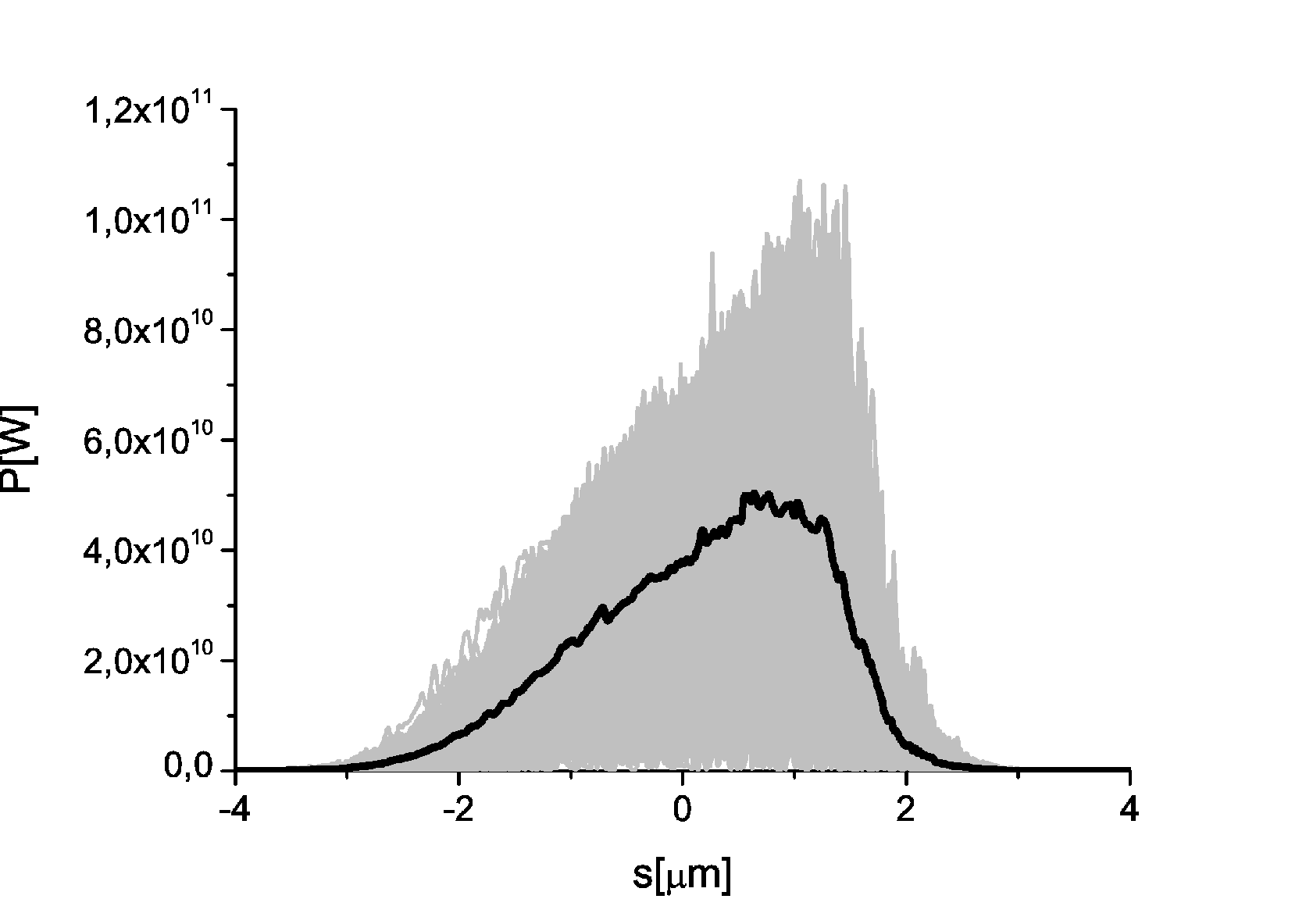}
\includegraphics[width=0.5\textwidth]{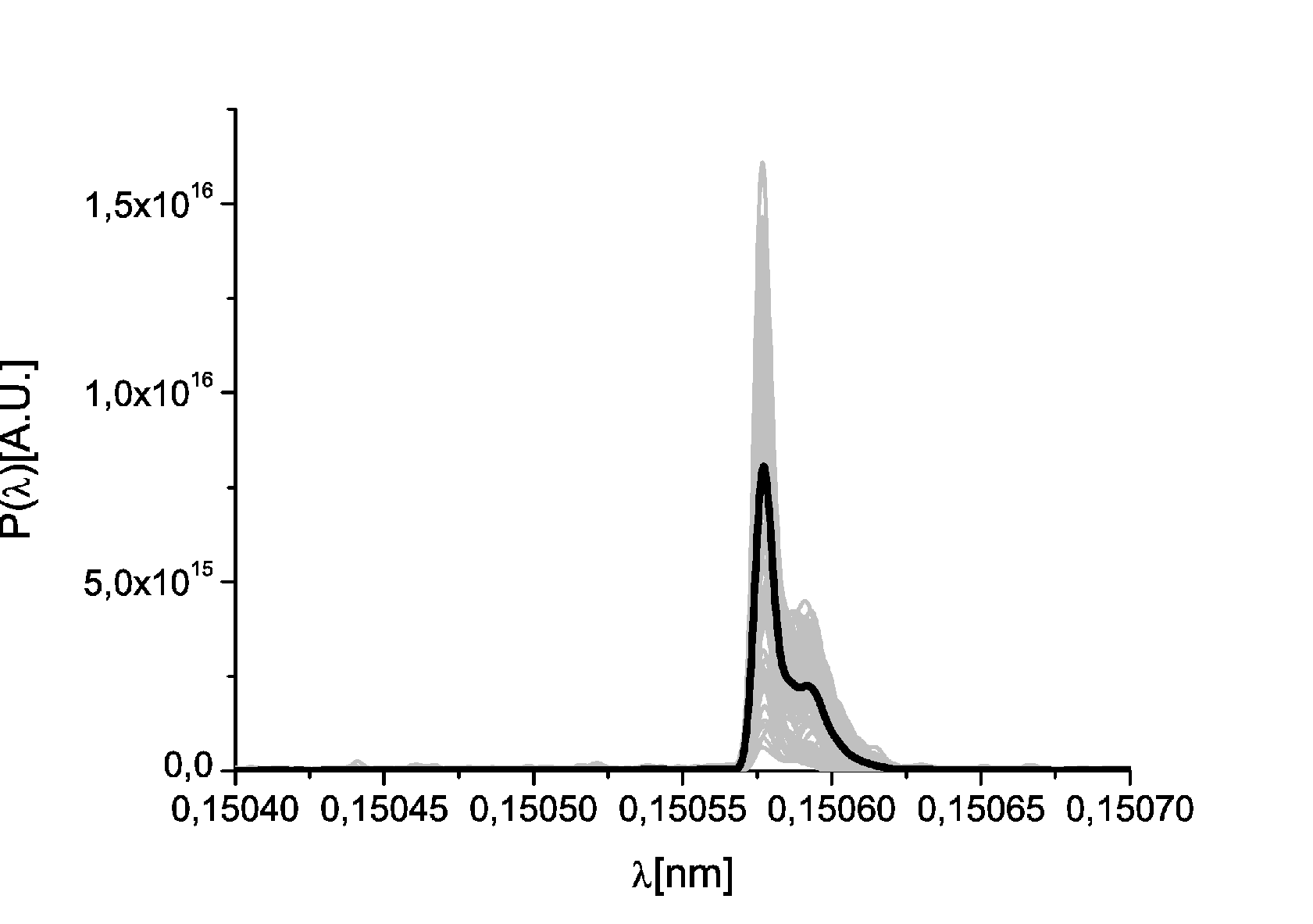}
\caption{Power and spectrum of output pulses at saturation. Grey
lines refer to single shot realizations, the black line refers to
the average over a hundred realizations.} \label{biofh2}
\end{figure}
\begin{figure}[tb]
\includegraphics[width=0.5\textwidth]{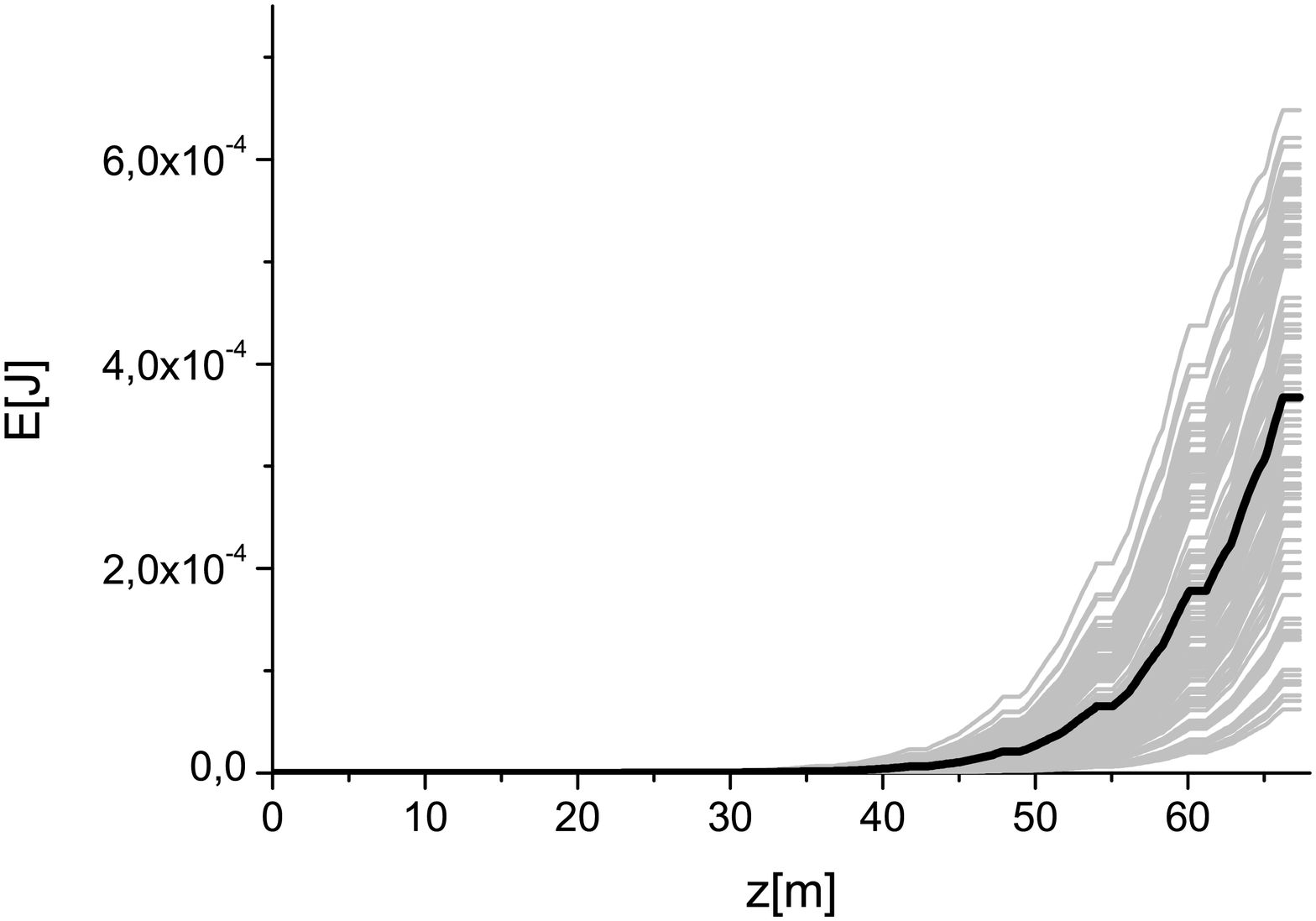}
\includegraphics[width=0.5\textwidth]{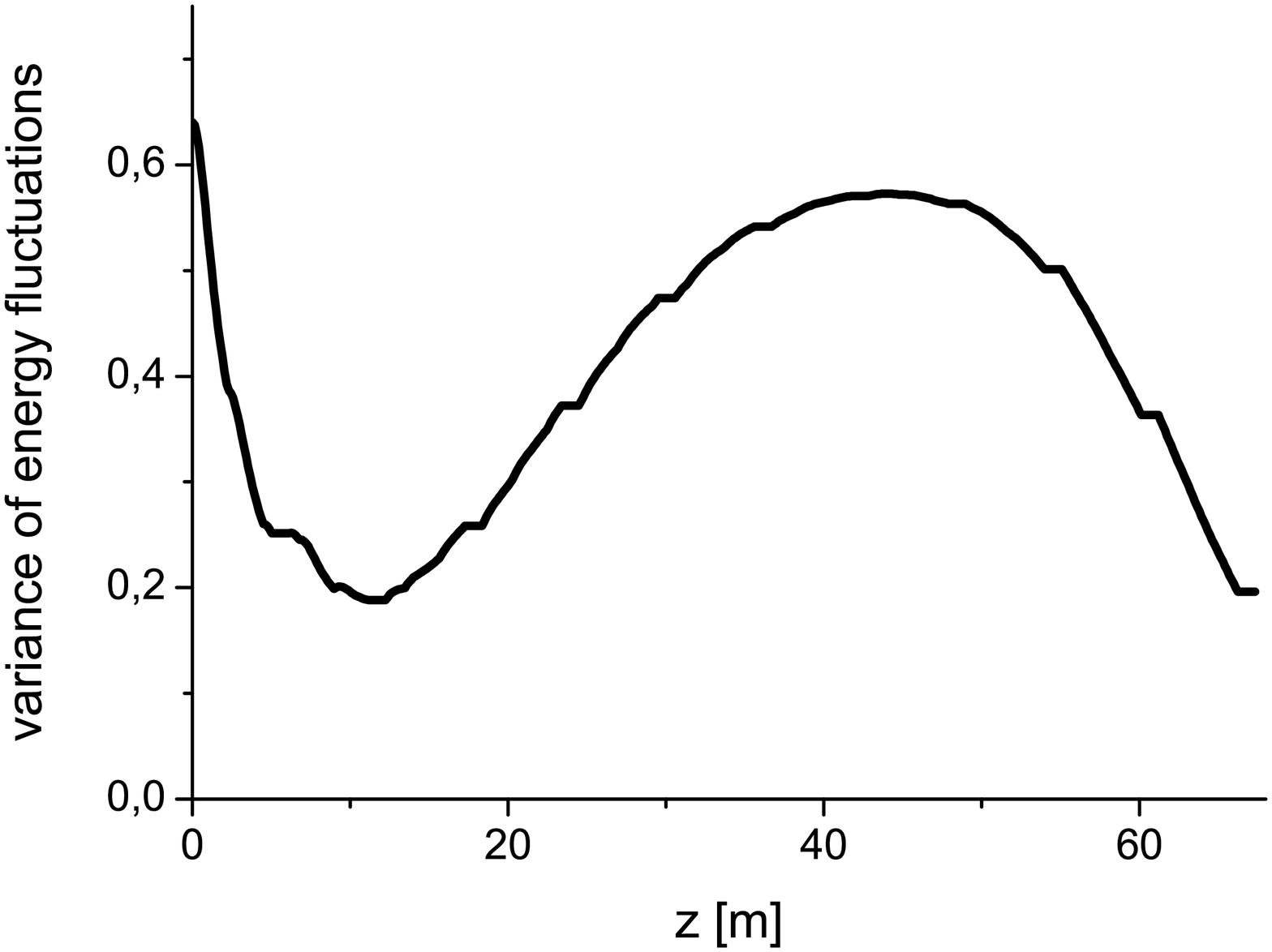}
\caption{Energy and energy variance of output pulses at saturation
for the case $\lambda = 0.15$ nm. In the left plot, grey lines refer
to single shot realizations, the black line refers to the average
over a hundred realizations.} \label{biofh1}
\end{figure}
The seed is amplified up to saturation in the output undulator. The
power and the spectrum of the output pulse at saturation are shown
in Fig. \ref{biofh2}, while the energy and the energy variance as a
function of the undulator length are plotted in Fig. \ref{biofh1}.
Saturation is reached after $11$ segments.

\begin{figure}[tb]
\begin{center}
\includegraphics[width=0.5\textwidth]{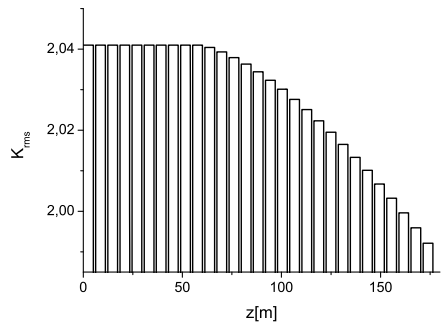}
\end{center}
\caption{Tapering law for the case $\lambda = 0.15$ nm.}
\label{biofh3}
\end{figure}
As already discussed, we can use post-saturation tapering to
increase the output power level. The tapering configuration in Fig.
\ref{biofh3} is optimized for maximum output power level. Note that
in this case tapering begins already after $10$ segments, while the
previously treated case with no tapering indicates that the optimal
output is reached after $11$ segments.

\begin{figure}[tb]
\includegraphics[width=0.5\textwidth]{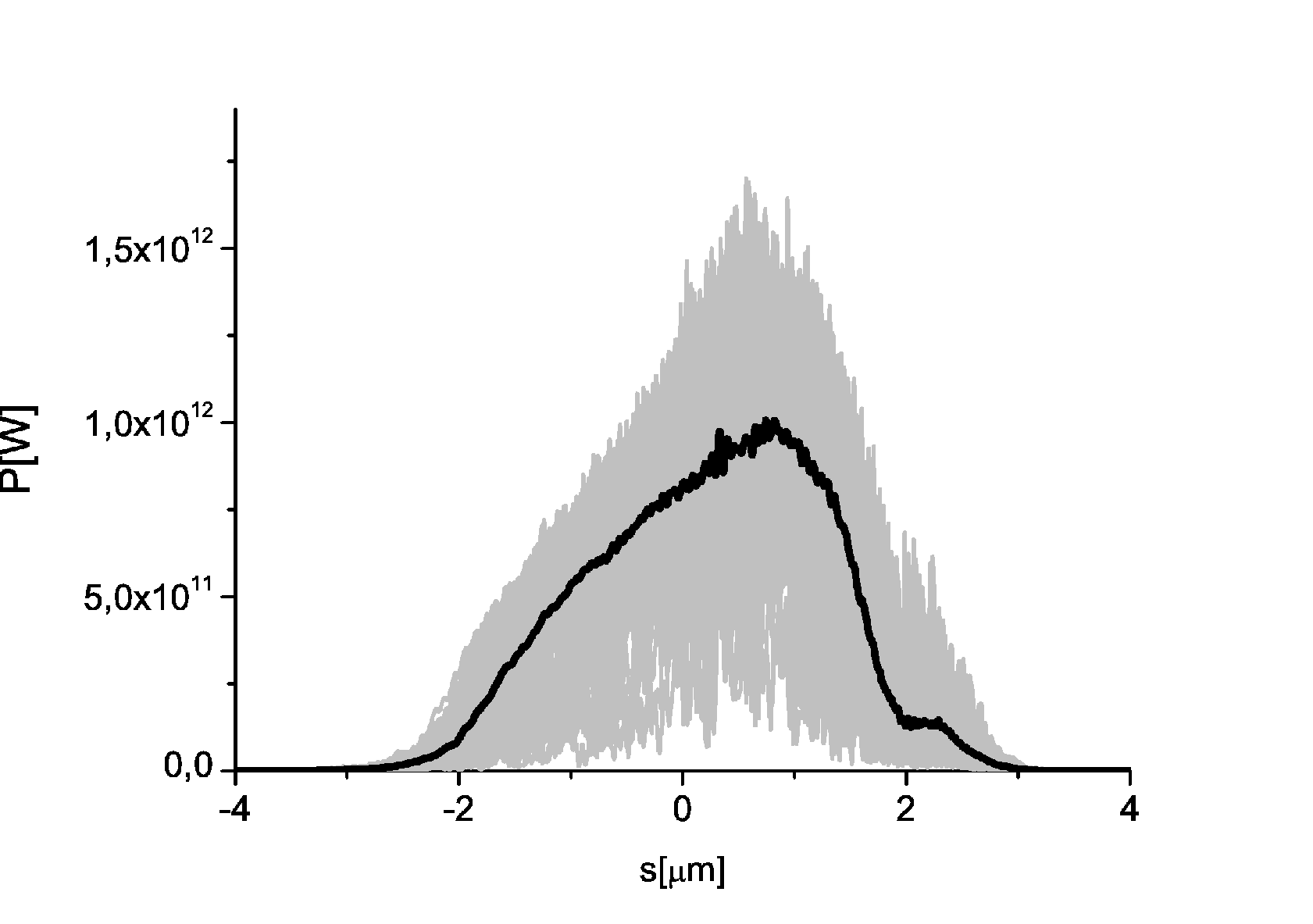}
\includegraphics[width=0.5\textwidth]{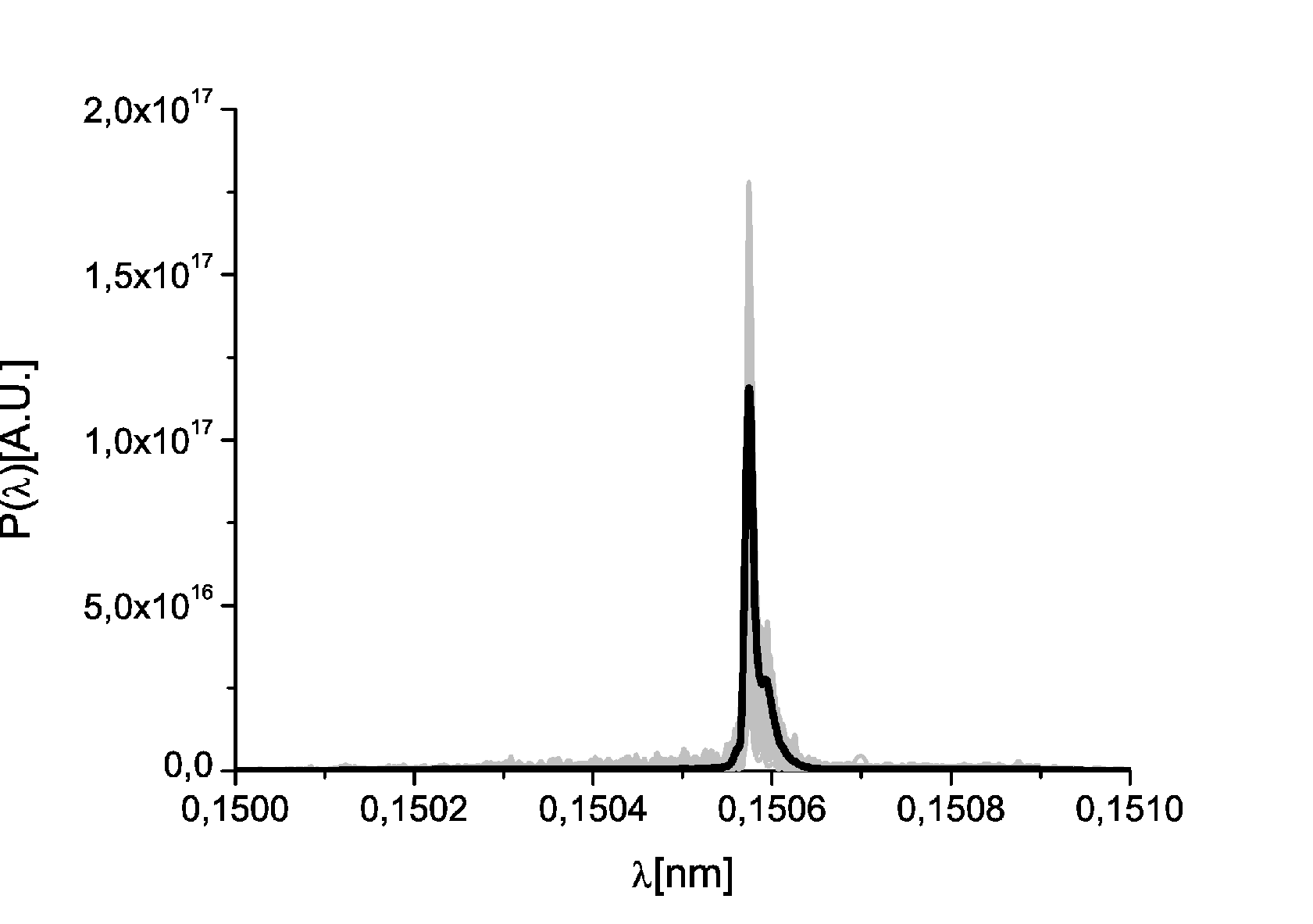}
\caption{Final output in the case of tapered output undulator for
$\lambda = 0.15$ nm. Power and spectrum are shown. Grey lines refer
to single shot realizations, the black line refers to the average
over a hundred realizations.} \label{biofh6}
\end{figure}

\begin{figure}[tb]
\includegraphics[width=0.5\textwidth]{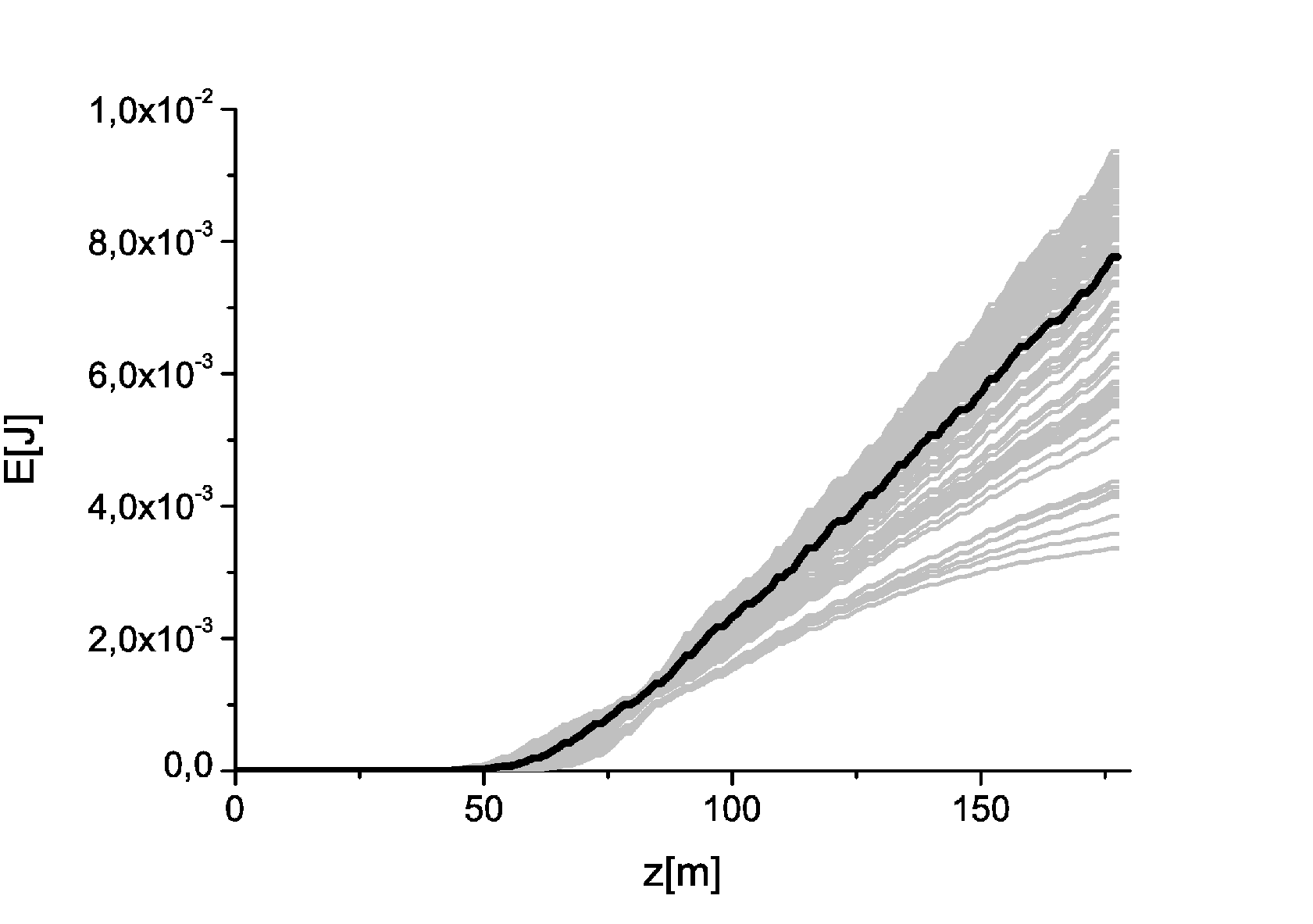}
\includegraphics[width=0.5\textwidth]{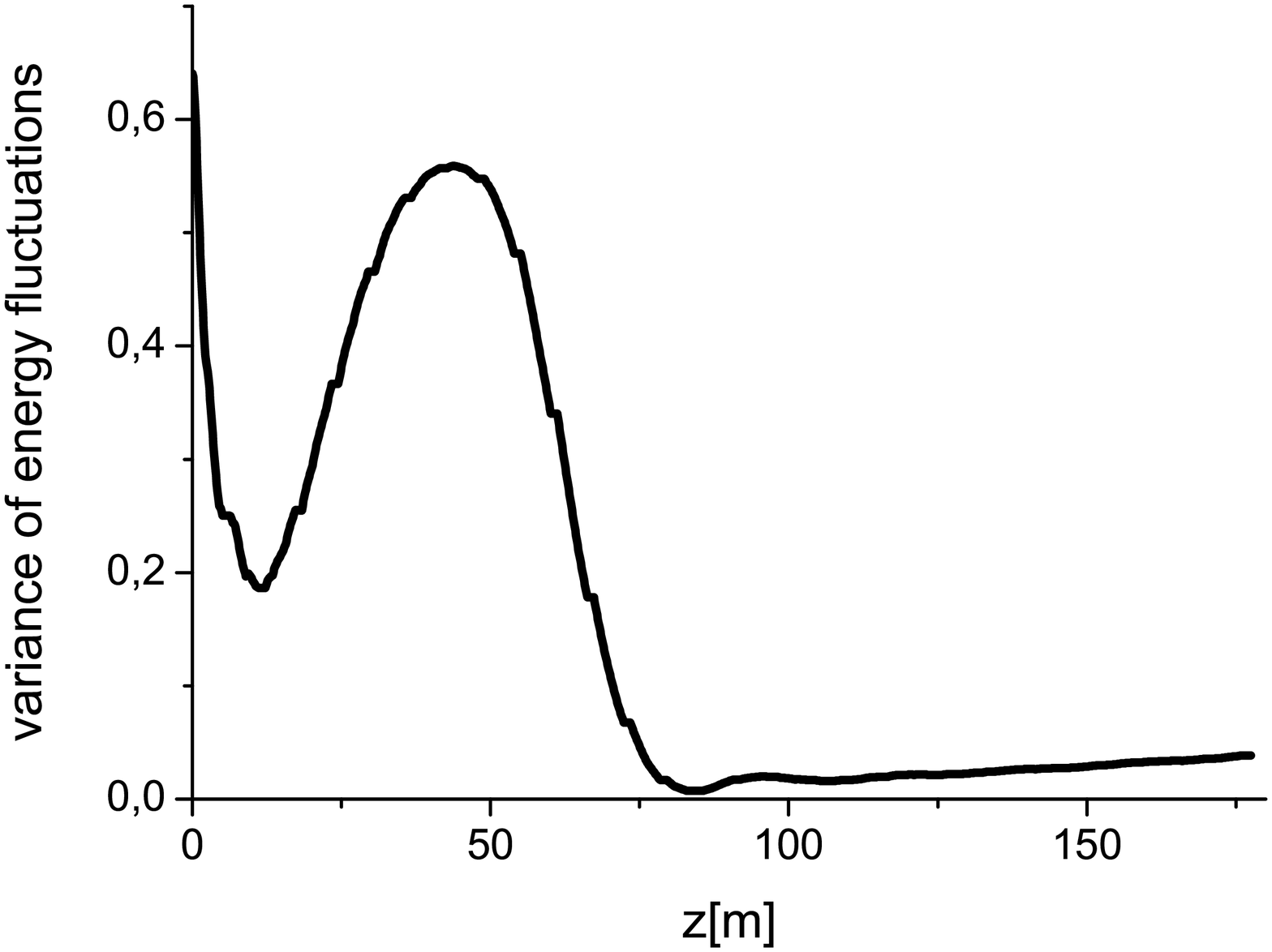}
\caption{Energy and energy variance of output pulses in the case of
tapered output undulator for $\lambda = 0.15$ nm. In the left plot,
grey lines refer to single shot realizations, the black line refers
to the average over a hundred realizations.} \label{biofh4}
\end{figure}

\begin{figure}[tb]
\includegraphics[width=0.5\textwidth]{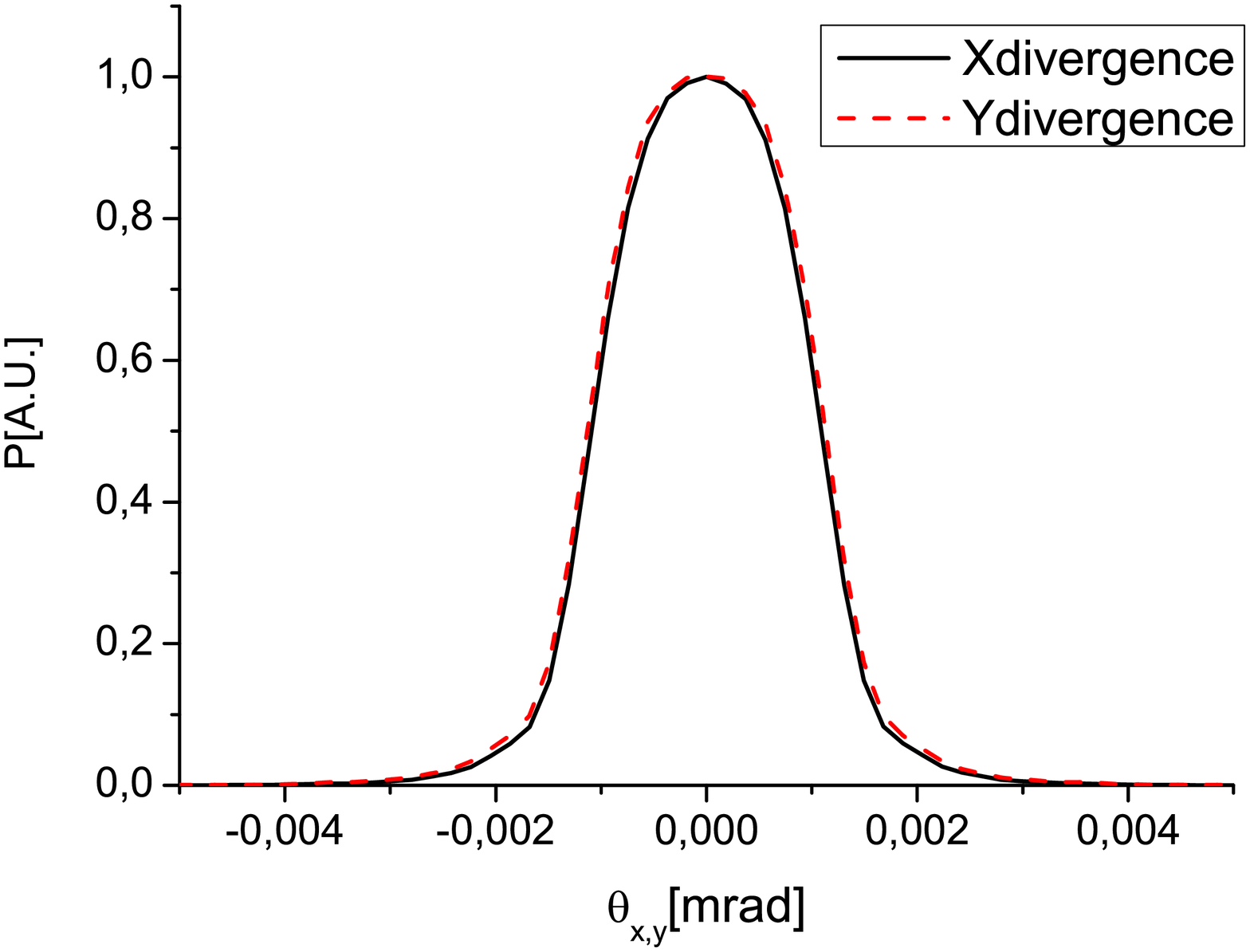}
\includegraphics[width=0.5\textwidth]{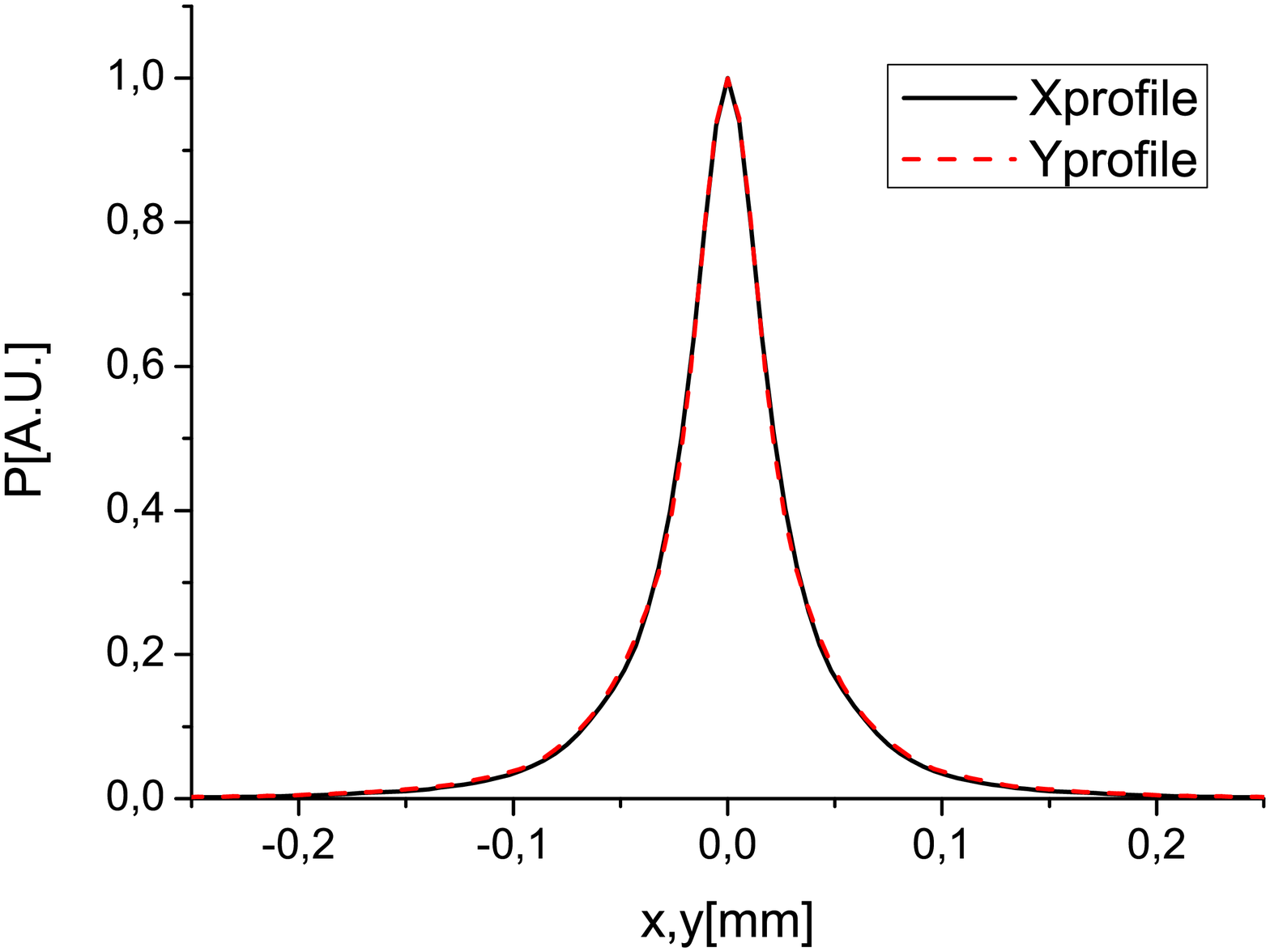}
\caption{Final output. X-ray radiation pulse energy distribution per
unit surface and angular distribution of the X-ray pulse energy at
the exit of output undulator for the case $\lambda = 0.15$ nm.}
\label{biofh5}
\end{figure}
The output characteristics, in terms of power and spectrum, are
shown in Fig. \ref{biofh6}. The output power is increased of about a
factor $20$, allowing one to reach about one TW. The spectral width
remains almost unvaried. The output level has been optimized by
changing the tapering law, resulting in Fig. \ref{biofh3}, and by
changing the electron beam transverse size along the undulator, as
suggested in \cite{LAST}, and shown in Fig. \ref{biofh7}.
Optimization was performed empirically. The evolution of the energy
per pulse and of the energy fluctuations as a function of the
undulator length are shown in Fig. \ref{biofh4}. Finally, the
transverse radiation distribution and divergence at the exit of the
output undulator are shown in Fig. \ref{biofh5}.

\section{\label{sec:conc} Conclusions}

This article describes a proposal for building a beamline dedicated
to bio-molecular imaging at the European XFEL. The present layout of
the European XFEL enables to accommodate such extension. The final
configuration foresees the new beamline located in the XTD4 tunnel,
which is now occupied by SASE3, while the SASE3 beamline can be
reinstalled inside the free XTD3 tunnel occupying the place instead
of the previously planned spontaneous emission undulator U1. This
final configuarion can be reached in two phases. During the low cost
 first phase, a soft X-ray self-seeding, a hard X-ray self-seeding
and a fresh bunch setup will be implemented at the SASE3 beamline in
XTD4. Initially, the SASE3 undulator (21 cells) can be kept as is,
and three additional magnetic chicanes respectively equipped with
grating monochromator, X-ray mirror delay line and crystal
monochromator will be installed as soon as possible. This first
phase constitutes a feasibility study of the system concept. Phase
$1$ experiments aim to demonstrate the combination of self-seeding,
fresh bunch and undulator tapering techniques at photon energy range
between $3$ keV and $5$ keV, and involve detailed benchmarking of
simulations, in order to provide confidence in extrapolating results
to higher power, that is to longer undulator length. During Phase
$1$, the peak power will be increased up to $1$ TW in the soft X-ray
range and up to $0.4$ TW at operation energies around $3$ keV. This
upgrade of the SASE3 beamline will fundamentally improve the
conditions for bio-imaging experiments. In particular, the concept
of Phase 1 upgrade permits single protein molecule structure
analysis in the most preferable photon energy range between $3$ keV
and $5$ keV. However, during Phase $1$, the maximum peak power in
the hard X-ray range will still be smaller than $0.1$ TW. In order
to obtain $1$ TW power in the photon energy range between $8$ keV
and $13$ keV the installation of additional $19$ undulator cells is
necessary, Phase 2. With this installation the bio-imaging beamline
will be optimized from the water window ($0.3$ keV) to K-edge of
Selenium ($13$ keV) and the power will be increased to $1$ TW in the
entire photon energy range. In a final step, a beamline identical to
SASE3 can be restored in XTD3.

The proposed upgrade program gives the possibility to build a
beamline optimized for life science maintaining a world-leading
position of the European XFEL in this field, by focusing
experimental activities on studies taking advantage of single
biomolecule imaging.

%
%\begin{figure}[tb]
%\includegraphics[width=1.0\textwidth]{pft6a.eps}
%\caption{Geometry of diffraction grating scattering.} \label{grat2}
%\end{figure}
%%

\section{Acknowledgements}

We are grateful to Massimo Altarelli, Reinhard Brinkmann, Henry
Chapman, Janos Haidu, Viktor Lamzin, Serguei Molodtsov and Edgar
Weckert for their support and their interest during the compilation
of this work.

\end{document}